\documentclass{IEEEtran}
\usepackage{amsmath,amssymb,graphicx,eucal,enumerate}%,doublespace,mathcomsteP}
\usepackage{pst-plot}
\usepackage{epsfig}
\usepackage{amsthm}
\usepackage{amsfonts}
\usepackage{subfigure}
\def \EE{\mathbb{E}}

\def \RR{\mathbb{R}}
\def \CC{\mathbb{C}}

\newcommand{\p}[1]{\begin{array}{l l l l l l l l l l} #1 \end{array} }

\newcommand{\ea} [1]{\begin{align}  #1 \end{align} }
\newcommand{\ean}[1]{\begin{align*} #1 \end{align*}}
\newcommand{\eas}[2]{\begin{subequations}\begin{align} #1 \end{align}{{#2}}\end{subequations}}
\newcommand{\htt}{\widetilde{h}}

\newcommand{\var}{{\rm Var}}
\newcommand{\cov}{{\rm Cov}}

\newcommand{\wt}{\widetilde{w}}

\newcommand{\bt}{\widetilde{b}}
\newcommand{\Xt}{\widetilde{X}}
\newcommand{\Zt}{\widetilde{Z}}
\newcommand{\Yt}{\widetilde{Y}}
\newcommand{\Pt}{\widetilde{P}}
\newcommand{\Sm}{\mathbf{S}}

\newcommand{\jj}{\mathrm{j}}
\newcommand{\eu}{\mathrm{e}}
\newcommand{\sgs}{\sigma^2}

\newcommand{\al}{\alpha}
\newcommand{\be}{\beta}
\newcommand{\ep}{\epsilon}
\newcommand{\f}[2]{\frac {#1} {#2}}

\newcommand{\lb}{\left(}
\newcommand{\rb}{\right)}
\newcommand{\lsb}{\left[}
\newcommand{\rsb}{\right]}
\newcommand{\lcb}{\left\{ }
\newcommand{\rcb}{\right\} }
\newcommand{\rnone}{\right. }
\newcommand{\lnone}{\left.  }

\newcommand{\labs}{\left|}
\newcommand{\rabs}{\right|}
\newcommand{\Cc}  {\mathcal{C}}
\newcommand{\Nc}  {\mathcal{N}}
\newcommand{\Ccal}{\mathcal{C}}
\newcommand{\Rcal}{\mathcal{R}}
\newcommand{\lag}{\quad}

\newtheorem{thm}{Theorem}[section]
\newtheorem{cor}[thm]{Corollary}

\theoremstyle{remark}
\newtheorem{rem}[thm]{Remark}
\theoremstyle{definition}

\begin{document}

%New Inner and Outer Bounds for the Interference Channel with a Cognitive Relay and Capacity in a Subset of the Strong Interference Regime
\title{On the Capacity of the Interference Channel with a Cognitive Relay}

\author{\IEEEauthorblockN{%
Stefano Rini\IEEEauthorrefmark{2},
Daniela Tuninetti\IEEEauthorrefmark{1},
Natasha Devroye\IEEEauthorrefmark{1}
and Andrea Goldsmith\IEEEauthorrefmark{2},\\}
\IEEEauthorblockA{%
\IEEEauthorrefmark{1}
Department of Electrical and Computer Engineering,
University of Illinois at Chicago, Chicago, IL 60607, USA,
Email: {\tt \{danielat, devroye\}@uic.edu} \\}
\IEEEauthorblockA{%
\IEEEauthorrefmark{2}
Department of Electrical Engineering,
Stanford University, Stanford, CA 94305, USA,
Email: {\tt \{stefano, andrea\}@wsl.stanford.edu}
}}

\maketitle

\begin{abstract}
The InterFerence Channel with a Cognitive Relay (IFC-CR) consists of the classical
interference channel with two independent source-destination pairs whose communication
is aided by an additional node, referred to as the cognitive relay,
that has \emph{a priori knowledge} of both sources' messages.
This a priori message knowledge is termed \emph{cognition} and idealizes the relay
learning the messages of the two sources from their transmissions over a wireless channel.
%In a wireless network,  the relay may overhear the sources' transmission : such ability is referred to as \emph{cognition}.
%
%This model addresses the rate advantages provided by cognition
%in a communicating network in the the idealized it allows full message knowledge at the relay.
%
%%Much remains unknown about the general IFC-CR whose capacity region would elucidate the role of cognition of two messages and how these should best be %%simultaneously used to aid two independent source-destination pairs.
%This channel has been little studied in the past years and capacity results for certain
%classes of channels have been proved.
%
This paper presents new inner and outer bounds for the capacity region of the general memoryless IFC-CR
that are shown to be tight for a certain class of channels.
%
%Previously proposed outer bounds are valid for Gaussian channels only or cannot be easily
%expressed in a single letter form.
%
The new outer bound follows from arguments originally devised for broadcast channels among which
Sato's observation that the capacity region of channels with non-cooperative receivers only depends
on the channel output conditional marginal distributions.
%The derived bound has a simple single letter expression.
%
The new inner bound is shown to include all previously proposed coding schemes and it is thus the largest known achievable rate region to date.
The new inner and outer bounds coincide for a subset of channel satisfying a \emph{strong interference} condition.
For these channels there is no loss in optimality if both destinations decode both messages.
This result parallels analogous results for the classical IFC and for the cognitive IFC
%is analogous to the ``very strong interference''  capacity
%result for the classical interference channel and for the cognitive interference channel,
and is the first known capacity result for the general IFC-CR.
Numerical evaluations of the proposed inner and outer bounds are presented for the Gaussian noise case.
\end{abstract}

\begin{IEEEkeywords}
Capacity;
Inner bound;
Interference channel with a cognitive relay;
Outer bound;
Strong interference;
Weak interference;
\end{IEEEkeywords}

\section{Introduction}
\label{sec:into}
%Cognition is a rapidly emerging new paradigm in wireless communication
%whereby a node changes its communication scheme to
%efficiently share the spectrum with other users in the network.
%Cooperation among smart and well-connected wireless devices has been
%recognized as  a key factor in improving the spectrum utilization
%and throughput of wireless networks~\cite{goldsmith_survey}.

\IEEEPARstart{T}{he} information theoretic study of cognitive networks -- networks in which a subset of the nodes has a priori knowledge of the messages of other subsets of nodes --  has focused mostly on the two user Cognitive InterFerence Channel (CIFC), i.e., a variation of the classical two-user IFC where one of the transmitters has \emph{non-causal a priori knowledge} of both messages to be transmitted.  While idealistic, this form of \emph{genie-aided} cognition has provided significant insights of the rate advantages obtainable through asymmetric or unilateral transmitter cooperation (please refer to~\cite{RTDjournal1} and~\cite{RTDjournal2}, and references therein, for an extensive summary of available results for the general and Gaussian CIFC, respectively).

In this paper we study a natural extension of the CIFC where the genie-aided cognition, instead of
being provided to only one of the sources of the IFC, is rather provided to a third node, referred to a the \emph{cognitive relay}, that aids the communication between both source-destination pairs.
One of the key challenges of this model is the issue of interference management at the cognitive relay.
Unlike in the Broadcast Channel (BC) and the CIFC, the cognitive relay in an IFC-CR has knowledge of the interference seen at each destination but has no control over the interfering signals that are sent by the sources.
Gel'fand-Pinsker binning~\cite{GamalArefSemiDetRelayChannel}, or Dirty Paper Coding (DPC) for Gaussian channels~\cite{costa1983writing}, is a celebrated well-known technique used to mitigate  interference known non-causally at a source through proper pre-coding of the message.  This strategy is known to be capacity achieving for certain classes of BCs and CIFCs.
In the IFC-CR, the cognitive relay
%has knowledge of the interference experienced at both decoders but cannot control the interfering signals sent by the sources. It
can only manage the interference experienced by the destinations through its own transmissions, begging the question of how this single transmission may best be used to simultaneously aid both source-destination pairs.

The IFC-CR model encompasses many previously studied multi-terminal networks as special cases: the BC, the classical IFC and the CIFC, none of whose capacity is known in general.
The generality of the IFC-CR model suggests a certain level of complexity in the analytical results, but also allows one to study whether and how results available for smaller networks may be incorporated into larger networks.
%The generality of this model implies a certain level complexity in the analytical results but also it offers the possibility to study how results available for smaller networks scale in larger models.
%
%Inner and outer bound for the IFC-CR often require a large number of equations and auxiliary random variables.
For instance, the derivation of inner and outer bounds for the general memoryless IFC-CR carefully combines ideas developed for simpler networks, such as Gel'fand-Pinsker binning and genie-aided outer bounds, adjusted to this more general network setting.  We seek to determine whether these extensions of previously proposed techniques to our more general channel is sufficient to achieve capacity (we answer this in the positive for a subset of the strong interference regime) or whether our model is sufficiently different such that it requires  new  transmission techniques to achieve capacity.

%new subtleties, perhaps necessitating new techniques, may arise.
%
%For this reason the study of the capacity of this channel would clearly indicate which how a particular network architecture compares to the theoretical %performance limits.
%
%
%This channel models the situation where two base station are transmitting to two users that lie in the transmission range of both base stations.
%In contemporary communication systems it is not possible to perform interference management  strategies but interference avoidance.
%As the number of users in the network increases, one can expect that interference avoidance becomes less and less viable.
%
%
%We consider the case where a cognitive relay is placed in the network to improve the performance of the system.
%The cognitive relay is connected with a high speed link to the base stations and can therefore have knowledge of the messages to be transmitted.
%
%To  understand the possible performance improvements of this architecture we take an information theoretic perspective on the system an study a statistical model of the system, determining inner and outer bound for the maximum achievable rates.
%
%We determine capacity in the very strong interference regime, a subspace a the strong interference regime. In the strong interference regime there is no loss of generality in having both decoders decode both messages.

\subsection{Past Work}
The information theoretic capacity of the general memoryless IFC-CR remains an open problem for the general case.
The IFC-CR was initially considered in~\cite{Sahin_2007_2} where the first achievable rate region was proposed, and was later improved upon in~\cite{sridharan2008capacity} for the Single-Input Single-Output (SISO) Gaussian channel. The authors of~\cite{sridharan2008capacity} also provided a sum-rate outer bound for the Gaussian channel based on an outer bound for the Multiple-Input Multiple-Output (MIMO) Gaussian CIFC.
%and, in general, has no closed form expression.
%
In~\cite{jiang-achievable-BCCR}%
\footnote{The authors of~\cite{jiang-achievable-BCCR} refer to the
IFC-CR as ``broadcast channel with cognitive relays'', arguing that the
model can also be obtained by adding two partially cognitive relays to
a broadcast channel.}
a general achievable rate region was derived that contains all previously known
achievable rate regions in~\cite{Sahin_2007_2,sridharan2008capacity}.
The first outer bound for a general (i.e. not necessarily Gaussian) IFC-CR was derived in~\cite{rini2010dublin} by using Sato's observation that the capacity region of channels with non-cooperative receivers depends only on the conditional marginal distribution of the channel outputs~\cite{sato1978outer}.
This general Sato-type outer bound was further tightened in~\cite{rini2010dublin} for a class of semi-deterministic channels in the spirit of~\cite{telatar2008bounds}.
%a model originally proposed in~\cite{etkin_tse_wang,elgamal_det_IC} for the classical IFC,
For the special case where the sources do not interfere at the non-intended destinations,
the tightened bound of~\cite{rini2010dublin} was shown to be capacity for the deterministic approximation of the Gaussian IFC-CR at high-SNR~\cite{Avestimehr:2007:allerton} and to be optimal to within 3~bits/sec/Hz for any finite SNR~\cite{us3channel}.
Furthermore, for a subset of parameters akin to the weak interference regime for the classical IFC, the tightened bound of~\cite{rini2010dublin} was shown to be capacity for the general deterministic approximation of the Gaussian IFC-CR at high-SNR; the achievability in this case suggests an interesting transmission strategy where the cognitive relay is able to ``pre-cancel'' the interference at both destinations simultaneously.

The channel model under consideration in this work is closely related to the interference relay channel: an IFC with an additional relay node which does not have a priori knowledge of the sources' messages, but rather learns these messages over the noisy channel between the sources and the relays~\cite{Sahin_2007_1}. Although more realistic than the IFC-CR considered here, the interference relay channel is harder to study due to the causal cognition. Recently new results were derived for the interference relay channel where the relay is assumed to operate {\em out-of-band}~\cite{TianOutOfBandRelay,SahinOutOfBandRelay}, i.e., a model in which the link between the relay and the destinations does not interfere with the underlaying IFC between the sources and the destinations; in this case, capacity is known to 1.15~bits/s/Hz in the symmetric Gaussian noise case~\cite{TianOutOfBandRelay}.

The IFC-CR subsumes several well studied channel models as special case.
The CIFC,%
\footnote{The CIFC has also been referred to as the cognitive channel~\cite{devroye_IEEE}, an interference channel with  ``unidirectional cooperation''~\cite{MaricUnidirectionalCooperation06} and an interference channel with  ``degraded message sets''~\cite{WuDegradedMessageSet}.}
that is, an IFC in which one transmitter has non-causal a-priori knowledge of the messages of both transmitters, may be obtained from the IFC-CR by eliminating the channel input of one of the sources.
The CIFC was first considered from an information theoretic perspective in~\cite{devroye_IEEE}, where the channel was formally defined and the first achievable rate region was obtained. The largest known achievable rate region is due to Rini {\em et al.}~\cite{rini2009state,RTDjournal1} and the tightest outer bound to Maric {\em et al.}~\cite{MaricGoldsmithKramerShamai07Eu}. Capacity has been established for channels with ``very weak interference'' in which (in Gaussian noise) treating interference at the primary receiver as noise is optimal~\cite{WuDegradedMessageSet, JovicicViswanath06}, for the ``very strong interference'' regime, where without loss of optimality both receivers can decode both messages and the cognitive channel reduces to a compound Multiple Access Channel (MAC)~\cite{MaricUnidirectionalCooperation06}, for the ``better cognitive decoding'' regime~\cite{Rini:Allerton2010,RTDjournal2}
%{\red DT: ADDED:
where the cognitive receiver can decode both messages without loss of optimality, for the semi-deterministic CIFC~\cite{Rini:ICC2010,RTDjournal1}
%{\red  DT: ADDED:
where a BC-type coding scheme is optimal, and for certain Gaussian CIFC without interference at the primary decoder~\cite{RTD-ISIT2011-Zchannel,RTDjournal2}. For the general Gaussian CIFC capacity is known to within 1~bit/s/Hz and to within a factor~2 regardless of channel parameters~\cite{Rini:ICC2010,RTDjournal2,Vaezi2011Superposition}.
%More complete surveys of this channel model are available in~\cite{RTDjournal1} for the general discrete memoryless CIFC, and in~\cite{RTDjournal2} for the Gaussian CIFC.

%{\red DT: ADDED:
The classical BC can be obtained from the IFC-CR by eliminating the channel inputs of both sources.  The capacity of the general BC is unknown.  The largest known achievable rate region is due to Marton~\cite{MartonBroadcastChannel} and the tightest outer bound to Nair and El Gamal~\cite{nair2007outer}. In all cases where capacity is known Marton's region is optimal (see~\cite{wang2010capacity} and references therein for an extensive discussion of all cases where capacity is known and for the challenges in determining capacity in the open cases).  Many techniques originally developed for the BC will prove useful for the derivations in this work.

Finally, the classical IFC can be obtained from the IFC-CR by eliminating the channel input of the cognitive relay. The largest known achievable rate region is due to Han and Kobayashi~\cite{Han_Kobayashi81}, which is optimal in all cases where capacity is known (see~\cite{Motahari09CapacityBounds} and references therein for an extensive discussion of all cases where capacity is known). In Gaussian noise, capacity is known only in strong interference~\cite{sato2002capacity,carleial1978interference,elgamal_det_IC} and known otherwise to within 1~bit/s/Hz~\cite{etkin_tse_wang}. Some techniques originally developed for the IFC, such as rate splitting and simultaneous decoding, will be adapted to the IFC-CR model in this work.

%}

%For the general memoryless cognitive interference channel the capacity region for both the discrete as well as Gaussian noise channel, remains unknown. Tools such as rate-splitting, binning, cooperation and superposition coding have been used to
%derive different achievable rate regions.
%%
%The authors of~\cite{MaricGoldsmithKramerShamai07Eu} proposed an achievable rate region that encompasses all the previously proposed inner bounds and derived a new outer bound  using an argument originally devised for the broadcast channel in~\cite{NairGamal06}.
%%
%A further improvement of the inner bound of~\cite{MaricGoldsmithKramerShamai07Eu} is provided in~\cite{cao2009interference} where the authors include a new feature in the transmission scheme allowing the cognitive transmitter to broadcast part of the message of the primary pair.
%This broadcast strategy is also encountered in the scheme derived in~\cite{jiang-achievable-BCCR} for the  broadcast channel with cognitive relays, which contains the cognitive interference channel as special case.

\subsection{Paper Main Contributions}

In this paper we determine:

\smallskip
\noindent 1) Outer Bound:

\begin{enumerate}[a)]
\item
{\bf Sato-type outer bound.} % in Section~\ref{subsec:satoout}.

This outer bound uses Sato's observation~\cite{sato1978outer}
that the capacity of a channel with non-cooperative receivers only depends on the channel
output conditional marginal distributions.  This bound does not contain any auxiliary random
variables and is thus computable in principle by determining the optimal distribution of the channel inpus .

\item
{\bf BC-type outer bound.} %in Section~\ref{subsec:bcout}.

This outer bound generalizes the tightest known outer bound for the general CIFC by Maric {\em et al.}~\cite{MaricGoldsmithKramerShamai07Eu}
to the general IFC-CR. It uses a technique originally developed to prove the converse for the ``more capable'' BC in~\cite{ElGamal1977capacity}
and later generalized to obtain an outer bound for the general BC in~\cite{nair2007outer}.
This BC-type outer bound is the tightest known to date for the general IFC-CR.
It is however expressed as a function of three auxiliary random variables
for which no cardinality bound exists on the corresponding alphabets.

%The first BC-type outer bound is inspired by an argument originally devised for the ``more capable''
%BC in~\cite{ElGamal1977capacity}, later generalized to the general BC in~\cite{nair2007outer},
%also used in deriving the capacity of the CIFC in ``weak interference''~\cite{WuDegradedMessageSet}.
%
%The second BC-type outer bound generalizes the first in the same way~\cite{MaricGoldsmithKramerShamai07Eu} generalized~\cite{WuDegradedMessageSet}.
%{\red DT: IS IT TRUE?  This latter is the tightest known outer bound for the general IFC-CR.}

\item
{\bf A simplification of the BC-type outer bound in the ``strong interference'' and ``weak interference'' regimes.} %in Section~\ref{subsec:outsimplification}.

The ``strong interference'' regime is defined as the regime where, loosely speaking, the
non-intended destination can decode more information than
the intended destination even after having removed the interfering signal.
This regime parallels the ``strong interference'' regime for the
IFC~\cite{CostaElGamal87} and for the CIFC~\cite{maric2005capacity}.

The ``weak interference'' regime is defined as the regime in which, loosely speaking,
treating interference as noise is optimal. This regime parallels the ``weak interference'' regime for the
IFC in~\cite{annapureddy2009gaussian,shang2009new,Motahari09CapacityBounds} and for the CIFC in~\cite{WuDegradedMessageSet}.

\end{enumerate}

\smallskip
\noindent 2) Inner Bound:
\begin{enumerate}[a)]

\item
{\bf Largest known inner bound.}

Our inner bound is shown to include all previously proposed inner bounds as special cases.
%The inner bound utilizes classical random coding techniques, such as rate splitting, superposition coding, Gel'fand Pinsker binning,  and simultaneous decoding.
%
This region equals the capacity region when the channel reduces to a simpler model (i.e. BC, IFC and CIFC ) for which capacity is known.  The novel ingredients are a rate-split in four parts of the source messages and a very structured nesting of superposition and binning. Although the expression of the inner bound is rather involved, it provides a unifying framework to evaluate the effect of different transmission strategies on the achievable rate region.

\item
{\bf The Fourier-Motzkin elimination of the proposed inner bound in several sub-cases.}
%{\bf closed form expressions for several simple achievable schemes.}

The Fourier-Motzkin elimination of our general inner bound region appears difficult to reduce to a manageable number of rate bounds. We therefore proceed to analyze several simpler achievability schemes. Besides being of use in numerically evaluating regions, the simpler regions are extensions of regions known to achieve capacity when the channel reduces to an IFC or a CIFC.
%From the Fourier-Motzkin elimination of these schemes we point out the difficulties of extending known capacity results from the simpler channel models subsumed by the general IFC-CR.

\end{enumerate}

\smallskip
\noindent 3) Capacity:

\begin{enumerate}[a)]
\item
{\bf Capacity in the ``very strong interference'' regime at one destination.}

This is a subset of the ``strong interference'' regime under which our general BC-type outer bound can be simplified.
In this regime both decoders can, without loss of optimality, decode both messages as in a compound MAC.
The ``strong interference'' outer bound may be achieved using superposition coding without rate splitting or binning.

\item
{\bf Capacity in the ``strong interference at both receivers'' regime.}

A corollary of the previous capacity result where both destinations
experience ``very strong interference''.

\end{enumerate}

\smallskip
\noindent 4) Gaussian Channels:

\begin{enumerate}[a)]

\item {\bf Capacity in the ``very strong interference'' regime at one destination and in the ``strong interference at both receivers'' .}

We determine the set of channel coefficients that satisfy the  condition of ``very strong interference'' at one destination and of ``very strong interference'' at both destinations, thereby establishing capacity in these cases.

\item {\bf Outer bound for the degraded IFC-CR.}

For a special class of channels that satisfies the ``weak interference'' condition  under which our general BC-type outer bound could be simplified,
we evaluate the outer bound in closed form. Unfortunately, we have not been able to find a transmission scheme that achieves this outer bound yet.

\item {\bf Numerical evaluations of the proposed simpler achievable rate regions.}

These evaluations %illustrate  the region (in terms of channel parameters) which falls under the ``very strong interference'' regime in which we have derived capacity. They also
visually illustrate the relationships between the derived inner and outer bounds for the cases where capacity is open.

\end{enumerate}

\subsection{Paper Organization}
In Section~\ref{sec:model} we formally define the general memoryless IFC-CR.
In Section~\ref{sec:OuterBound} we proceed to derive our new outer bounds, two of which hold in general, and two of which are valid under ``strong interference'' and ``weak interference'' conditions, respectively.
In Section~\ref{sec:InnerBound} we derive a general achievable rate region for the IFC-CR and analytically show
that this contains all other known inner bounds; we further simply our general inner bound in a number of simpler sub-cases with a limited number of auxiliary random variables and rate splits.
In Section~\ref{sec:Capacity} we prove capacity for the IFC-CR in the ``very strong interference'' regime; this is the first general capacity result for the IFC-CR and parallels results for similar regimes for the IFC and the CIFC.
In Section~\ref{sec:The Gaussian Case} we numerically illustrate the ``very strong interference'' capacity region and the ``weak interference'' outer bound for the Gaussian IFC-CR, as well as numerical results comparing several of the simplified inner bounds.
We conclude the paper in Section~\ref{sec:Conclusion and Future Work}.

\section{Channel Model}
\label{sec:model}

\begin{figure}
\centering
\includegraphics[width= 8 cm]{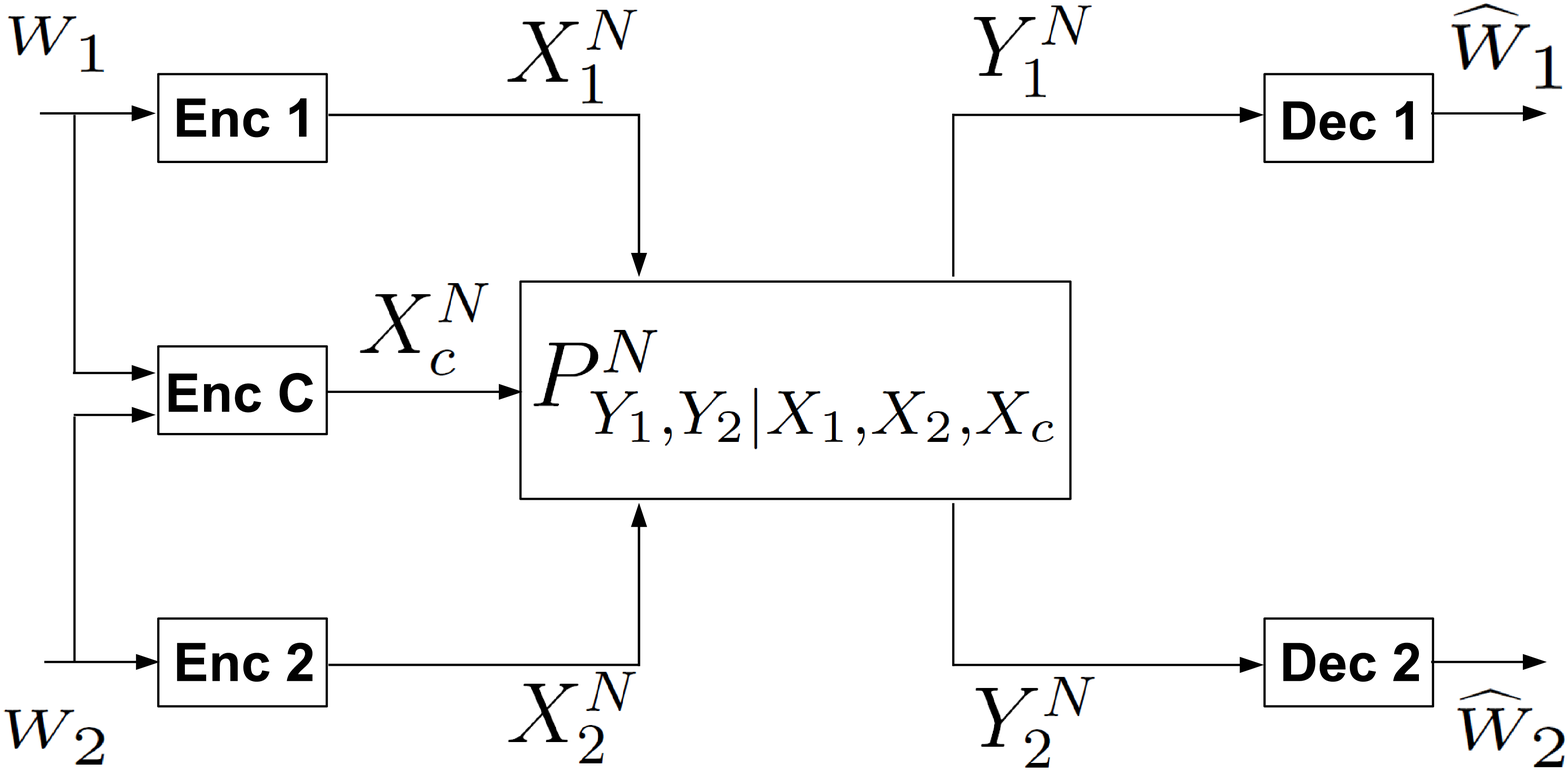}
\caption{The general memoryless IFC-CR channel model.}
\label{fig:channel}
\end{figure}

We consider the channel model depicted in Fig.~\ref{fig:channel}.
In the IFC-CR the transmission of the two independent messages
$W_{i}$ uniformly distributed on $[1:2^{N R_i}]$, $i \in \{1,2\}$,
block-length $N\in {\mathbb Z}^+$, and rates $R_i \in \mathbb{R}^+$,
is aided by a single {\em cognitive  relay}, whose input to the channel has subscript $c$.
%The notation $P^{N}_{Y_1,Y_2|X_1,X_2,X_c}$ represents the $N$-fold
We define $X_{i,n}$ and $Y_{i,n}$ to be the input and output of the channel for the $i$-th  source-destination pair  at the $n$-th  channel use, $i \in \{1,2\}$, $n \in [1:N]$,
and define $X_{i,j}^k : = [X_{i,j}, X_{i,j+1}, \cdots , X_{i,k}]$ for $k\geq j$, and similarly for $Y_{i,j}^k$.
%To reduce notation slightly, we also define  $X_{i}^k : = X_{i,1}^k$ and $Y_{i}^k := Y_{i,1}^k$.
The channel %with input provided from the sources and the cognitive relay and outputs at the destinations
is assumed to be memoryless with transition probability $P_{Y_1,Y_2|X_1,X_2,X_c}$. Since the destinations do not cooperate, the capacity of the memoryless IFC-CR is only a function of the output conditional marginal distributions $P_{Y_1|X_1,X_2,X_c}$ and $P_{Y_2|X_1,X_2,X_c}$.
%The inputs $X_i$, $i \in \{1,2,c\}$, are from alphabets  $\Xc_i$ and the outputs $Y_i$, $i \in \{1,2\} $ from $\Yc_i$.
%We use the shorthand RV to denote a Random Variable.

A non-negative rate pair $(R_1, R_2)$ is said to be achievable if there exists a sequence of encoding functions
\ean{
X_1^N&=X_1^N(W_1), \\
X_2^N&=X_2^N(W_2), \\
X_c^N&=X_c^N(W_1,W_2),
}
and a sequence of decoding functions
\ean{
\widehat{W}_1&=\widehat{W}_1(Y_1^N), \\
\widehat{W}_2&=\widehat{W}_2(Y_2^N),
}
such that
\ean{
\lim_{N \to \infty } \max_{i\in\{1,2\}} \Pr \lsb \widehat{W}_i \neq W_i \rsb  =  0.
}
The capacity region is defined as the closure of the region of all
achievable $(R_1,R_2)$ pairs.

Note that the IFC-CR subsumes three well-studied channels as special cases:
\begin{itemize}
\item IFC:  for $X_c=\emptyset$,
\item CIFC: for $X_1=\emptyset$ or $X_2=\emptyset$, and
\item BC:   for $X_1=X_2=\emptyset$.
\end{itemize}

The capacity region of the general IFC-CR
is unknown in general.

\section{Outer Bounds}
\label{sec:OuterBound}

\begin{table*}
\centering
\caption{The outer bounds presented in this work and their relationship to underlying simpler channels where capacity is known.}
\label{table:outer}

\begin{tabular}{|l|l|l|l|}
\hline
Outer bound and Theorem in this work & Capacity result & Reference \\
\hline
\hline
Sato-type outer bound %, Thm.~\ref{thm:general outer ITW dublin}
 & ``strong interference'' IFC-CR & Thm.~\ref{th:ifc-cr very strong int} \\
 & Gaussian ``strong interference'' CIFC  &\cite[Thm.6]{maric2005capacity} \\
 & Gaussian ``primary decodes cognitive'' CIFC &\cite[Thm.3.1]{Rini:Allerton2010} \\
 &  ``strong interference'' IFC &\cite{CostaElGamal87,sato2002capacity,carleial1978interference} \\
\hline
BC-type outer bound %s, Thm.~\ref{thm:more capable BC outer} and Thm.~\ref{thm:generalBC outer}
 & ``very weak interference'' CIFC    &\cite[Thm. 3.2]{WuDegradedMessageSet}   \\
 &  ``better cognitive decoding'' CIFC & \cite[Thm. 7.1]{RTDjournal2} \\
 & semi-deterministic CIFC & \cite[Thm. 8.1]{RTDjournal2}\\
 & more capable BC        & \cite[Sec. 3]{ElGamal1977capacity} \\
 & semi-deterministic BC & \cite{gelfand_pinsker_BC} \cite{MartonBroadcastChannel}\\
 %S. I. Gelfand and M. S. Pinsker, ÒCapacity of a broadcast channel with one deterministic component,Ó Probl. Inf. Transm., vol. 16, no. 1, pp. 24Ð34, 1980
\hline
%\hline
%``Strong interference'' outer bound, Cor.~\ref{cor:strong int outer bound}
% & ``Very strong interference'' IFC-CR & This paper Thm.~\ref{th:ifc-cr very strong int} \\
% & {\red DT: WHY? Gaussian} ``very strong interference'' CIFC  &\cite[Thm.6]{maric2005capacity} \\
%\hline
%``Weak interference'' outer bound, Cor.~\ref{cor:weak int outer bound} & ``Very weak interference'' CIFC &\cite{WuDegradedMessageSet} \\
%\hline
%BC-type outer bound, Thm.~\ref{thm:generalBC outer} & ``less noisy'' {\red more capable} BC &~\cite{kornermarton} \\
%& ``weak interference'' CIFC &~\cite{WuDegradedMessageSet} \\
%IFC  & strong interference & $U_{1  {\rm c}}, U_{2  {\rm c}}$ &~\cite{sato2002capacity} \\
%IFC  & deterministic IFC & $U_{1  {\rm c}}, X_1, U_{2  {\rm c}}, X_2$ &~\cite{elgamal_det_IC} \\
%CIFC & Gaussian, weak interference & $X_2,U_{1  {\rm pb}}$ &~\cite{WuDegradedMessageSet} \\
%CIFC & very strong  interference & $U_{2  {\rm c}},U_{0  {\rm cb}}$  with $R_{2  {\rm cb}}=0$ &~\cite{maric2005capacity} \\
%CIFC & better cognitive decoding & $U_{2  {\rm c}},U_{0  {\rm cb}},U_{1  {\rm pb}}$ with $R_{2  {\rm cb}}=0$ &~\cite{RTDjournal1} \\
%CIFC & Gaussian, primary decodes cogintive regime & $X_2,U_{0  {\rm cb}}$  with $R_{2  {\rm cb}}=0$ &~\cite{Rini:Allerton2010} \\
%CIFC & Semi-deterministic CIFC & $X_2,U_{1  {\rm pb}},U_{2  {\rm pb}}$ &~\cite{RTDjournal1} \\
\end{tabular}
\end{table*}

%The first outer bounds for a general IFC-CR were derived in~\cite{rini2010dublin}; we re-state the most general outer bound of ~\cite{rini2010dublin} here for completeness.
In this section we present two new outer bounds which we term
the Sato-type and the BC-type outer bound.
The names of these bounds reflect the channels and/or techniques which inspired them.
We then proceed to simplify the expression of these bounds in
the ``strong interference'' and ``weak interference''  regime.%
\footnote{
We note that our naming convention is not entirely consistent
with past uses of the term ``strong/weak interference''. Here, as in
our previous work on the CIFC~\cite{RTDjournal1,RTDjournal2}, we use
``strong/weak interference'' to denote regimes inspired by similar
results for the IFC under which we may obtain either a tighter
or simpler outer bound for the channel of interest, and use
the terms ``very strong/very weak'' to denote regimes
in which additional conditions (therefore forming subsets of
the ``strong/weak'' regimes) are imposed on top of the
``strong/weak'' conditions that allow these outer bounds
to be achieved.
}
As the IFC-CR generalizes a number of multi-user channels such as the CIFC, the IFC and the BC, one expects techniques relevant in those channels to be of use in the IFC-CR, and conversely, the IFC-CR outer bounds should reduce to capacity of the simpler sub-channels when they are known. Indeed, our outer bounds generalize the underlying sub-channels, as shown in Table~\ref{table:outer}.

\subsection{Sato-type Outer Bound}
\label{subsec:satoout}

We start with the outer bound for the general IFC-CR first derived by the Rini, Tuninetti and Devroye in~\cite[Thm.3.1]{rini2010dublin}.   It uses Sato's argument~\cite{sato1978outer} that the capacity region of the IFC-CR only depends on the channel output conditional marginal distributions since the destinations do not cooperate.

\begin{thm}
%(Rini, Tuninetti and Devroye)
% of \cite[Thm.3.1]{rini2010dublin}}
\label{thm:general outer ITW dublin}
If $(R_1,R_2)$ lies in the capacity region of the IFC-CR, then the following must hold for
any $\widetilde{Y}_1$ and $\widetilde{Y}_2$ having the same conditional marginal distributions
as $Y_1$ and $Y_2$, respectively, but otherwise arbitrarily correlated:
\eas{
R_1    &\leq I(Y_1;X_1,X_c|X_2,Q),\label{eq:gen sato r1}\\
R_2    &\leq I(Y_2;X_2,X_c|X_1,Q),\label{eq:gen sato r2}\\
R_1+R_2&\leq I(Y_2;X_1,X_2,X_c|Q)+I(Y_1;X_1,X_c|\widetilde{Y_2},X_2,Q),\label{eq:gen sato r1+r2 a}\\
R_1+R_2&\leq I(Y_1;X_1,X_2,X_c|Q)+I(Y_2;X_2,X_c|\widetilde{Y_1},X_1,Q),\label{eq:gen sato r1+r2 b}
}{\label{eq:thm:general outer ITW dublin}}
for some input distribution that factors as
\begin{align}
P_{Q,X_1,X_2,X_c}=P_{Q}P_{X_1|Q}P_{X_2|Q}P_{X_c|X_1,X_2,Q}.
\label{eq:factorization thm:general outer ITW dublin}
\end{align}
\end{thm}

\begin{IEEEproof}
The proof may be found in Appendix~\ref{app:thm:general outer ITW dublin}.
\end{IEEEproof}

%{ DT: THE FOLLOWING IS NOT TRUE! FOR ``CIFC in the weak interference'' WE NEED AN AUXILIARY RV!!!
%The outer bound of Thm.~\ref{thm:general outer ITW dublin} for the general memoryless
%IFC-CR  is capacity when the
%channel reduces to a Gaussian CIFC in the ``weak interference'' regime
%\cite[Lem.3.6]{WuDegradedMessageSet}, in the ``very strong interference''
%regime \cite[Thm.6]{maric2005capacity} and in the ``primary decodes cognitive''
%regime \cite[Thm.3.1]{Rini:Allerton2010}. However, it does not reduce to the
%outer bound in \cite[Thm. 3.2]{WuDegradedMessageSet}, which is
%capacity for the CIFC in the ``very weak interference'' regime
%\cite[Thm.3.4]{WuDegradedMessageSet}, and for the semi-deterministic
%CIFC \cite[Thm.8.1]{RTDjournal1}.
%}

\medskip
The outer bound of Thm.~\ref{thm:general outer ITW dublin}
has the appealing feature that is does not contain any auxiliary Random Variable (RV)
and is thus computable.  For example (see Section~\ref{sec:The Gaussian Case})
the ``Gaussian maximizes entropy'' principle suffices to show that a jointly Gaussian input
exhausts the outer bound of Thm.~\ref{thm:general outer ITW dublin} for the Gaussian noise channel.
It also gives the capacity in several cases (please refer to Table~\ref{table:outer}).
%of the CIFC in the ``very strong interference'' regime~\cite[Thm.6]{maric2005capacity},
%of the IFC in the ``strong interference'' regime~\cite{CostaElGamal87,sato2002capacity,carleial1978interference},
%and of the Gaussian CIFC in the ``primary decodes cognitive'' regime~\cite[Thm.3.1]{Rini:Allerton2010}.
%
However it does not reduce to the other cases where capacity is known for simpler channels subsumed by the IFC-CR
(please refer to Table~\ref{table:outer}) nor to the tightest known outer bounds for the general CIFC and BC.
To remedy this, we next derive an outer bound by using a bounding technique originally developed for the BC~\cite{ElGamal1977capacity}.
The derived bound indeed reduces to the tightest known outer bounds for the general CIFC~\cite{MaricGoldsmithKramerShamai07Eu} and
the general BC~\cite{nair2007outer} when the IFC-CR reduces to these channel models.

\subsection{BC-type outer bound}
\label{subsec:bcout}

The outer bound in~\cite{MaricGoldsmithKramerShamai07Eu} for the CIFC
and in~\cite{nair2007outer} for the BC use in their bounding steps
the Csisz\'{a}r's  sum identity~\cite{csiszar1982information}.
We extend this technique here to the general IFC-CR.

\begin{thm}
%(Rini, Tuninetti, Devroye and Goldsmith)
\label{thm:more capable BC outer}
If $(R_1,R_2)$ lies in the capacity region of the IFC-CR then
%in addition to the constraints in~\eqref{eq:thm:general outer ITW dublin}
the following must hold
\eas{
R_1
%\leq I(Y_1;V,U_1,X_1,X_c|U_2,X_2) \nonumber
&\leq I(Y_1;X_1,X_c|U_2,X_2), \label{eq:gen r1 1st}\\
R_2    %\leq I(Y_2;V,U_2,X_2,X_c|U_1,X_1)\nonumber
&\leq I(Y_2;X_2,X_c|U_1,X_1), \label{eq:gen r2 1st}\\
R_1    &\leq I(Y_1;V,U_1,X_1),             \label{eq:gen r1 2nd}\\
R_2    &\leq I(Y_2;V,U_2,X_2),             \label{eq:gen r2 2nd}\\
R_1+R_2&\leq I(Y_2;V,U_2,X_2)+I(Y_1;,X_1,X_c|V,U_2,X_2),\label{eq:gen r1+r2 1st}\\
R_1+R_2&\leq I(Y_1;V,U_1,X_1)+I(Y_2;,X_2,X_c|V,U_1,X_1),\label{eq:gen r1+r2 2nd}
}{\label{eq:thm:generalBC outer}}
such that
\ea{
V \to (U_1,U_2) \to (X_1,X_2,X_c) \to (Y_1,Y_2)
\label{eq:markov chain thm:generalBC outer}
}
for some input distribution that factors as
\ea{
  &P_{U_1,U_2,V,X_1,X_2,X_c} \nonumber
\\&=P_{U_1} P_{U_2} P_{V| U_1, U_2} P_{X_1|U_1} P_{X_2|U_2} P_{X_c|U_1,U_2}.
\label{eq:factorization thm:generalBC outer}
}
\end{thm}

\begin{IEEEproof}
The proof may be found in Appendix~\ref{app:thm:more capable BC outer}.
\end{IEEEproof}

%{ DT: I DO NOT THINK WE NEED A Q HERE BECAUSE WE HAVE A V THAT CAN INCLUDE THE Q.
%NOTICE THAT CONDITIONED ON (U1,U2) WE HAVE THAT V AND THE X'S AND THE Y'S ARE INDEPENDENT -- WHY
%HAS NOBODY REMARKED THAT FOR CIFC AND BC? OR AM I MISSING SOMETHING? THINK ABOUT IT
%BECAUSE WE WANT TO BE CAREFUL WHEN WE CLAIM THE TIGHTEST KNOWN OUTER BOUND FOR THE GENERAL BC AND THE GENERAL CIFC ...}

\medskip

\begin{rem}
Thm.~\ref{thm:more capable BC outer} is the tightest known outer bound for a general IFC-CR and
\begin{enumerate}

\item
it reduces to the tightest known outer bound for the general BC without common rate~\cite{nair2007outer} when $X_1=X_2=\emptyset$,
which is tight for all cases where capacity is known.

%{ DT: THE BOUNDS IN~\eqref{eq:gen r1 1st}-\eqref{eq:gen r2 1st} ARE NOT IN THE NAIR-ELGAMAL REGION, WHY? CAN WE SHOW THAT THEY ARE REDUNDANT? LET'S THINK ABOUT IT BECAUSE WE WANT TO BE CAREFUL WHEN WE CLAIM THE TIGHTEST KNOWN OUTER BOUND FOR THE GENERAL BC ...}

\item
it reduces to  the tightest known outer bound for the general CIFC~\cite[Thm.4]{MaricGoldsmithKramerShamai07Eu} when $X_1=\emptyset$,
which is tight for all cases where capacity is known.
%{ DT: SAME AS FOR THE BC CASE ABAOVE:  LET'S THINK ABOUT IT BECAUSE WE WANT TO BE CAREFUL WHEN WE CLAIM THE TIGHTEST KNOWN OUTER BOUND FOR THE GENERAL CIFC ...}
The outer bound in~\cite[Thm.4]{MaricGoldsmithKramerShamai07Eu} is tighter than the one in~\cite[Thm. 3.2]{WuDegradedMessageSet} (see~\cite[Remark 6]{MaricGoldsmithKramerShamai07Eu}). We can obtain the equivalent of the outer bound in~\cite[Thm. 3.2]{WuDegradedMessageSet} by defining in Thm.~\ref{thm:more capable BC outer} a new pair of auxiliary RVs $U_2':=[V,U_2], U_1':=[V,U_2]$ and then reasoning as in~\cite[Remark 6]{MaricGoldsmithKramerShamai07Eu}.

\item
it is tighter than Thm.~\ref{thm:general outer ITW dublin}.
In fact, the region in Thm.~\ref{thm:more capable BC outer}
can be enlarged by dropping~\eqref{eq:gen r1 2nd}-\eqref{eq:gen r2 2nd}.
Moreover, the bound in~\eqref{eq:gen r1 1st} is tighter than
the one in~\eqref{eq:gen sato r1} by the ``conditioning reduces entropy''
principle.
%DT: PROOF
%I(Y_1;X_1,X_c|U_2,X_2,Q)
%= H(Y_1|U_2,X_2,Q)-H(Y_1|X_1,X_c,U_2,X_2,Q)
%= H(Y_1|U_2,X_2,Q)-H(Y_1|X_1,X_c,X_2,Q)
%< H(Y_1|X_2,Q)-H(Y_1|X_1,X_c,X_2,Q)
%= I(Y_1;X_1,X_c|X_2,Q)
Similarly, to~\cite[Remark IV.2]{RTDjournal1}
the sum-bound in~\eqref{eq:gen r1+r2 1st}
is tighter than the bound in~\eqref{eq:gen sato r1+r2 a}.
%{ DT: CAN U PLEASE DOUBLE CHECK?}
%DT: PROOF: Y_2 same marginal of Y_2'
%\[
%sato-BC
%= I(Y_2;X_1,X_2,X_c|Q)
%+ I(Y_1;X_1,X_c|Q,Y_2',X_2)
%- I(Y_2;V,U_2,X_2|Q)
%- I(Y_1;U_1,X_1,X_c|V,U_2,X_2,Q)
%= I(Y_2;X_1,X_c|V,U_2,X_2,Q)
%+ H(Y_1|Y_2',X_2,Q)-H(Y_1|V,U_2,X_2,Q)
%= ???
%\]
However, the region in Thm.~\ref{thm:more capable BC outer} is expressed as a function
of three auxiliary RVs for which we have not obtained cardinality bounds on the respective alphabets,
while the looser region in Thm.~\ref{thm:general outer ITW dublin}
is expressed only as a function of the inputs and is thus computable in principle.

\item
Thm.~\ref{thm:more capable BC outer} neither reduces to the
capacity region of a class of deterministic IFCs studied in~\cite{elgamal_det_IC}
nor reduces to the outer bound for the semi-deterministic IFC in~\cite{telatar2008bounds} when $X_c=\emptyset$.
The difficulty in deriving outer bounds for the general IFC-CR that are tight
when it reduces to an IFC is also noted in~\cite{rini2010dublin}.
The authors of \cite[Thm.3.2]{rini2010dublin} were able to derive tight bounds
in this scenario by imposing additional constraints on the effect of interference
on the channel outputs.

\end{enumerate}

\end{rem}

\subsection{Simplified BC-type outer bound in the ``weak interference'' and ``strong interference'' regimes}
\label{subsec:outsimplification}

We next proceed to simplify the proposed BC-type outer bound under specific
``strong interference'' and  ``weak interference'' conditions.

\begin{cor}{\bf ``Strong interference at Rx~1'' outer bound.}
\label{cor:strong int outer bound}
If
\ea{
I(Y_2 ; X_2 , X_c |  X_1 ) \leq I(Y_1 ; X_2 , X_c | X_1 )
\label{eq:strong int. cond. at Rx1}
}
%\ea{
%I(Y_1 ; X_1 , X_c | X_2 ) \leq I(Y_2 ; X_1 , X_c | X_2 )
%\label{eq:strong int. cond. at Rx2}
%}
for all distributions that factor as
\ea{
P_{X_1,X_2,X_c}=P_{X_1}P_{X_2}P_{X_c|X_1,X_2},
\label{eq:factorization condition cor:strong int outer bound}
}
then, if $(R_1,R_2)$ lies in the capacity region of the IFC-CR, the
following must hold
\eas{
R_1      & \leq I(Y_1; X_1, X_c | X_2, Q), \label{eq:cor:strong int outer bound R1 1}\\
R_2      & \leq I(Y_2; X_2, X_c | X_1, Q), \label{eq:cor:strong int outer bound R2 1}\\
R_1 +R_2 & \leq I(Y_1; X_1, X_2,  X_c| Q), \label{eq:cor:strong int outer bound R1+R2 1}
%\\
%R_1 +R_2 & \leq & I(Y_2; X_1, X_2 , X_c |Q ) \label{eq:cor:strong int outer bound R1+R2 2},
}{\label{eq:cor:strong int outer bound}}
for some distribution that factors as in~\eqref{eq:factorization thm:general outer ITW dublin}.
%{eq:factorization condition cor:strong int outer bound}.
\end{cor}

\begin{IEEEproof}
The proof follows from showing that under the condition in~\eqref{eq:strong int. cond. at Rx1}
the sum-rate bounds in Thm.~\ref{thm:more capable BC outer}
simplify to~\eqref{eq:cor:strong int outer bound R1+R2 1}.
The details of the proof may be found in Appendix~\ref{app:cor:strong int outer bound}.
\end{IEEEproof}

%\label{thm:general outer ITW dublin}
%\label{thm:more capable BC outer}
%\label{thm:generalBC outer}

\medskip
Note that, given the symmetry of the channel model, Cor.~\ref{cor:strong int outer bound} also holds by reversing the role of the sources.
Although not valid for a general IFC-CR, Cor.~\ref{cor:strong int outer bound} is expressed only as a function of the channel inputs and does not contain auxiliary RVs as Thm.~\ref{thm:general outer ITW dublin}, which simplifies both the calculation of the outer bound and the derivation of a capacity achieving encoding strategy.

\medskip
\begin{cor}{\bf ``Weak interference at Rx~2'' outer bound.}
\label{cor:weak int outer bound}
If
\ea{
I(Y_2 ; U |X_2) \leq  I(Y_1 ; U |X_2)
\label{eq:weak int. cond. at Rx1}
}
holds for all distributions
\ea{
P_{U,X_1,X_2,X_c}=P_{X_1}P_{X_2}P_{X_c|X_1,X_2}P_{U| X_1,X_2,X_c},
\label{eq:factorization condition cor:weak int outer bound}
}
such that $U\to (X_1,X_2,X_c) \to (Y_1,Y_2)$,
then, if $(R_1,R_2)$ lies in the capacity region of the IFC-CR, the
following must hold:
\eas{
R_1 & \leq  I(Y_1 ; X_1 , X_c|  X_2, U)
\label{eq:DM ifc-cr weak int outer bound R1}\\
R_2 & \leq  I(Y_2 ; X_2, U)
\label{eq:DM ifc-cr weak int outer bound R2}\\
R_2 & \leq  I(Y_2 ; X_2, X_c | X_1)
\label{eq:DM ifc-cr weak int outer bound R2-2}
}{\label{eq:DM ifc-cr weak int outer bound}}
for some distribution that factors as in~\eqref{eq:factorization condition cor:weak int outer bound}.
%\ea{
%P_{X_1, X_2, X_c, U}=P_{X_1} P_{X_2} P_{X_c| X_1,X_2} P_{U| X_1,X_2,X_c}.
%\label{eq:factorization cor:weak int outer bound}
%}
\end{cor}

\begin{IEEEproof}
The proof may be found in Appendix~\ref{app:cor:weak int outer bound}.
\end{IEEEproof}

\medskip
Again, given the symmetry of the channel model,
Cor.~\ref{cor:weak int outer bound} also holds when the sources
are reversed.
% Although not valid for a general IFC-CR, Thm.~\ref{cor:strong int outer bound}
%is expressed only as a function of one auxiliary RV. { DT: HOW ABOUT CARDINALITY BOUNDS?}

%{\blue We can also put a  broadcast channel outer bound obtained by allowing full cooperation among the transmitters.
%This may be tight in some regimes for the Gaussian case (as it is for the CIFC ). I'm not sure it makes sense for the DMC case though. Natasha: I suggest to leave this out unless we can show a capacity result in the Gaussian case (need to simulate this first, if tight try to analytically get it....), Seems like too much given our limited time.}

\section{Inner Bounds}
\label{sec:InnerBound}

In this section we derive an inner bound for a general IFC-CR,
then analytically show that this region contains all other previously derived regions, and
finally derive simple and easy-to-understand expressions for a number of sub-schemes of our general inner bound.
Given the generality of the IFC-CR channel model, the coding scheme we propose contains a large number of rate bounds and several auxiliary RVs. Unfortunately, this is unavoidable if one wishes the achievable scheme to be capacity in all the cases when the channel reduces to one where capacity is known.
Our aim in deriving this achievable rate region is therefore mainly to provide a unified framework to efficiently investigate the rate advantages provided by different transmission strategies.

\subsection{General achievable rate region}

The achievable scheme is obtained as  a combination of the following well established random coding techniques:
\begin{itemize}
%\item
%DT :DO WE REALLY NEED TO MOTIVATE `Time-sharing'???
%{\bf Time-sharing:}
%Time sharing refers to the use of different transmission schemes at different times. This is achieved by employing a RV, $Q$, which indicates which achievable scheme is used at a given time. Time sharing is known to yield a larger rate region than taking the convex hull over all rate points achieved by different schemes for some channels.
%%convex hull operation for some channels.
%%This strategy is known to corresponds to the convex hull operation for the DM-IFC~\cite{chong2008han} but numerical simulations for the Gaussian IFC suggests that it outperforms the convex hull operation in this scenario {\blue which reference?}.

\item
{\bf Rate-splitting:}
This refers to splitting the  message of a source into different independent sub-messages, one for each possible subset of destinations. Rate splitting was first introduced by Han and Kobayashi for the classical IFC~\cite{Han_Kobayashi81} (referred to as the Han and Kobayashi region or rate-splitting from now on)  and is a fundamental tool in achieving capacity in a number of cases when combined with superposition coding and binning.
In our achievable scheme we rate split each message into
private and public parts at the intended transmitter and at the cognitive relay.

\item
{\bf Superposition coding:}
Superposition coding was first introduced in~\cite{CoverCapacityDetBC} for the degraded BC and intuitively consists of generating codewords conditional on other ones, or ``stacking'' codewords on top of each other.
Destinations in the system decode (some of the) codewords starting from the bottom of the stack, while treating the remaining codewords as noise. Thus, a given message may be decoded at one destination but treated as noise at another.
Here we superpose public messages to broadcast messages and
the messages known at the cognitive relay over the messages at the two sources.

\item
{\bf Gel'fand-Pinsker binning:}
Often simply referred to as binning~\cite{GelFandPinskerClassic}, it allows a transmitter to ``pre-code'' (portions of) the message against the interference that message is known to experience at a destination.  Binning is also used in Marton's largest known achievable rate region for the general memoryless BC~\cite{MartonBroadcastChannel}. It is also a crucial element in other channels, usually with some form of ``broadcast'' element, including the CIFC~\cite{RTDjournal1}.
In this achievable scheme the cognitive relay performs binning against the private messages of the sources.

\item
{\bf Simultaneous decoding:}
As at the destination of a MAC, a destination jointly decodes its intended message and some of the sub-messages of non-intended sources with the objective to reduce the level of interference.  Simultaneous or joint decoding is optimal in many cases of ``strong'' interference.

\end{itemize}

We next derive a transmission scheme that contains a general combination these encoding techniques.
By removing certain features from this general scheme, one can quickly obtain simpler and analytically more tractable sub-schemes that can be compared to each other and to outer bounds, as we shall do in the next subsection. We shall also show that this general inner bound includes all known to-date achievable rate regions.
The novelty of our proposed region, which will allow us to show inclusion in all known regions, is a rate split into four parts for each source message (as opposed to the classical rate split in two parts for the classical IFC~\cite{Han_Kobayashi81} and to the rate split in three parts for the CIFC~\cite{RTDjournal1}).

\begin{thm} {\bf Region $\Rcal^{\rm(RTDG)}$.}
\label{thm:general inner bound}
The region $\Rcal^{\rm(RTDG)}$ is defined as the set of non-negative rate pairs $(R_1,R_2)$
% in is achievable for the general IFC-CR
for which there exists a non-negative rate vector
\ea{
&(R_{1  {\rm c}}, R_{2  {\rm c}},R_{1  {\rm p}}, R_{2  {\rm p}},R_{1  {\rm cb}}, R_{2  {\rm cb}},R_{1  {\rm pb}}, R_{2  {\rm pb}},
R_{0  {\rm cb}}', %% R_{1  {\rm cb}}', R_{2  {\rm cb}}',
R_{1  {\rm pb}}', R_{2  {\rm pb}}')
\nonumber
\\&\in \bigcup_{P} \big\{ \Rcal_{0} \cap \Rcal_{1} \cap \Rcal_{2} \big\}
%\in \Rcal^{\rm(RTDG)}
\label{eq:total inner bound}
} %P_{U_i, i \in \{ 1p, 1c,1pb, 1cp, 2p, 2c, 2pb, 2cp\}}
such that
\begin{align}
R_i&=R_{i  {\rm c}}+R_{i  {\rm p}}+R_{i  {\rm cb}}+R_{i  {\rm pb}}, \ i \in \{1,2\},
\label{eq:ratespliintofourparts}
\end{align}
where the union in~\eqref{eq:total inner bound}
is over all input distributions $P$ given by
\ea{
%P_{U_i, i \in \{ 1p, 1c,1pb, 2p, 2c, 2pb, cp\}}
%=P_{U_{1  {\rm p}},U_{1  {\rm c}}}
% P_{U_{2  {\rm p}},U_{2  {\rm c}}}
% P_{U_{1  {\rm pb}},U_{2  {\rm pb}},U_{0  {\rm cb}}|U_{1  {\rm c}},X_1,U_{2  {\rm c}},X_2},
P=&
P_{Q}
P_{U_{1  {\rm c}},X_1|Q}
P_{U_{2  {\rm c}},X_2|Q}
\nonumber\\&
P_{U_{1  {\rm pb}},U_{2  {\rm pb}},U_{0  {\rm cb}},X_c|U_{1  {\rm c}},X_1, U_{2  {\rm c}},X_2, Q},
%P_{Y_1,Y_2|X_1,X_2,X_c}
\label{eq:inner bound distirbution}
}
where
the ``binning rate region'' $\Rcal_{0}$ in~\eqref{eq:total inner bound} is given in~\eqref{eq:binning rates} and
the ``decoding rate region at destination~1'' $\Rcal_{1}$ in~\eqref{eq:total inner bound} is given in~\eqref{eq:inner bound} for
\ean{
L_{i  {\rm pb}}&=R_{i  {\rm pb}}+R_{i  {\rm pb}}', \quad i \in \{1,2\}, \\
L_{0  {\rm cb}}&=R_{1  {\rm cb}}+R_{2  {\rm cb}}+R_{0  {\rm cb}}',
}
and where the ``decoding rate region at destination~2'' $\Rcal_{2}$ in~\eqref{eq:total inner bound}
is obtained permuting the indices $1$ and $2$ in the
``decoding rate region at destination~1'' $\Rcal_{1}$ in~\eqref{eq:inner bound}.

\begin{figure*}

%\hline
\rule[0.0em]{\textwidth}{0.5pt}

\eas{
    R_{0  {\rm cb}}' & \geq  I(X_1, X_2; U_{0  {\rm cb}}| U_{1  {\rm c}},U_{2  {\rm c}}, Q) \label{eq:binning rates 0}\\
    R_{1  {\rm pb}}' & \geq  I(X_2; U_{1  {\rm pb}}| U_{1  {\rm c}},X_1,U_{2  {\rm c}},U_{0  {\rm cb}}, Q) \label{eq:binning rates 1}\\
    R_{2  {\rm pb}}' & \geq  I(X_1; U_{2  {\rm pb}}| U_{1  {\rm c}},X_2,U_{2  {\rm c}},U_{0  {\rm cb}}, Q) \label{eq:binning rates 2}\\
    R'_{1  {\rm pb}}+R'_{2  {\rm pb}} &\geq
          I(X_2; U_{1  {\rm pb}}| U_{1  {\rm c}},U_{2  {\rm c}},X_1,U_{0  {\rm cb}}, Q) %\nonumber \\
         +I(X_1; U_{2  {\rm pb}}| U_{1  {\rm c}},U_{2  {\rm c}},X_2,U_{0  {\rm cb}}, Q) \nonumber \\
  &\lag  +I(U_{1  {\rm pb}}; U_{2  {\rm pb}}| U_{1  {\rm c}},X_1,U_{2  {\rm c}},X_2,U_{0  {\rm cb}}, Q), \label{eq:binning rates 12}
}{\label{eq:binning rates}}

%\hline
\rule[0.0em]{\textwidth}{0.5pt}

\eas{
%1
R_{1  {\rm c}}+R_{1  {\rm p}}+R_{2  {\rm c}}+
 L_{0  {\rm cb}} %L_{1  {\rm cb}}+L_{2  {\rm cb}}+R_{0  {\rm cb}}'
+L_{1  {\rm pb}} & \leq I(U_{0  {\rm cb}}; X_1 | U_{1  {\rm c}},U_{2  {\rm c}}, Q)+
% \nonumber \\
%&\lag +
I(Y_1; U_{1  {\rm c}},U_{2  {\rm c}},X_1,U_{0  {\rm cb}},U_{1  {\rm pb}}, Q)
\label{eq:inner bound 1}\\
%2
R_{1  {\rm c}}+R_{1  {\rm p}} \phantom{+R_{2  {\rm c}}} +
 L_{0  {\rm cb}} %L_{1  {\rm cb}}+L_{2  {\rm cb}}+R_{0  {\rm cb}}'
+L_{1  {\rm pb}} & \leq I(U_{0  {\rm cb}}; X_1 | U_{1  {\rm c}},U_{2  {\rm c}}, Q)+
%\nonumber \\
%&\lag +
I(Y_1; U_{1  {\rm c}},X_1,U_{0  {\rm cb}},U_{1  {\rm pb}}|U_{2  {\rm c}}, Q)
\label{eq:inner bound 2}\\
%3
R_{1  {\rm p}}+R_{2  {\rm c}}+
 L_{0  {\rm cb}} %L_{1  {\rm cb}}+L_{2  {\rm cb}}+R_{0  {\rm cb}}'
+L_{1  {\rm pb}} & \leq I(U_{0  {\rm cb}}; X_1 | U_{1  {\rm c}},U_{2  {\rm c}}, Q)+
% \nonumber \\
%& \lag +
I(Y_1; U_{2  {\rm c}},X_1,U_{0  {\rm cb}},U_{1  {\rm pb}}|U_{1  {\rm c}}, Q)
\label{eq:inner bound 3}\\
%5
\phantom{R_{1  {\rm c}}+} R_{1  {\rm p}} \phantom{+R_{2  {\rm c}}} +
 L_{0  {\rm cb}} %L_{1  {\rm cb}}+L_{2  {\rm cb}}+R_{0  {\rm cb}}'
+L_{1  {\rm pb}} & \leq I(U_{0  {\rm cb}}; X_1 | U_{1  {\rm c}},U_{2  {\rm c}}, Q)+
%\nonumber \\
%& \lag +
I(Y_1;X_1,U_{0  {\rm cb}},U_{1  {\rm pb}}|U_{1  {\rm c}},U_{2  {\rm c}}, Q)
\label{eq:inner bound 4}\\
%4
R_{2  {\rm c}}+
 L_{0  {\rm cb}} %L_{1  {\rm cb}}+L_{2  {\rm cb}}+R_{0  {\rm cb}}'
+L_{1  {\rm pb}} & \leq I(U_{0  {\rm cb}}; X_1| U_{1  {\rm c}},U_{2  {\rm c}}, Q)+
%\nonumber \\
%&\lag +
I(Y_1; U_{2  {\rm c}},U_{0  {\rm cb}},U_{1  {\rm pb}}| U_{1  {\rm c}},X_1, Q)
\label{eq:inner bound 7}\\
%7
 L_{0  {\rm cb}} %L_{1  {\rm cb}}+L_{2  {\rm cb}}+R_{0  {\rm cb}}'
+L_{1  {\rm pb}} & \leq I(U_{0  {\rm cb}}; X_1  | U_{1  {\rm c}},U_{2  {\rm c}}, Q)+
%\nonumber \\
%& \lag +
I(Y_1;U_{0  {\rm cb}},U_{1  {\rm pb}}|U_{1  {\rm c}},U_{2  {\rm c}},X_1, Q)
 \label{eq:inner bound 5}\\
%6
R_{1  {\rm p}}
+L_{1  {\rm pb}} & \leq I(U_{0  {\rm cb}}; X_1 | U_{1  {\rm c}},U_{2  {\rm c}}, Q)+
%\nonumber \\
%& \lag +
I(Y_1; X_1,U_{1  {\rm pb}}|U_{1  {\rm c}},U_{2  {\rm c}},U_{0  {\rm cb}}, Q)
\label{eq:inner bound 6})\\
%8
L_{1  {\rm pb}} &\leq \phantom{I(U_{0  {\rm cb}}; X_1 | U_{1  {\rm c}},U_{2  {\rm c}}, Q)+}
I(Y_1;U_{1  {\rm pb}}|U_{1  {\rm c}},U_{2  {\rm c}},X_1,U_{0  {\rm cb}}, Q),
  \label{eq:inner bound 8}
}{\label{eq:inner bound}}

%\hline
\rule[0.0em]{\textwidth}{0.5pt}

\end{figure*}

Moreover, in the ``decoding rate region at destination~1'' $\Rcal_{1}$ in~\eqref{eq:inner bound} (and similarly for $\Rcal_{2}$ but with the role of the sources swapped) the following rate bounds can be dropped
\begin{itemize}
  \item  \eqref{eq:inner bound 1} and~\eqref{eq:inner bound 2}: when $R_1=R_{1  {\rm c}}=R_{1  {\rm p}}=R_{1  {\rm cb}}=R_{1  {\rm pb}}=0$,
  \item  \eqref{eq:inner bound 3} and~\eqref{eq:inner bound 4}: when $R_{1  {\rm p}}=R_{1  {\rm cb}}=R_{1  {\rm pb}}=0$,
  \item  \eqref{eq:inner bound 7} and~\eqref{eq:inner bound 5}: when $R_{1  {\rm cb}}=R_{1  {\rm pb}}=0$,
  \item  \eqref{eq:inner bound 6}: when $R_{1  {\rm p}} =R_{1  {\rm pb}}=0$,
  \item  \eqref{eq:inner bound 8}: when $R_{1  {\rm pb}}=0$,
\end{itemize}
because these bounds correspond to an error event in which a non-intended common message or a bin index is incorrectly decoded and no other intended message is incorrectly decoded.
\end{thm}

\begin{IEEEproof}
The achievable rate region in~\eqref{eq:total inner bound} may be obtained using the result in~\cite{TuniGraph} by specifying how rate splitting, binning and superposition coding are performed. The details of the proof are reported in Appendix~\ref{app:inner bound} for completeness.  In what follows we sketch the main elements of the encoding and decoding procedures and we give an intuitive explanation about the proposed choices. We do not consider the time sharing RV $Q$ to simplify the description.

\smallskip
{\bf Rate Splitting:}
%We only describe the encoding operations for message $W_1$.
%By permuting the message indices one obtains the encoding for message $W_2$.
The  message $W_i$, $i\in\{1,2\}$, is split into four sub-messages:
    \begin{itemize}
      \item
      Private message $W_{i  {\rm p}}$ of rate $R_{i  {\rm p}}$,
      \item
      Common message  $W_{i  {\rm c}}$ of rate $R_{i  {\rm c}}$,
      \item
      Common Broadcasted message  $W_{i  {\rm cb}}$  of rate $R_{i  {\rm cb}}$, and
      \item
      Private Broadcasted message $W_{i  {\rm pb}}$  of rate $R_{i  {\rm pb}}$,
    \end{itemize}
      so that~\eqref{eq:ratespliintofourparts} holds.  %$R_i=R_{i  {\rm p}}+R_{i  {\rm c}}+R_{i  {\rm cb}}+R_{i  {\rm pb}}$.

\smallskip
{\bf Codebook Generation:}
The sources and the cognitive relay generate the following codebooks:
    \begin{itemize}
      \item
      Common message: $w_{i  {\rm c}}\in[1:2^{N R_{i  {\rm c}}}]$ is encoded into $U_{i  {\rm c}}^N(w_{i  {\rm c}})$ with iid distribution $P_{U_{i  {\rm c}}}$, $i\in\{1,2\}$.

      \item
      Private message: for a given $w_{i  {\rm c}}$, $w_{i  {\rm p}}\in[1:2^{N R_{i  {\rm p}}}]$ is encoded into $X_i^N(w_{i  {\rm p}}|w_{i  {\rm c}})$ with iid distribution $P_{X_i|U_{i  {\rm c}}}$ (i.e., $X_i^N$ is superimposed to $U_{i  {\rm c}}^N$), $i\in\{1,2\}$.

      \item
      Common broadcasted messages: for a given pair $(w_{1  {\rm c}},w_{2  {\rm c}})$, the pair $(w_{1  {\rm cb}},w_{2  {\rm cb}})\in[1:2^{N R_{1  {\rm cb}}}] \times [1:2^{N R_{2  {\rm cb}}}]$ is encoded into $U_{0  {\rm cb}}^N(w_{1  {\rm cb}},w_{2  {\rm cb}},b_{0  {\rm cb}}| w_{1  {\rm c}},w_{2  {\rm c}})$, $b_{0  {\rm cb}}\in[1:2^{N R_{0  {\rm cb}}'}]$, with iid distribution $P_{U_{0  {\rm cb}}|U_{1  {\rm c}},U_{2  {\rm c}}}$.

      %{ DT: I FEEL THAT BINNING HERE IS NOT NEEDED HERE AS BOTH DESTINATIONS DECODE ALL COMMON MGS ... RIGHT? CAN WE SEE THIS FROM F-M ELIMINATION?}

      \item
      Private broadcasted message: for a given $(w_{1  {\rm c}},w_{2  {\rm c}},w_{1  {\rm cb}},w_{2  {\rm cb}},b_{0  {\rm cb}}, ,w_{i  {\rm p}})$, $w_{i  {\rm pb}}\in[1:2^{N R_{i  {\rm pb}}}]$ is encoded into
      $U_{i  {\rm pb}}^N(w_{i  {\rm pb}},b_{i  {\rm pb}}|w_{1  {\rm c}},w_{2  {\rm c}},w_{1  {\rm cb}},w_{2  {\rm cb}},b_{0  {\rm cb}}, w_{i  {\rm p}})$,
      $b_{i  {\rm pb}}\in[1:2^{N R_{i  {\rm pb}}'}]$, with distribution $P_{U_{i  {\rm pb}}|U_{1  {\rm c}},U_{2  {\rm c}},U_{0  {\rm cb}}, X_i}^N$, $i\in\{1,2\}$.

    \end{itemize}

\smallskip
{\bf Encoding:}
The cognitive relay has knowledge of both messages $W_1,W_2$ and is thus able to perform binning with the goal
to create the most general distribution among conditionally independent RVs/codebooks. It does the following:
    \begin{itemize}
      \item
       $U_{0  {\rm cb}}^N$ was generated only based on $(U_{1  {\rm c}}^N,U_{2  {\rm c}}^N)$.
       The cognitive relay bins $U_{0  {\rm cb}}^N$ against $(X_1^N,X_2^N)$,
       as for channel with states known non-causally at the encoder~\cite{GelFandPinskerClassic},
       to make it look like it were generated iid with distribution $P_{U_{0  {\rm cb}}|X_1,X_2,U_{1  {\rm c}},U_{2  {\rm c}}}$.
       For this to be possible, the ``binning rate'' $R_{0  {\rm cb}}'$ must satisfy~\eqref{eq:binning rates 0}.

      \item
      %.  If the binning of $U_{0  {\rm cb}}$is successful,
      $U_{1  {\rm pb}}^N$, resp. $U_{2  {\rm pb}}^N$, was generated independently of $(X_2^N,U_{2  {\rm pb}}^N)$, resp. $(X_1^N,U_{1  {\rm pb}}^N)$, conditioned on the ``common'' RVs $(U_{1  {\rm c}}^N,U_{2  {\rm c}}^N,U_{0  {\rm cb}}^N)$.
      The cognitive relay bins $U_{1  {\rm pb}}^N$ and $U_{2  {\rm pb}}^N$ against each other, as in Marton's region for the general BC~\cite{MartonBroadcastChannel},  and against $(X_1^N,X_2^N)$ to make them look like they were generated iid with distribution $P_{U_{1  {\rm pb}},U_{2  {\rm pb}}|X_1,X_2,U_{1  {\rm c}},U_{2  {\rm c}},U_{0  {\rm cb}}}$.   For this to be possible, the ``binning rate'' pair $(R_{1  {\rm pb}}',R_{2  {\rm pb}}')$ must satisfy~\eqref{eq:binning rates 1}-\eqref{eq:binning rates 12}.

      \item to send $w_i=(w_{i  {\rm c}},w_{i  {\rm p}},w_{i  {\rm cb}},w_{i  {\rm pb}})$ source $i$ sends $X_i^N(w_{i  {\rm p}}|w_{i  {\rm c}})$, $i\in\{1,2\}$.

      \item to send $(w_1, w_2)= \big(
      (w_{1  {\rm c}},w_{1  {\rm p}},w_{1  {\rm cb}},w_{1  {\rm pb}}),$
      $(w_{2  {\rm c}},w_{2  {\rm p}},w_{2  {\rm cb}},w_{2  {\rm pb}}) \big)$ the cognitive relay sends $X_c^N$ obtained as a deterministic function of the tuplet $(U_{1  {\rm c}}^N,U_{2  {\rm c}}^N,X_1^N,X_2^N,U_{0  {\rm cb}}^N,U_{1  {\rm pb}}^N,U_{2  {\rm pb}}^N)$ found after the different binning operations.

    \end{itemize}

\begin{figure}%[h]
\centerline{\includegraphics[width=8cm ]{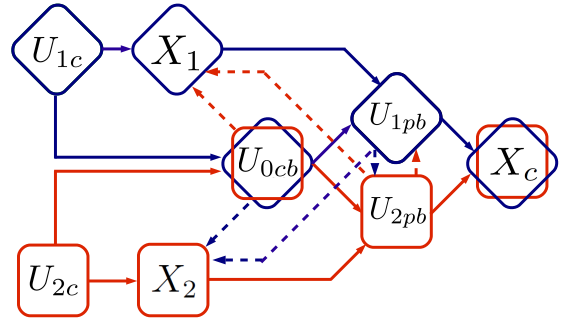}}
\caption{A graphical representation of the coding scheme for the inner bound region in Section~\ref{sec:InnerBound}.
The RVs for message~1 are in blue diamond boxes while
the RVs for message~2 are in red square boxes.
A solid line among RVs indicates that the RVs are superposed while a dashed line
that the RVs are binned against each other.
%{ DT: NEED TO ADD A SUBSCRIPT 0 FOR COMMON BROADCASTED MSG.}
}
\label{fig:superposition step}
\end{figure}

\smallskip
Fig.~\ref{fig:superposition step} is a graphical representation of the proposed achievable scheme.
Each box represents an auxiliary RV/codebook carrying the sub-message with the same subscript
(note that the RVs $X_1$ and $X_2$ carry the sub-messages $W_{1  {\rm p}}$ and $W_{2  {\rm p}}$, respectively,
and $U_{0  {\rm cb}}$ carries the pair of sub-messages $(W_{1  {\rm cb}}, W_{2  {\rm cb}})$).
%
%The RVs for message~1 are in blue diamond boxes while
%the RVs for message~2 are in red square boxes.
%%
%A solid line among RVs indicates that the RVs are superposed while a dashed line
%that the RVs are binned against each other.
%%
%We note that due to the message knowledge structure, the cognitive RVs may be binned against the private RVs but not vice-versa.
%%In addition, we note that a RV can be binned only against a RV over which it is {\it not} superposed.

\smallskip
{\bf Decoding:}
Destination $i$, $i\in\{1,2\}$, simultaneously decodes all RVs/codebooks except $(X_{\overline{i}}^N,U_{\overline{i}  {\rm pb}}^N)$ with $\overline{i}\not=i$. This is successful with high probability if the rates belong to the ``decoding rate region at destination~$i$'' $\Rcal_{i}$ defined in~\eqref{eq:inner bound}, $i\in\{1,2\}$.
\end{IEEEproof}

\medskip
\begin{rem}
[Intuitive interpretation of the proposed coding scheme]
\label{rem:intuitive interpretation of the proposed coding scheme}
%We now provide some intuition on the structure of the achievable scheme in Thm.~\ref{thm:general inner bound}
%using the graph representation in Fig.~\ref{fig:superposition step} and highlight the key features of the scheme.
Loosely speaking the achievable rate region is obtained by considering a Han and Kobayashi transmission scheme for the IFC among the two source-destination pairs and extending this coding scheme with the scheme for the CIFC~\cite{rini2009state} for each source-destination pair.
The RVs $U_{1  {\rm c}},U_{2  {\rm c}},X_1,X_2$ correspond to the Han and Kobayashi scheme~\cite{Han_Kobayashi81} for the IFC.
% this scheme is known to achieve capacity for the IFC in the ``strong interference'' regime ~\cite{sato2002capacity} and for the deterministic IFC~\cite{elgamal_det_IC}.
%
The common broadcasted message $U_{0  {\rm cb}}$ is superposed to both the common messages $U_{1  {\rm c}}, U_{2  {\rm c}}$ and carries the common broadcasted messages for both users, $W_{1  {\rm cb}}$ and $W_{2  {\rm cb}}$.  Since these messages are to be decoded at both decoders, there is no rate advantage in assigning a different RV to each rate split.
%message at Since the common parts $U_{1  {\rm c}}, U_{2  {\rm c}}$ are decoded at both transmitters, we can superposed the broadcasted both common parts $U_{1  {\rm cb}},U_{2  {\rm cb}}$ to the set $[U_{1  {\rm c}},U_{2  {\rm c}}]$.
Note that $U_{0  {\rm cb}}$  cannot be stacked over to the private messages $(X_1,X_2)$ since these messages are not decoded at the non-intended destinations.
To achieve the most general input distribution, the cognitive relay performs binning of $U_{0  {\rm cb}}$ against the known interfering signals $(X_1,X_2)$.
The private broadcasted message $U_{1  {\rm pb}}$ is stacked onto $(U_{1  {\rm c}}, U_{2  {\rm c}},U_{0  {\rm cb}},X_1)$ -- this can be done since this RV is to be decoded only at destination~1 which also decodes $X_1$.
The same procedure is applied to $U_{2  {\rm pb}}$.
At the last encoding step at the cognitive relay, $U_{1  {\rm pb}}$ and $U_{2  {\rm pb}}$  are binned against each other and against the non-intended private messages to achieve the most general distribution.

Finally, note that the proposed scheme with only the ``broadcast'' RVs $(U_{0  {\rm cb}},U_{1  {\rm pb}},U_{2  {\rm pb}})$ corresponds to Marton's achievable rate region for the general BC~\cite{MartonBroadcastChannel},
without the ``broadcast'' RVs it corresponds to Han and Kobayashi's achievable rate region for the general IFC~\cite{Han_Kobayashi81},
and with the ``broadcast'' RVs only for one source it corresponds to  Rini {\em et al}'s achievable rate region for the general CIFC~\cite{RTDjournal1}.
Therefore, our proposed achievable rate region reduces to the largest known achievable rate regions for the simpler channels subsumed by the IFC-CR, which are capacity-achieving for all cases where capacity is known.
\end{rem}

\subsection{Inclusion of the Jiang et al. region \cite{jiang-achievable-BCCR} for the IFC-CR: scheme with $(U_{1  {\rm c}},X_1,U_{2  {\rm c}},X_2,U_{1  {\rm pb}},U_{2  {\rm pb}})$
\label{sec:fig:JiangComparison}}

We now show that the achievable rate region in Thm.~\ref{thm:general inner bound} includes all previously proposed achievable rate regions for the IFC-CR by showing that the region in Thm.~\ref{thm:general inner bound} includes the region in~\cite{jiang-achievable-BCCR} as a special case, which is currently the largest known region for this channel and contains the regions of~\cite{Sahin_2007_2} and~\cite{sridharan2008capacity} .
\begin{thm}
\label{thm:inclusion Jiang}
The achievable rate region in Thm.~\ref{thm:general inner bound} contains the achievable rate region in~\cite[(21)-(31)]{jiang-achievable-BCCR}.
\end{thm}

\begin{IEEEproof}
Set $U_{0  {\rm cb}}=\emptyset$ in Thm.~\ref{thm:general inner bound}.
The resulting achievable rate region includes the region in~\cite[(20)-(31)]{jiang-achievable-BCCR}
(which includes the region in~\cite[(1)-(19)]{jiang-achievable-BCCR})
as shown in Appendix~\ref{app:inclusion of jiang region in our region}.
\end{IEEEproof}

%A representation of the achievable scheme used in Thm.~\ref{thm:inclusion Jiang}   is provided in Fig.~\ref{fig:JiangComparisonScheme}.
%\begin{figure}[h]
%\centering
%\caption{A graphical representation of the scheme used to show inclusion of the scheme in~\cite{jiang-achievable-BCCR} in Sec.~\ref{sec:fig:JiangComparison}.}
%\label{fig:JiangComparisonScheme}
%\end{figure}

%We now investigate some new schemes that cannot  be implemented in a BC, IFC or CIFC.
%We propose to investigate the different forms of collaborations and cooperation that we can implement among the three transmitters.

\subsection{Sub-schemes from the general achievable rate region in Thm.~\ref{thm:general inner bound}}
\label{sec:subschemes}

\begin{figure*}
\centering
\subfigure[Scheme ``all private messages'' in Section~\ref{sec:All private messages}]{
   \includegraphics[width =8cm] {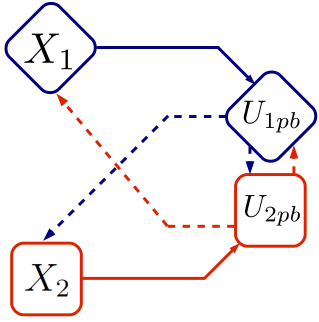}
   \label{fig:allPrivate}
 }
 \hfill
 \subfigure[Scheme ``all common messages'' in Section~\ref{sec:All public messages}]{
   \includegraphics[width =8cm] {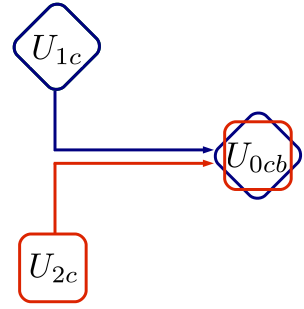}
   \label{fig:AllPublic}
 }
 \\
 \subfigure[Scheme ``one common and one private message'' in Section~\ref{sec:OnePrivateOnePublic}]{
   \includegraphics[width =8cm] {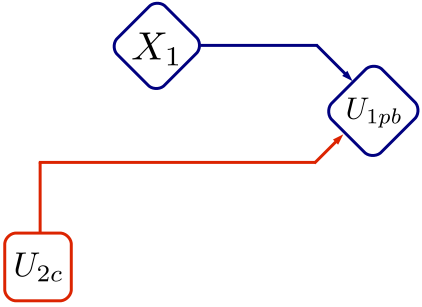}
   \label{fig:OnePrivateOnePublic2}
 }
  \hfill
 \subfigure[Scheme ``common from sources and private from relay messages'' in Section~\ref{sec:OnePrivateOnePublic2}]{
   \includegraphics[width =8cm] {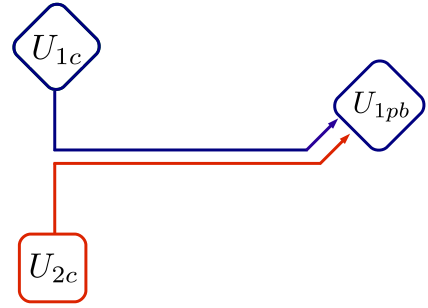}
   \label{fig:OnePrivateOnePublic}
 }
\label{myfigure}
\caption{Specific choice of RVs for the general coding scheme in Fig.~\ref{fig:superposition step}.
The missing nodes in each figure indicates that the associated auxiliary RV has rate zero.  The remaining nodes are encoded as prescribed by  Th. \ref{thm:general inner bound}.
%The nodes in each figure indicate those RV whose rate
 }
\end{figure*}

\begin{table*}
\centering
\caption{The capacity results available for  BC, IFC and CIFC and the assignment of RVs in the region in~\eqref{eq:total inner bound} that achieve the corresponding region.
%{ DT: line 3 and 4 are confusing; det.IFC was wrong and corrected in 2nd to last line; for completeness also added cap.result for BC; our semi.det. CIFC was missing.}
}
\label{tab:subschemes}
\begin{tabular}{|l|l|l|l|}
\hline
Sub-scheme \# & RVs used & Capacity result & Reference \\
\hline
\hline
1 (all private) & $X_1,X_2,U_{1  {\rm pb}},U_{2  {\rm pb}}$ & semi-deterministic BC, semi-det. CIFC &\cite{gelfand_BC,RTDjournal1}  \\
\hline
2 (all common)  & $U_{1  {\rm c}}, U_{2  {\rm c}}, U_{0  {\rm cb}}$ & very strong interference CIFC, IFC, IFC-CR &\cite{maric2005capacity, sato2002capacity} \\
\hline
3 (one common, one private) & $U_{1  {\rm p}},U_{2  {\rm c}},U_{1  {\rm pb}}$ & very weak CIFC &\cite{WuDegradedMessageSet} \\
\hline
4 (common from sources, private from relay) & $U_{1  {\rm c}}, U_{2  {\rm c}}, U_{1  {\rm pb}}$ & very weak CIFC &\cite{WuDegradedMessageSet}\\
\hline
Han and Kobayashi region  & $X_1,X_2,U_{1  {\rm c}},U_{2  {\rm c}}$  & a class of deterministic IFC       &\cite{elgamal_det_IC}  \\
\hline
Marton region &  $U_{0  {\rm cb}},U_{1  {\rm pb}},U_{1  {\rm pb}}$  & a More capable BC, BC with degraded message set &\cite{ElGamal1977capacity,korner1977general}  \\
\hline
\end{tabular}
\end{table*}

The inner bound of Thm.~\ref{thm:general inner bound} provides a unified framework from which
we may derive simpler inner bounds that may  be more easily manipulated and understood.
In particular one would like an achievable rate region to be expressed in terms of the rate bounds directly on $R_1$ and $R_2$ rather than  on the rates corresponding to the rate-split messages.
Such a region may be obtained by eliminating the sub-rates from the rate region expression using the Fourier-Motzkin elimination procedure. Fourier-Motzkin elimination yields an analytically manageable number of rate bounds only for a relatively small number of rate splits.
In this section we introduce a series of sub-schemes containing a limited number of auxiliary RVs and derive the corresponding Fourier-Motzkin eliminated rate regions (resulting in $(R_1,R_2)$ rate regions) which are then compared to the outer bounds derived in Section~\ref{sec:OuterBound}. In addition to these sub-schemes being more analytically tractable due to the small number of auxiliary random variables and rate-splits, these particular sub-schemes were chosen as they are natural extensions of schemes that achieve capacity when the IFC-CR reduces to specific classes of CIFC, IFC and BC channels. Table~\ref{tab:subschemes} illustrates the different sub-schemes and for which classes of channels this reduces to capacity.

\subsubsection{All private messages: scheme with only $(X_1,X_2,U_{1  {\rm pb}},U_{2  {\rm pb}})$}
% in Thm.~\ref{thm:general inner bound}}
\label{sec:All private messages}

This sub-scheme is obtained by setting the rate of the common messages to zero.
It illustrates the effect of binning performed at the cognitive relay
to pre-code against the interference due to the non-intended message at
each destination.

\begin{cor}
\label{cor:all private messages}
By considering %only $(X_1,X_2,U_{1  {\rm pb}},U_{2  {\rm pb}})$
$U_{1  {\rm c}}=U_{2  {\rm c}}=U_{0  {\rm cb}}=\emptyset$
in Thm.~\ref{thm:general inner bound} the following rate region is achievable
\eas{
R_1 & \leq  I(Y_1; X_1,U_{1  {\rm pb}}|Q)-I(X_2; U_{1  {\rm pb}}|X_1, Q) \label{eq:all private messages R1}\\
%    &  =   I(Y_1; X_1)+I(Y_1,U_{1  {\rm pb}}|X_1)-I(X_2; U_{1  {\rm pb}}| X_1)\\
R_2 & \leq  I(Y_2; X_2,U_{2  {\rm pb}}|Q)-I(X_1; U_{2  {\rm pb}}|X_2, Q) \label{eq:all private messages R2}\\
%    &   =  & I(Y_2; X_2) + I(Y_2; U_{2  {\rm pb}}|X_2)-I(X_1; U_{2  {\rm pb}}| X_2)\\
R_1 + R_2 &\leq   I(Y_1; X_1,U_{1  {\rm pb}}|Q)+ I(Y_2; X_2,U_{2  {\rm pb}}|Q)
\nonumber\\ &- I(X_2; U_{1  {\rm pb}}| X_1, Q) - I(X_1; U_{2  {\rm pb}}| X_2, Q)
\nonumber\\ & -I(U_{1  {\rm pb}};U_{2  {\rm pb}}|X_1,X_2, Q) \label{eq:all private messages R1+R2}
}{\label{eq:all private messages}}
for all the distributions that factors as
\[
P_{Q}P_{X_1|Q}P_{X_2|Q}P_{X_c,U_{1  {\rm pb}},U_{2  {\rm pb}}|X_1,X_2,Q}.
\]
%{   say connection with DET CIFC and DET-BC }
\end{cor}

\begin{IEEEproof}
The proof may be found in Appendix~\ref{app:cor:all private messages}.
\end{IEEEproof}

The graphical representation of the achievable scheme in Cor.~\ref{cor:all private messages} is provided in Fig~\ref{fig:allPrivate}.

The scheme in Cor.~\ref{cor:all private messages}
achieves capacity (see Table~\ref{tab:subschemes})
when the channel reduces to a semi-deterministic BC~\cite{marton1977capacity,GelFandPinskerClassic}
and to a semi-deterministic CIFC~\cite{RTDjournal1}; in these two cases the private broadcasted
RV for the destination with noiseless output must equal the noiseless channel output;
if both destination outputs are noiseless, the optimal assignment is $U_{1  {\rm pb}}=Y_1$ and $U_{2  {\rm pb}}=Y_2$.

\subsubsection{All common messages: scheme with only $(U_{1  {\rm c}},U_{2  {\rm c}},U_{0  {\rm cb}})$}
% in Thm.~\ref{thm:general inner bound}}
\label{sec:All public messages}

We now  consider an achievability scheme where both decoders decode both messages and where, therefore, no binning or rate splitting is necessary.
%This scheme is obtained by considering { DT: $X_1=U_{1  {\rm c}}, X_2=U_{2  {\rm c}}, X_c=U_{0  {\rm cb}}$}
%only the RVs $(U_{1  {\rm c}},U_{2  {\rm c}},U_{1  {\rm cb}},U_{2  {\rm cb}})$  with $U_{0  {\rm cb}}=(U_{1  {\rm c}},U_{2  {\rm cb}}$ in the scheme
%in  Thm.~\ref{thm:general inner bound}.

%{ DT: CHANGED THE COR.: USE THE INPUTS RATHER THAN THE AUX.RV. WE ALSO ANNA USE THE FORM BELOW FOR CAP. IN VIS}
\begin{cor}
\label{cor:all public messages}
By considering $X_1=U_{1  {\rm c}}, X_2=U_{2  {\rm c}}, X_c=U_{0  {\rm cb}}$
and $U_{1  {\rm pb}}=U_{2  {\rm pb}}=\emptyset$
in Thm.~\ref{thm:general inner bound} the following rate region is achievable
%The scheme with $(U_{1  {\rm c}},U_{2  {\rm c}},U_{0  {\rm cb}})$ achieves the rate region
%\eas{
%R_1 & \leq  I(Y_1;U_{1  {\rm c}}|U_{2  {\rm c}}), \\
%R_2 & \leq  I(Y_2; U_{2  {\rm c}}| U_{1  {\rm c}}), \\
%R_1 + R_2 &\leq  I(Y_1; U_{1  {\rm c}}, U_{2  {\rm c}}), \\
%R_1 + R_2 &\leq   I(Y_2; U_{1  {\rm c}}, U_{2  {\rm c}}),
%}{\label{eq:all public messages}}
%for distributions that factor as
%\[
%P_{U_{1  {\rm c}}}P_{U_{2  {\rm c}}}P_{X_1|U_{1  {\rm c}}}P_{X_2|U_{2  {\rm c}}}P_{X_c|U_{1  {\rm c}},U_{2  {\rm c}}}.
%\]
\eas{
R_1 & \leq  I(Y_1; X_1, X_c|X_2,Q), \\
R_2 & \leq  I(Y_2; X_2, X_c| X_1,Q), \\
R_1 + R_2 &\leq  I(Y_1; X_1, X_2, X_c|Q), \label{eq:all public messages sumrate 1}\\
R_1 + R_2 &\leq  I(Y_2; X_1, X_2, X_c|Q), \label{eq:all public messages sumrate 2}
}{\label{eq:all public messages}}
for all distribution that factors as
\[
P_{Q}P_{X_1|Q}P_{X_2|Q}P_{X_c|X_1,X_2,Q}.
\]
\end{cor}

\begin{IEEEproof}
The proof may be found in Appendix~\ref{app:cor:all public messages}.
\end{IEEEproof}

A graphical representation of the achievable rate region in Cor.~\ref{cor:all public messages} is depicted in Fig.~\ref{fig:AllPublic}.

This scheme achieves capacity (see Table~\ref{tab:subschemes})
when the channel reduces to a CIFC in the ``very strong interference'' regime of~\cite{maric2005capacity} and to a IFC in the ``strong interference'' regime of~\cite{sato2002capacity}.
%In this scheme the sole role of the cognitive relay is to relay the messages of the two transmitters.
%It does so by sending  $X_c$, a deterministic function of the two messages. IT ALWAYS DOES IT!!!
%In this scenario the achievable rate region may be obtained by considering the equivalent IFC channel
%\[
%\sum_{x_c} P_{Y_1,Y_2| X_1,X_2,X_c}P_{X_c|X_1,X_2}
%\]
%obtained by fixing an encoding strategy $P_{X_c|X_1,X_2}$ at the cognitive relay.
%{ STEFANO: not sure what you mean by ``virtual interference channel'' -- not very precise. Perhaps we can just leave out this comment?}

\subsubsection{One common and one private message:  scheme with only $(X_{1},U_{2  {\rm c}},U_{1  {\rm pb}})$}
% in Thm.~\ref{thm:general inner bound}}
\label{sec:OnePrivateOnePublic}

For a CIFC in the ``very weak interference'' regime, capacity is achieved by a fully common primary message and full private cognitive message~\cite{WuDegradedMessageSet}. We extend this transmission strategy to the IFC-CR by considering the case where one of the two source messages is private while the other is common. %The cognitive relay relays part of the private message and the public message.

%
%{ DT: I'Vie CHANGED THE COR.: USE THE INPUTS RATHER THAN THE AUX.RV. WARE IS THE BINNING RATE FOR U1BP?
%AND WHY IS THERE A BOUND ON R2 DEPENDING OPT Y1? QUESTION: WHY DO NOT USE U2PB TO PRECANCEL THE PRIVATE MESSAGE OF SOURCE 1? TO ME THIS REGION SEEMS SMALLER THAN THE ``ALL COMMON'' ONE ....

\begin{cor}
\label{cor:OnePrivateOnePublic}
By considering $U_{1  {\rm c}}=\emptyset, X_2=U_{2  {\rm c}}=U_{0  {\rm cb}}, U_{2  {\rm pb}}=\emptyset, U_{1  {\rm pb}}=X_c$ in Thm.~\ref{thm:general inner bound} the following rate region is achievable
%The scheme with $(X_1,U_{2  {\rm c}},U_{1  {\rm pb}})$ achieves the rate region
\eas{
%% &\text{DT NEW}\\
%%R_1    & \leq  I(Y_1 ; X_1, X_c| X_2, Q), \\
%%R_2    & \leq  I(Y_2; X_2|Q), \\
%%R_1+R_2& \leq  I(Y_1 ; X_2, X_1, X_c|Q),
%%\\
%% &\text{SR OLD}\\
%R_1    & \leq  I(Y_1 ; X_1, U_{1  {\rm pb}}| U_{2  {\rm c}}, Q), \\
%R_2    & \leq  \min\{I(Y_2; U_{2  {\rm c}}|Q), \ I(Y_1; U_{1  {\rm pb}}, U_{2  {\rm c}} | X_1, Q) \}, \\
%R_1+R_2& \leq  I(Y_1 ; U_{2  {\rm c}}, X_1, U_{1  {\rm pb}}|Q),
R_1    & \leq  I(Y_1 ; X_1, X_c| X_2, Q), \\
R_2    & \leq  I(Y_2; X_2|Q), \\
R_2    & \leq  I(Y_1; X_c, X_2| X_1, Q) \} \}, \\
R_1+R_2& \leq  I(Y_1 ; X_2, X_1, X_c|Q),
}{\label{eq:OnePrivateOnePublic}}
for all distribution that factors as
\[
P_{Q}P_{X_1|Q}P_{X_2|Q} P_{X_c,U_{1  {\rm pb}}| X_1,X_2,Q}.
%P_{Q}P_{X_1|Q}P_{X_2,U_{2  {\rm c}}|Q} P_{X_c,U_{1  {\rm pb}}| X_1,X_2,U_{2  {\rm c}},Q}.
\]
\end{cor}
%}
\begin{IEEEproof}
The proof may be found in Appendix~\ref{app:cor:OnePrivateOnePublic}.
\end{IEEEproof}

A graphical representation of the achievable rate region of Cor.~\ref{cor:OnePrivateOnePublic} is depicted in Fig.~\ref{fig:OnePrivateOnePublic2}.

This scheme achieves capacity (see Table~\ref{tab:subschemes})
when the channel reduces to a CIFC in the very weak interference regime~\cite{WuDegradedMessageSet}.

%\subsection{ One private and one public message with unidirectional cognitive cooperation $U_{1  {\rm p}},U_{2  {\rm c}},U_{1  {\rm pb}}$ }
%
%{\blue come on...find a better name}
%
%\eas{
%R_{2  {\rm c}}+R_{1  {\rm p}}+R_{1  {\rm pb}} \leq I(Y_1; U_{1  {\rm p}},U_{2  {\rm c}},U_{1  {\rm pb}}) \\
%R_{2  {\rm c}}+ \ \ \ \quad R_{1  {\rm pb}} \leq I(Y_1; U_{2  {\rm c}},U_{1  {\rm pb}}|U_{1  {\rm p}}) \\
%      R_{1  {\rm p}}+R_{1  {\rm pb}} \leq I(Y_1; U_{1  {\rm p}},U_{1  {\rm pb}}|U_{2  {\rm c}}) \\
%R_{1  {\rm pb}} \leq I(Y_1; U_{1  {\rm pb}}| U_{1  {\rm p}},U_{2  {\rm c}}) \\
%\nonumber \\
%R_2 \leq I(Y_2 ; U_{2  {\rm c}})
%}
%
%which can be FME as
%\eas{
%R_1 & \leq & I(Y_1 ; U_{1  {\rm c}}, U_{1  {\rm p}}| U_{2  {\rm c}}) \\
%R_2 & \leq &  I(Y_2; X_2) \\
%R_2 & \leq &  I(Y_1; U_{1  {\rm pb}}, X_2 | U_{1  {\rm c}}) \\
%R_1+R_2  & \leq & I(Y_1 ; U_{2  {\rm c}}, U_{1  {\rm p}}, U_{1  {\rm pb}})
%}

\subsubsection{Common messages for the sources and private messages from the cognitive relay: scheme with only $(U_{1  {\rm c}},U_{2  {\rm c}},U_{1  {\rm pb}})$}
% only in Thm.~\ref{thm:general inner bound}}
\label{sec:OnePrivateOnePublic2}

Here we aim to expand the scheme that achieves capacity it the ``very weak interference'' regime for the CIFC~\cite{WuDegradedMessageSet} (see Table~\ref{tab:subschemes}) by having the two sources transmit common messages while the cognitive relay sends part of a private message for source~1.
%{ ND: so both this and case 3 can achieve capacity for very weak interference CIFC, you just tried to extend it in different ways?}

\begin{cor}
\label{cor:OnePrivateOnePublic2}
By considering  $X_1=U_{1  {\rm c}}$, $X_2=U_{2  {\rm c}}$, $X_c=U_{1  {\rm pb}}$, $U_{0  {\rm cb}}=U_{1  {\rm pb}}$, $U_{2pb}=\emptyset$ in Thm.~\ref{thm:general inner bound} the following rate region is achievable
\eas{
R_1 & \leq   I(Y_1;X_1,X_c|X_2,Q) \\
R_1 & \leq   I(Y_1; X_c|X_1,X_2)+I(Y_2;X_1|X_2,Q) \\
R_2 & \leq   I(Y_1;X_2,X_c|X_1,Q)\\
R_2 & \leq   I(Y_2;X_2|X_1,Q) \\
R_1+R_2 & \leq   I(Y_1;X_1,X_2,X_c,Q) \\
R_1+R_2 & \leq   I(Y_1;X_2,X_c|X_1,Q)+I(Y_2;X_1|X_2,Q) \\
R_1+R_2 & \leq   I(Y_1; X_c|X_1,X_2,Q)+ I(Y_2;X_2,X_2,Q)\\
R_1 + 2 R_2 & \leq  I(Y_1;X_2,X_c|X_1)+I(Y_2;X_1,X_2,Q)
}{\label{eq:OnePrivateOnePublic2}}
for some distributions that factor as
\[
P_{Q}P_{X_1|Q}P_{X_1|Q} P_{X_c| X_1,X_2,Q}.
\]
%{ DT: AGAIN, WHERE ARE THE INPUTS? WITHOUT BINNING $U_{1  {\rm c}}=X_1,U_{2  {\rm c}}=X_2,U_{1  {\rm pb}}=X_c$ IS OPTIMAL, RIGHT?}
\end{cor}

\begin{IEEEproof}
The proof may be found in Appendix~\ref{app:cor:OnePrivateOnePublic2}.
\end{IEEEproof}

A graphical representation of the achievable rate region of Cor.~\ref{cor:OnePrivateOnePublic2} is depicted in Fig.~\ref{fig:OnePrivateOnePublic}.

\section{Capacity in ``very strong interference at Rx~1'' and in ``strong interference at both Rxs''}
\label{sec:Capacity}

In this section we show the achievability of the outer bound in Cor.~\ref{cor:strong int outer bound}
in the ``very strong interference at Rx~1'' and the ``strong interference at both Rxs'' regime
(to be defined later), which are two subsets of the ``strong interference'' regime defined by~\eqref{eq:strong int. cond. at Rx1}.
These results parallel the ``very strong interference'' capacity result
for the IFC~\cite{CostaElGamal87} and the CIFC~\cite{maric2005capacity}, where,
the channel reduces to a compound two-user MAC.
For this class of channels the interfering signal at each receiver can
be decoded without loss of optimality.
Since the interference can always be distinguished from the intended
signal, there is no need to perform interference pre-coding at the
cognitive relay.
This greatly simplifies the achievable scheme required to match the
outer bound in Cor.~\ref{cor:strong int outer bound} and the
simple superposition coding scheme in Cor.~\ref{cor:all public messages}
will be shown to be optimal.

\begin{thm} {\bf Capacity in   ``very strong interference at Rx~1''.}
\label{th:ifc-cr very strong int}
If
%$I(Y_2 ; X_2 , X_c |  X_1 ) \leq I(Y_1 ; X_2 , X_c | X_1 )$ (``strong interference'' condition of~\eqref{eq:strong int. cond. at Rx1}) holds together with the ``very strong interference'' condition
\eas{
I(Y_2 ;      X_2, X_c |  X_1 ) &\leq I(Y_1 ;      X_2 , X_c | X_1 )\label{eq:very strong int. condition 1  same} \\
I(Y_1 ; X_1, X_2, X_c )        &\leq I(Y_2 ; X_1 , X_2, X_c )      \label{eq:very strong int. condition 1 extra}
}{\label{eq:very strong int. condition 1}}
holds for all distributions that factor as $P_{X_1,X_2,X_c}=P_{X_1}P_{X_2}P_{X_c|X_1,X_2}$
(same factorization as in~\eqref{eq:factorization condition cor:strong int outer bound}),
then the region in Cor.~\ref{cor:strong int outer bound} is capacity.
\end{thm}

\begin{IEEEproof}
Under the condition in~\eqref{eq:very strong int. condition 1  same}
(which is the same as the ``strong interference at Rx~1'' condition in~\eqref{eq:strong int. cond. at Rx1})
the region in~\eqref{eq:cor:strong int outer bound}
is an outer bound for the considered IFC-CR.
Consider now the achievable rate region in Cor.~\ref{cor:all public messages}
given by~\eqref{eq:all public messages}.
Under the condition in~\eqref{eq:very strong int. condition 1 extra}
the sum-rate bound in~\eqref{eq:all public messages sumrate 2} is redundant
and the resulting region coincides with
the outer bound in~\eqref{eq:cor:strong int outer bound}.
\end{IEEEproof}

\begin{thm} {\bf Capacity in  ``strong interference at both Rxs''.}
\label{th:ifc-cr double very strong int}
If
\eas{
I(Y_2 ;      X_2, X_c |  X_1 ) &\leq I(Y_1 ;      X_2 , X_c | X_1 )\label{eq:very strong int. condition 1  samesame} \\
I(Y_1 ;      X_1, X_c |  X_2 ) &\leq I(Y_2 ;      X_1 , X_c | X_2 )\label{eq:very strong int. condition 2  samesame}
}{\label{eq:very strong int. condition at both}}
holds for all distributions that factor as $P_{X_1,X_2,X_c}=P_{X_1}P_{X_2}P_{X_c|X_1,X_2}$
(same factorization as in~\eqref{eq:factorization condition cor:strong int outer bound}),
then the region in~\eqref{eq:all public messages} is capacity.
% $I(Y_2 ; X_2 , X_c |  X_1 ) \leq I(Y_1 ; X_2 , X_c | X_1 )$ (``strong interference at Rx~1'' condition of~\eqref{eq:strong int. cond. at Rx1}) holds together with
%the  $I(Y_1 ; X_1 , X_c |  X_2 ) \leq I(Y_2 ; X_1 , X_c | X_2 )$ (``strong interference at Rx~2'' condition of~\eqref{eq:strong int. cond. at Rx1})
%%
%for all distributions that factor as $P_{X_1,X_2,X_c}=P_{X_1}P_{X_2}P_{X_c|X_1,X_2}$ (equation~\eqref{eq:factorization condition cor:strong int outer bound}),
%then the region in~\eqref{eq:cor:strong int outer bound} of Thm.~\ref{cor:strong int outer bound} is the capacity region.
% capacity.
\end{thm}

\begin{IEEEproof}
The proof follows similarly to that of Thm.~\ref{th:ifc-cr very strong int}.
\end{IEEEproof}

\section{The Gaussian Case}
\label{sec:The Gaussian Case}
In the following, to obtain more of a feel for the channel model
and the conditions under which capacity holds, we evaluate
the ``strong interference'' outer bound conditions and the region in Cor.~\ref{cor:strong int outer bound}, as well as
the ``very strong interference''  capacity conditions and the region in Thm.~\ref{th:ifc-cr very strong int}
for the Gaussian IFC-CR (G-IFC-CR).

\subsection{Channel Model}
The G-IFC-CR is shown in Fig.~\ref{fig:Gaussian IFC-CR}.
Without loss of generality
(see Appendix~\ref{app:standard form})
we can restrict our attention to the G-IFC-CR in \emph{standard form}
%in which the noise variance and the second moment constraint for the
%channel inputs are normalized to one, i.e.,:
%where the input output relationship is expressed as
given by:
\eas{
Y_1=|h_{11}| X_1+ | h_{2c}| X_c + h_{12} X_2 + Z_1,\\
Y_2=|h_{22}| X_2+ | h_{2c}| X_c + h_{21} X_1 + Z_2,
}{\label{eq:standard form IFC-CR}}
where $h_{i} \in \CC$, $i \in \{11,1c,12,22,2c,21\}$, are constant and
known to all terminals, $Z_i \sim \Nc_\CC(0, 1)$, $i \in \{1,2\}$,
and $\EE[|X_i|^2] \leq 1$, $i \in \{1,2,c\}$.
The channel links $h_{i}, i \in \{11,22,1c,2c\}$
%direct link gains $h_{11}$ and $h_{22}$
can be taken to be real-valued without loss of generality because receivers and transmitters
can compensate for the phase of the signals.
%One of the coefficients $h_{2c}$ or $h_{2c}$ can also be fixed to be real-valued, but we prefer a the expression in~\eqref{eq:standard form channel definition} to obtain a fully symmetric channel.
The correlation among
the noises is irrelevant because the capacity  of the channel without
receiver cooperation only depends on the noise marginal distributions.

\begin{figure}
\centering
\includegraphics[width=8cm]{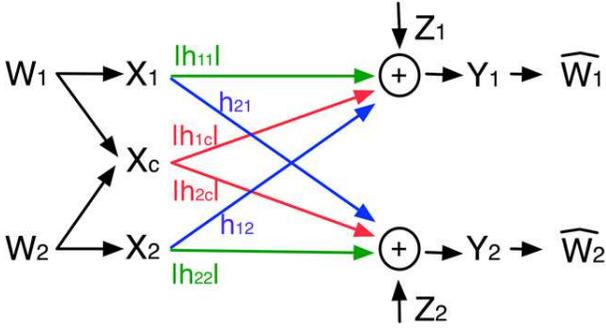}
%\vspace{-0.4in}
\caption{The Gaussian IFC-CR in standard form.}
\label{fig:Gaussian IFC-CR}
\end{figure}

\subsection{Gaussian Channel under ``strong interference at Rx 1''}
We now evaluate Cor.~\ref{cor:strong int outer bound}  and Thm.~\ref{th:ifc-cr very strong int} for the G-IFC-CR.

\begin{thm}{\bf The ``strong interference at Rx 1'' outer bound for the G-IFC-CR.}
\label{thm:strong int. outer bound Gaussian}
If
%\eas{
%&|h_{2c}|^2  \leq  |h_{2c}|^2 \label{eq:strong interfernce gaussian condition 1},\\
%& |h_{22}|^2 +|h_{2c}|^2-| h_{12}|^2 -|h_{2c}|^2  \leq 2 \labs |h_{22}| |h_{2c}|-h_{12}|h_{2c}|    \rabs ,
%}{\label{eq:strong int. conditions Gaussian}}`
%\ea{
%\max_{|\beta_2| \leq 1 }\labs |h_{22}|+\beta_2 |h_{2c}| \rabs^2-\labs h_{12}+\beta_2|h_{2c}| \rabs^2 \leq 0
%%|h_{2c}| \rabs^2 \nonumber \\
%%\quad -(|h_{2c}|^2-|h_{2c}|^2)  (1-|\beta_2|^2) \leq 0.
\ea{
\Big| |h_{22}|+\widetilde{\be}_{2} \ |h_{2c}| \Big|^2 \leq
\Big|  h_{12} +\widetilde{\be}_{2} \ |h_{1c}| \Big|^2
\label{eq:strong int. conditions Gaussian}
}
for
\eas{
\measuredangle \widetilde{\be}_{2}& = \measuredangle\big( |h_{2c}||h_{22}|-|h_{1c}|h_{12} \big),  \\
|\widetilde{\be}_{2}|^2 &= \lcb \p{
1              & {\rm if} \  |h_{2c}| \geq |h_{1c}| \\
\min \lcb 1, \f{\big||h_{2c}||h_{22}|-|h_{1c}|h_{12}\big|}{\big| |h_{2c}|^2-|h_{1c}|^2\big|}
\rcb              & {\rm if} \  |h_{2c}| < |h_{1c}|
             }  \rnone
}{\label{eq:max solution strong int}}
the capacity of a G-IFC-CR is contained in the set:
\eas{
R_1 &\leq  \Ccal \lb \labs|h_{11}| +\be_{1}^*|h_{1c}| \rabs^2  + |h_{1c}|^2(1-|\be_{1}|^2+ |\be_{2}|^2) \rb ,\\
R_2 &\leq  \Ccal \lb \labs|h_{22}| +\be_{2}^*|h_{2c}| \rabs^2  + |h_{2c}|^2(1-|\be_{1}|^2+ |\be_{2}|^2)\rb ,\\
R_1+ R_2  &\leq  \Ccal \lb
||h_{11}|+ \be_{1}^*|h_{1c}||^2+| h_{12} + \be_{2}^*|h_{1c}| |^2 \right.\nonumber\\&\left.\quad
+ |h_{1c}|^2(1-|\be_{1}|^2+ |\be_{2}|^2)
 \rb,
}{\label{eq:strong int outer bound region }}
taken over the union of all $(\beta_1,\beta_2)\in \CC^2 : |\beta_1|^2+|\beta_2|^2 \leq 1$,
where  $\Ccal(x) : = \log(1+x)$.
\end{thm}

\begin{IEEEproof}
The proof may be found in Appendix~\ref{app:proof eq:strong int. conditions Gaussian}.
\end{IEEEproof}

\begin{thm}{\bf Capacity  in ``very strong interference at Rx 1''  for
the Gaussian IFC-CR.}
\label{th:Capacity in the ``very strong interference'' regime for the G-IFC-CR}

If in addition to the condition in~\eqref{eq:strong int. conditions Gaussian} the following also holds
\ea{
\Big\{
&(|h_{11}|^2+|h_{1c}|^2+|h_{12}|^2)-(|h_{21}|^2+|h_{2c}|^2+|h_{22}|^2)
\nonumber \\&
 + 2\sqrt{ \big| |h_{11}||h_{1c}|- h_{21} |h_{2c}|\big|^2
 +         \big|  h_{12} |h_{1c}|-|h_{22}||h_{2c}|\big|^2 }
\Big\}
\leq 0,
\label{eq:VSI}
}
then the region in~\eqref{eq:strong int outer bound region } is capacity.
\end{thm}

\begin{IEEEproof}
The proof may be found in Appendix~\ref{app:proof eq:VSI}.
\end{IEEEproof}

\begin{rem}
Thm.~\ref{th:Capacity in the ``very strong interference'' regime for the G-IFC-CR}
reduce to known capacity results in the ``very strong interference'' regime when the IFC-CR
reduces to a simpler channel:
\begin{itemize}
\item
When the IFC-CR reduces to an IFC, i.e., $|h_{1c}|=|h_{2c}|=0$,
the condition in~\eqref{eq:strong int. conditions Gaussian}
reduces to the well-known ``strong interference at Rx 1''  $|h_{22}|^2 \leq |h_{12}|^2$,
and the condition in~\eqref{eq:VSI} to
$|h_{11}|^2+|h_{12}|^2
\leq
 |h_{21}|^2+|h_{22}|^2$
(larger total received power at Rx 2 than at Rx 1).

\item
When the IFC-CR reduces to a C-IFC with user~1 as primary user, i.e., $|h_{22}|=h_{12}=0$,
the condition in~\eqref{eq:strong int. conditions Gaussian} reduces to
$|h_{2c}|^2 \leq |h_{1c}|^2$ (strong interference at the primary receiver)
and the condition in~\eqref{eq:VSI} to
\ean{
&|h_{11}|^2+|h_{1c}|^2 -|h_{21}|^2-|h_{2c}|^2
\\&\lag
+ 2 \big| |h_{11}||h_{1c}|- h_{21} |h_{2c}|\big|
\leq 0,
%\label{eq:strong int CIFC}
}
which is the same as the condition in~\cite[Thm.II.3]{RTDjournal2}.

\item
When the IFC-CR reduces to a C-IFC with user~2 as primary user, i.e., $|h_{11}|=h_{21}=0$,
the conditions in~\eqref{eq:strong int. conditions Gaussian} and ~\eqref{eq:VSI} are
equivalent to $I(Y_1;X_2,X_c)=I(Y_2;X_2,X_c)$ for all input distributions, that is,
\ean{
&\{h_{12}=|h_{22}|, \ |h_{1c}|=|h_{2c}|\}
\\&
\quad\text{or}\quad
\{h_{12}=|h_{2c}|, \ |h_{22}|=|h_{1c}|\}.
}
%\[
% |h_{1c}|^2+|h_{12}|^2=
% |h_{2c}|^2+|h_{22}|^2, \quad
%    h_{12} |h_{1c}|
% = |h_{22}||h_{2c}|.
%\]

\item
When the IFC-CR reduces to a BC. i.e., $|h_{11}|=h_{21}=|h_{22}|=h_{12}=0$
the conditions in~\eqref{eq:strong int. conditions Gaussian}
and ~\eqref{eq:VSI} are equivalent to $I(Y_1;X_c)=I(Y_2;X_c)$ for all input distributions, that is,
a BC with statistically equivalent receivers, i.e., $|h_{2c}| = |h_{1c}|$.

\end{itemize}
\end{rem}

\begin{thm}{\bf Capacity in  ``strong interference at both Rxs'' for the G-IFC-CR.}
\label{thm:double strong interference capacity}
When the condition in~\eqref{eq:strong int. conditions Gaussian}
along with the symmetric condition for source-destination pair~2
hold, the region
\eas{
R_1 &\leq  \Ccal \lb \labs|h_{11}| +\be_{1}^*|h_{1c}| \rabs^2  + |h_{1c}|^2(1-|\be_{1}|^2+ |\be_{2}|^2) \rb ,\\
R_2 &\leq  \Ccal \lb \labs|h_{22}| +\be_{2}^*|h_{2c}| \rabs^2  + |h_{2c}|^2(1-|\be_{1}|^2+ |\be_{2}|^2) \rb ,\\
R_1+ R_2  &\leq  \Ccal \lb
||h_{11}|+ \be_{1}^*|h_{1c}||^2+| h_{12} + \be_{2}^*|h_{1c}| |^2  \right.\nonumber\\&\left.\quad
+ |h_{1c}|^2(1-|\be_{1}|^2+ |\be_{2}|^2)
 \rb,\\
R_1+ R_2  &\leq  \Ccal \lb
| h_{21} + \be_{1}^*|h_{2c}||^2+||h_{22}|+ \be_{2}^*|h_{2c}| |^2  \right.\nonumber\\&\left.\quad
+ |h_{2c}|^2(1-|\be_{1}|^2+ |\be_{2}|^2)
 \rb,
}{\label{eq:strong int at both Rx Gaussian region}}
taken over the union of all $(\beta_1,\beta_2)\in \CC^2 : |\beta_1|^2+|\beta_2|^2 \leq 1$
is capacity.
\end{thm}

\begin{IEEEproof}
The proof follows similarly to the one of Thm.~\ref{thm:strong int. outer bound Gaussian}.
\end{IEEEproof}

\begin{rem}
Thm.~\ref{thm:double strong interference capacity}
reduce to known capacity results when the IFC-CR
reduces to a simpler channel:
\begin{itemize}

\item
When the IFC-CR reduces to an IFC, i.e., $|h_{1c}|=|h_{2c}|=0$,
the condition in~\eqref{eq:strong int. conditions Gaussian}
reduces to the well-known ``strong interference'' regime,
$\{ |h_{22}|^2 \leq |h_{12}|^2, \ |h_{11}|^2 \leq |h_{21}|^2\}$.

\item
When the IFC-CR reduces to a C-IFC with user~1 as primary user, i.e., $|h_{22}|=h_{12}=0$ or $X_2=\emptyset$,
%the condition in~\eqref{eq:strong int. conditions Gaussian}  reduces to
%$|h_{2c}|^2 \leq |h_{1c}|^2$ (strong interference at the primary receiver)
%while the equivalent condition for destination~2 (which is
%given by~\eqref{eq:strong int. conditions Gaussian} with the role of the users swapped) is
%equivalent to~\eqref{eq:strong int CIFC}. I
interestingly,
the ``very strong interference at Rx~1'' condition is equivalent to
the ``strong interference at both Rx's'' condition.
This can be seen by noticing that for $X_2=\emptyset$ the conditions in~\eqref{eq:very strong int. condition 1}
coincide with the conditions in~\eqref{eq:very strong int. condition at both}.

\item
When the IFC-CR reduces to a C-IFC with user~2 as primary user, i.e., $|h_{11}|=h_{21}=0$,
we have the equivalent of case $|h_{22}|=h_{12}=0$ in the above bullet point but with the
role of the users swapped.

\item
When the IFC-CR reduces to a BC. i.e., $|h_{1c}|=|h_{2c}|=|h_{22}|=h_{12}=0$
the ``strong interference at both Rx's'' condition and the
``very strong interference at Rx~1'' conditions are the same and are
%conditions in~\eqref{eq:strong int. conditions Gaussian} for both receivers are
equivalent to $I(Y_1;X_c)=I(Y_2;X_c)$ for all input distributions, that is, $|h_{1c}| = |h_{2c}|$.

\end{itemize}
\end{rem}

\subsection{Gaussian Channel under ``weak interference''}
\label{sec:Gaussian Channel under ``weak interference''}
The condition in~\eqref{eq:weak int. cond. at Rx1} for the ``weak interference at Rx~2'' outer bound in Cor.~\ref{cor:weak int outer bound} is, in general, very hard to verify as it must hold for a large set of distribution involving an auxiliary RV. In this section we restrict attention to a special class of G-IFC-CR in which the condition in~\eqref{eq:weak int. cond. at Rx1} is easily verified, namely a class of ``degraded'' G-IFC-CR defined by
\ea{
\f{h_{21}}{|h_{11}|} =\f{|h_{2c}|}{|h_{1c}|} := |\rho| \in[0,1],
%\f{|h_{1c}|}{|h_{11}|}= \f{|h_{2c}|}{h_{21}}:=|\al_c|, \quad |h_{11}|\geq |h_{21}|,
\label{eq:degraded gaussian IFC-CR condition}
}
so that the channel input/output relationship becomes
\eas{
Y_1 &=       |h_{11}| X_1+ |h_{1c}| X_c  + h_{12}  X_2+Z_1 \\
Y_2 &=|\rho|(|h_{11}| X_1+ |h_{1c}| X_c) +|h_{22}| X_2+Z_2.
%Y_1 &=  |h_{11}| ( X_1+|\al_c| X_c )+h_{12}   X_2+Z_1 \\
%Y_2 &=  |h_{21}| ( X_1+|\al_c| X_c )+|h_{22}| X_2+Z_2.
}{\label{eq:degraded Gaussian IFC-CR outputs}}
Since the noise correlation among the noises is irrelevant for capacity,
conditioned on $X_2$ we have the following Markov chain
\ea{
 &X_{\rm eq} \to Y_1 \to Y_2, \label{eq:degraded Gaussian IFC-CR markov chain}
\\&\quad \quad X_{\rm eq} := |h_{11}| X_1+ |h_{1c}| X_c, \nonumber
\\&\quad \quad Y_2 \sim |\rho| Y_1 + \sqrt{1-|\rho|^2} Z_0, \nonumber
\\&\quad \quad Z_0 \sim \Nc_{\CC}(0,1) \ \text{independent of everything else}, \nonumber
%\\&\quad \quad X_{\rm eq} := X_1+|\al_c| X_c, \nonumber
%\\&\quad \quad Y_2 \sim \f{|h_{21}| }{|h_{11}| } Y_1 + \sqrt{1-\f{|h_{21}|^2}{|h_{11}|^2}} Z_0, \nonumber
%\\&\quad \quad Z_0 \sim \Nc_{\CC}(0,1) \ \text{independent of everything else}, \nonumber
}
in other words, conditioned on $X_2$, the channel in~\eqref{eq:degraded gaussian IFC-CR condition}
is equivalent to a SISO degraded BC with input $X_{\rm eq}$.
From~\eqref{eq:degraded Gaussian IFC-CR markov chain} and for any $P_{U,X_1,X_2,X_c}$ such that
$U \to (X_1,X_2,X_c) \to (Y_1,Y_2)$ we have that
\[
I(U; Y_2|X_2) \leq I(U; Y_1|X_2),
\]
which is exactly the ``weak interference at Rx~2'' condition in~\eqref{eq:weak int. cond. at Rx1}.

\begin{thm} {\bf The ``weak interference at Rx~2'' outer bound for the degraded G-IFC-CR.}
\label{th:Weak interference outer bound for the degraded Gaussian IFC-CR}
For the degraded G-IFC-CR in~\eqref{eq:degraded gaussian IFC-CR condition}
the capacity region is contained into the region
\eas{
R_1 & \leq \Ccal \lb  \labs |h_{11}|+ |h_{1c}| \beta_{1}^* \rabs ^2\al \rb
%\lb  |h_{11}|^2 \labs 1+ |\al_c| \beta_{1}^* \rabs ^2\al \rb
\label{eq:weak int outer bound expression degraded Gaussian IFC-CR R1}\\
R_2 & \leq \Ccal \lb |\rho|^2\labs |h_{11}|+ |h_{1c}| \beta_{1}^* \rabs ^2
           +\lb |h_{22}| +|\rho||h_{1c}| \beta_{2}^* \rb^2 \rb
\nonumber\\
&\quad - \Ccal \lb |\rho|^2\labs |h_{11}|+ |h_{1c}| \beta_{1}^* \rabs ^2\al  \rb
%\Ccal \lb |h_{21}|^2 \labs 1+ |\al_c| \beta_{1}^* \rabs^2
%            +\lb |h_{22}| +|h_{21}| |\al_c| \beta_{2}^* \rb^2 \rb
%\nonumber\\
%&\quad - \Ccal \lb |h_{21}|^2   \labs 1+ |\al_c| \beta_{1}^* \rabs^2 \al  \rb
\label{eq:weak int outer bound expression degraded Gaussian IFC-CR R2-1} \\
R_2 & \leq \Ccal \lb \lb |h_{22}| +|\rho||h_{1c}| \beta_{2}^* \rb^2 \rb,
%\lb |h_{22}| +|h_{21}| |\al_c| \beta_{2}^* \rb^2 \rb
\label{eq:weak int outer bound expression degraded Gaussian IFC-CR R2-2}
}{\label{eq:weak int outer bound expression degraded Gaussian IFC-CR}}
taken over the union of all $\al \in [0,1]$ and $(\beta_{1},\beta_{2})$ such that $|\beta_{1}|^2+|\beta_{2}|^2 =  1$.
\end{thm}
% DT: LOOK AT THE APPENDIX FOR NEW SIMPLIFIED LEANER PROOF. DOUBLE CHECK IT!!!

\begin{IEEEproof}
%By using the extremal inequality~\cite{liu2007extremal}
%we show the optimality of jointly Gaussian inputs for the degraded G-IFC-CR in~\eqref{eq:degraded gaussian IFC-CR condition}
%and explicitly compute the outer bound Thm.~\ref{cor:weak int outer bound}.
The proof can be found in Appendix~\ref{app:weak interference outer bound for the degraded Gaussian IFC-CR}.
\end{IEEEproof}

\begin{rem}
Special cases for the outer bound in Thm.~\ref{th:Weak interference outer bound for the degraded Gaussian IFC-CR}:
\begin{itemize}

\item
When  $|h_{1c}|=0$, %=|h_{2c}|
the channel in~\eqref{eq:degraded Gaussian IFC-CR outputs}
reduces to an IFC with ``weak interference'' at receiver~2
%\ean{
%Y_1 &=       |h_{11}| X_1 +h_{12}  X_2+Z_1 \\
%Y_2 &= |\rho||h_{11}| X_1+|h_{22}| X_2+Z_2,
%%Y_2 &=  |h_{21}| X_1+|h_{22}| X_2+Z_2.
%}
whose capacity is not known.
The outer bound in Thm.~\ref{th:Weak interference outer bound for the degraded Gaussian IFC-CR} in this case
%becomes:
%\pp{
%R_1 & \leq \Ccal \lb |h_{11}|^2 \al \rb \\
%R_2 & \leq \Ccal \lb \f{(1-\al) |h_{21}|^2+|h_{22}|^2} {|h_{21}|^2\al +1}\rb    \\
%R_2 & \leq \Ccal( |h_{22}|^2)
%}
%which is, in general,
is looser than the outer bounds in~\cite{kramer2004outer,etkin_tse_wang}.
However, the Sato-type outer bound in Thm.~\ref{thm:general outer ITW dublin}
reduces to~\cite{kramer2004outer} and the tightened outer bound in~\cite{rini2010dublin}
reduces to~\cite{etkin_tse_wang}.

\item
When the IFC-CR reduces to a C-IFC with user~1 as primary user, i.e., $|h_{22}|=h_{12}=0$,
the channel in~\eqref{eq:degraded Gaussian IFC-CR outputs}
reduces to a Gaussian degraded CIFC~\cite{RTDjournal1}
%\ean{
%Y_1 &=         |h_{11}| X_1+ h_{12} X_c +Z_1 \\
%Y_2 &=  |\rho|(|h_{11}| X_1+ h_{12} X_c)+Z_2.
%%Y_2 &=  |h_{21}| ( X_1+|\al_c| X_c )+Z_2.
%}
whose capacity is not known.
The outer bound in Thm.~\ref{th:Weak interference outer bound for the degraded Gaussian IFC-CR} in this case
%becomes:
%\pp{
%R_1 & \leq \Ccal \lb  |h_{11}|^2 \labs 1+ |\al_c| \beta_{1}^* \rabs ^2\al \rb \\
%R_2 & \leq \Ccal \lb |h_{21}|^2 \labs 1+ |\al_c| \beta_{1}^* \rabs^2+\lb |h_{21}|^2 |\al_c|^2 |\beta_{2}|^2 \rb^2 \rb \\
%&\quad - \Ccal \lb |h_{21}|^2   \labs 1+ |\al_c| \beta_{1}^* \rabs^2 \al  \rb\\
%R_2 & \leq \Ccal \lb \lb |h_{21}|^2 |\al_c|^2 |\beta_{2}|^2 \rb^2 \rb
%}
%which is, in general,
is looser that the outer bound in~\cite[Cor. 3.5]{RTDjournal1}.
In this case, the best known outer bound in~\cite[Cor. 3.5]{RTDjournal1}
is still of BC-type, from a MIMO BC with degraded message set however.

\item
When the IFC-CR reduces to a C-IFC with user~2 as primary user, i.e., $|h_{11}|=0$, %=h_{21}
the channel in~\eqref{eq:degraded Gaussian IFC-CR outputs}
reduces to a Gaussian CIFC in weak interference~\cite{WuDegradedMessageSet}
%\eas{
%Y_1 &=      |h_{1c}| X_c  + h_{12}  X_2+Z_1 \\
%Y_2 &=|\rho||h_{1c}| X_c  +|h_{22}| X_2+Z_2,
%}
whose capacity is known~\cite{WuDegradedMessageSet,JovicicViswanath06}.
The outer bound in Thm.~\ref{th:Weak interference outer bound for the degraded Gaussian IFC-CR} in this case
reduces to capacity.
%%we have the degenerate case where $Y_1$ and $Y_2$ are
%%noisy versions of $X_2$:
%%%{ DT:
%%\ean{
%%Y_1 &=  h_{12}   X_2+Z_1 \\
%%Y_2 &=  |h_{22}| X_2+Z_2.
%%}
%%The outer bound of Thm. \ref{th:Weak interference outer bound for the degraded Gaussian IFC-CR} is $(R_1=0, R_2=\Ccal(|h_{22}|^2)$ which is capacity.
%%%}

\item
When the IFC-CR reduces to a BC. i.e., $|h_{11}|=h_{21}=|h_{22}|=0$, %=|h_{22}|=h_{12}
the channel in~\eqref{eq:degraded Gaussian IFC-CR outputs} reduces to a
degraded SISO BC whose capacity is known~\cite{bergmans1973random}.
The outer bound in Thm.~\ref{th:Weak interference outer bound for the degraded Gaussian IFC-CR} in this case
 reduces to capacity.

\end{itemize}

\end{rem}

\section{Numerical Evaluations}

%{\blue
%Stefano is working on it as of now
%%
%%Put comment that strong interference
%%
%%what happens when we have both strong interference for both users, then the outer bound wcorresponds to the achievable region
%}

In this section we present a series of numerical evaluations of the results presented in the paper for the G-IFC-CR with real-valued inputs and real-valued channel coefficients. Using numerical examples, we investigate the relationship between inner and outer bounds as well as the position and extension of the ``strong'', ``weak'' and ``very strong'' interference regimes.

\begin{figure*}%[ht]
\centering
\subfigure[ The ``strong'' (blue, hatched) and the ``very strong interference at Rx~1'' (blue, cross-hatched) regimes ]{
\includegraphics[width=8.5 cm]{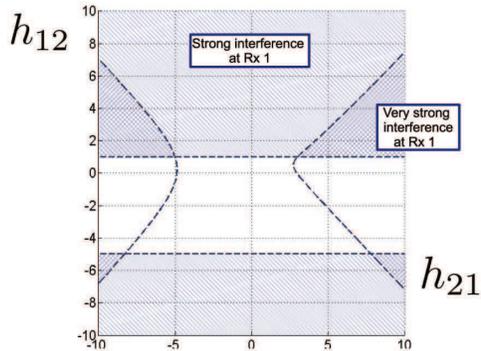}
   \label{fig:strongIntRegionRX1}
 }
 \hfill
\subfigure[The ``strong'' (green, hatched) and the ``very strong interference at Rx~2'' (green, cross-hatched) regimes.]{
\includegraphics[width=8.5 cm]{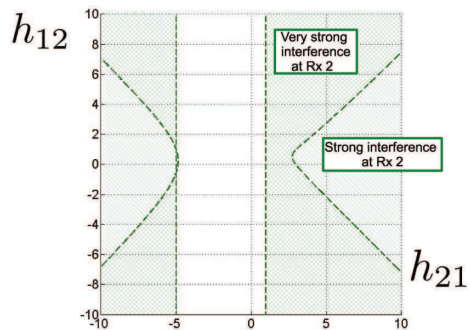}
   \label{fig:strongIntRegionRX2}
 }
 \\
 \subfigure[The ``strong interference at Rx~1'' (green hatched) and the ``strong interference at Rx~2'' (blue-hatched) regimes.]{
\includegraphics[width=8.5 cm]{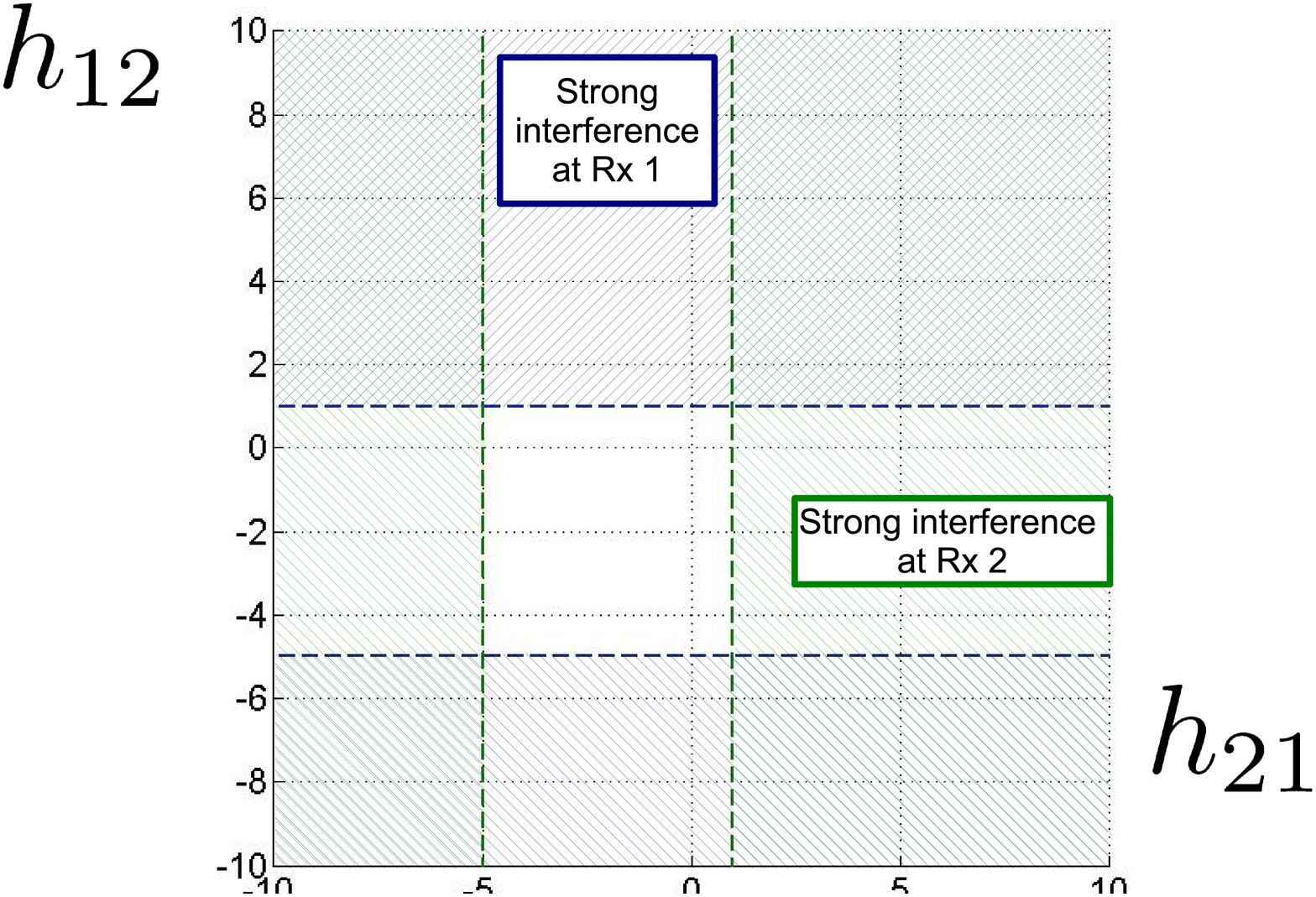}
   \label{fig:doubleStrong}
 }
 \hfill
 \subfigure[The degraded the G-IFC-CR for Rx~1(blue, dotted) and Rx~2 (green, dotted ) and the ``weak interference'' regime for Rx~1 (blue solid) and Rx~2 (green solid) ]{
  \includegraphics[width=8.5 cm]{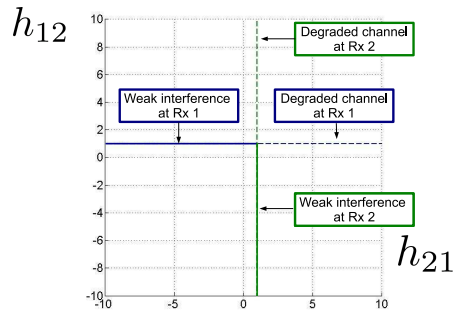}
   \label{fig:degraded}
 }
\caption{Different parameter regimes for G-IFC-CR with $h_{11}=h_{22}=1$, $h_{1c}=h_{2c}=2$ and $[h_{12},h_{21}] \in [-10,10]\times[-10,10]$. }
\label{fig:all regimes}
\end{figure*}

In Fig.~\ref{fig:all regimes} we depict
\begin{itemize}
  \item Fig.~\ref{fig:strongIntRegionRX1}:
  the ``strong interference at Rx~1'' regime  of~\eqref{eq:strong int. conditions Gaussian} and the ``very strong interference at Rx~1'' regime of~\eqref{eq:VSI},
  \item Fig.~\ref{fig:strongIntRegionRX2}:
  the ``strong interference at Rx~2'' regime  of~\eqref{eq:strong int. conditions Gaussian} and the ``very strong interference at Rx~2'' regime of~\eqref{eq:VSI},
  \item Fig.~\ref{fig:doubleStrong}:
  the ``strong interference'' regime  of~\eqref{eq:strong int. conditions Gaussian}  at Rx~1 and at Rx~2 and the ``strong interference at both Rxs'' regime of Thm.~\ref{thm:double strong interference capacity},
  \item Fig.~\ref{fig:degraded}:
  the degraded G-IFC-CR of~\eqref{eq:degraded gaussian IFC-CR condition} and the  ``weak interference'' regime of Thm.~\ref{th:Weak interference outer bound for the degraded Gaussian IFC-CR},
\end{itemize}
for fixed $h_{11}=h_{22}=h_{1c}=h_{2c}=1$ on the plane  $[h_{12},h_{21}] \in [-10,10]\times[-10,10].$

Since $|h_c|=|h_{1c}|=|h_{2c}|$, from~\eqref{eq:max solution strong int} we have that  the ``strong interference'' condition becomes linear in $h_{21}$ and $h_{12}$,
i.e.
condition~\eqref{eq:strong int. conditions Gaussian} becomes:
\ea{
\Big| |h_{11}|+ |h_{2c}| \Big|^2 \leq \Big|  h_{21} + |h_{2c}| \Big|^2 \\
\lb |h_{11}|-h_{21} \rb \lb |h_{11}|+h_{21}+2 |h_c| \rb \leq 0
\label{eq:strong int. conditions Gaussian hc=h1c=h2c}
}
Similarly, since $|h_c|=|h_{1c}|=|h_{2c}|$, the degraded condition at destination~1 in~\eqref{eq:degraded gaussian IFC-CR condition} coincides with
 $|h_{11}|=h_{21}$: from this consideration and given~\eqref{eq:strong int. conditions Gaussian hc=h1c=h2c},     we have that the degraded channel at destination~1
 is also in ``strong interference'' at destination~2.
Given the symmetry of the channel,  we also have that the degraded channel at destination~2 is also in ``strong interference'' at destination~1.

%\begin{figure*}%[ht]
%\centering
%\includegraphics[width=16cm]{plotStrongHcVarynig-1}
%\caption{The conditions in~\eqref{eq:boundaries different regimes} for $|h_{11}|=|h_{22}|=1$,
%$|h_{1c}|=|h_{2c}|\in[1:5]$ and $(h_{12},h_{21}) \in [-10:10]\times [-10:10]$.}
%\label{fig:plotStrongHcVarynig}
%\end{figure*}

\begin{figure*}
\centering
\subfigure[The condition $I(Y_1; X_1,X_2,X_c)=I(Y_2; X_1,X_2,X_c)$ for increasing $h_{1c}=h_{2c}$]{
   \includegraphics[width =12 cm] {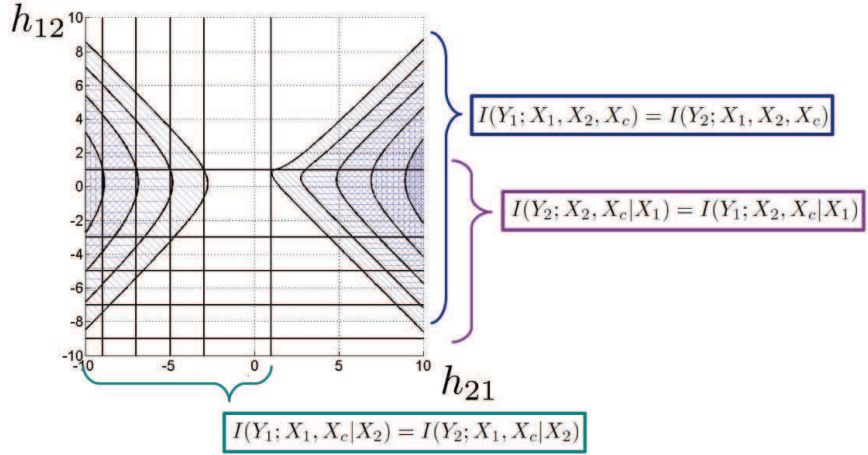}
   \label{fig:F1}
 }
 \subfigure[The condition $I(Y_1; X_2,X_c|X_1)=I(Y_2; X_2,X_c|X_1)$ for increasing $h_{1c}=h_{2c}$ ]{
   \includegraphics[width =12 cm] {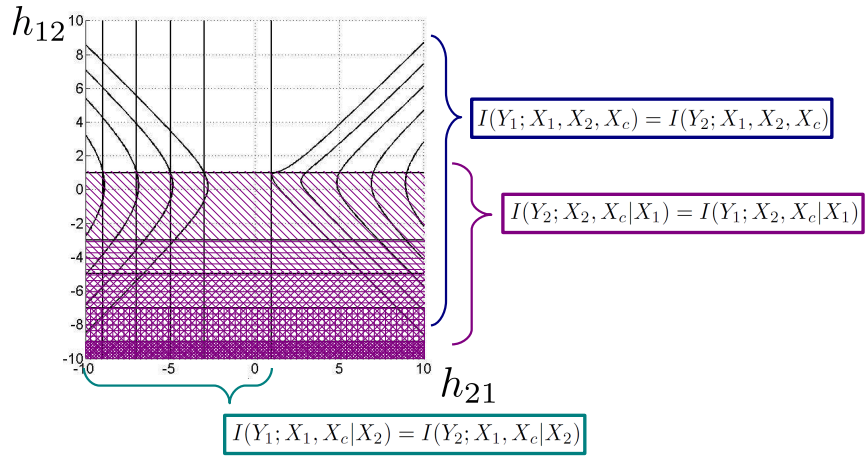}
   \label{fig:F2} }
 \subfigure[The condition $I(Y_1; X_1,X_c|X_2)=I(Y_2; X_1,X_c|X_2)$ for increasing $h_{1c}=h_{2c}$ ]{
   \includegraphics[width =12 cm] {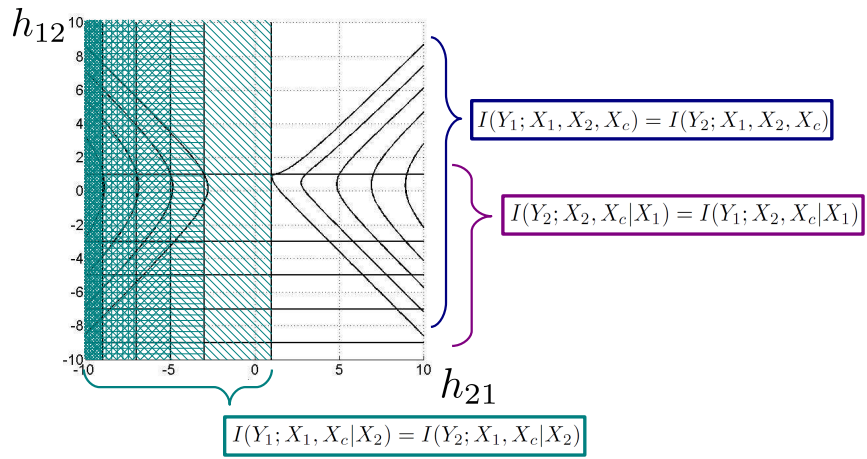}
   \label{fig:F3}
 }
\caption{The conditions in~\eqref{eq:boundaries different regimes} for $|h_{11}|=|h_{22}|=1$ and $h_{1c}=h_{2c} \in\{1 \ldots 5\} $.}
\label{fig:plotStrongHcVarynig}
\end{figure*}

In Fig.~\ref{fig:plotStrongHcVarynig} we plot the conditions
\eas{
I(Y_1; X_1,X_2,X_c) &= I(Y_2; X_1,X_2,X_c)\\
I(Y_2 ; X_2 , X_c |  X_1 ) &= I(Y_1 ; X_2 , X_c | X_1 )\\
I(Y_1 ; X_1 , X_c |  X_2 ) &= I(Y_2 ; X_1 , X_c | X_2 )
}{\label{eq:boundaries different regimes}}
for increasing values of $|h_c|=|h_{1c}|=|h_{2c}|\in[1,5]$ for fixed $|h_{11}|=|h_{22}|=1$ on the plane  $[h_{12},h_{21}] \in [-10,10]\times[-10,10]$.
The line corresponding to each condition marks the boundary of the ``strong interference'' and the ``very strong interference'' conditions at destination~1 and~2.
The darker hues are associated with smaller values of $|h_c|$ while lighter hues with larger values.
While the boundaries of the ``strong interference'' regime are always linear in $h_{12},h_{21}$ for any $|h_c|$, the ``very strong interference''  condition is approximated by an hyperbole for large $h_{21}$ and $h_{12}$.

\begin{figure*}%[ht]
\centering
\includegraphics[width=16 cm]{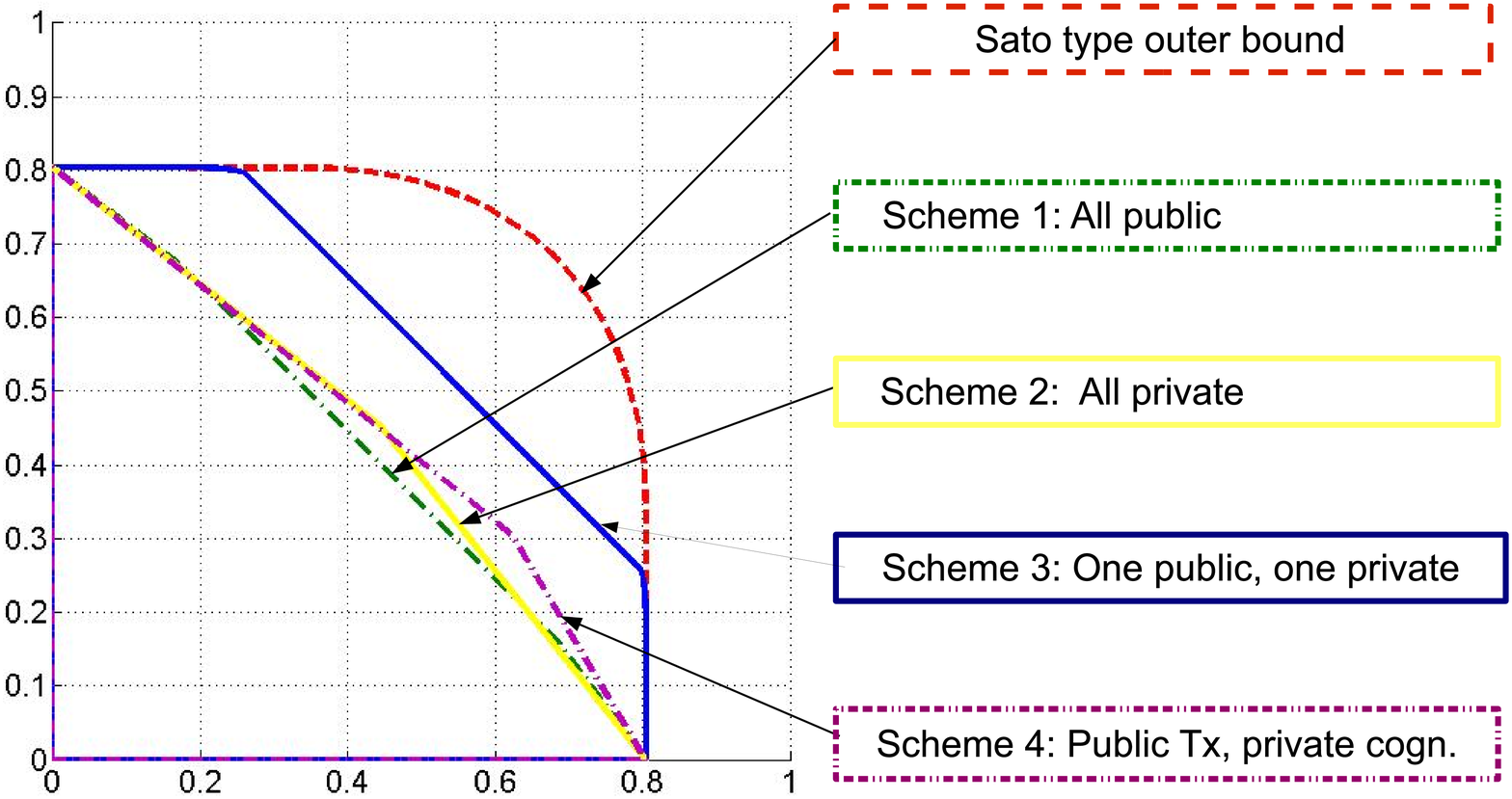}
\caption{A plot for $|h_{11}|=|h_{22}|=|h_{1c}|=|h_{2c}|=1$ and $h_{12}=h_{21}=-2$.}
\label{fig:plot-2-2}
\end{figure*}

\begin{figure*}%[ht]
\centering
\includegraphics[width=16 cm]{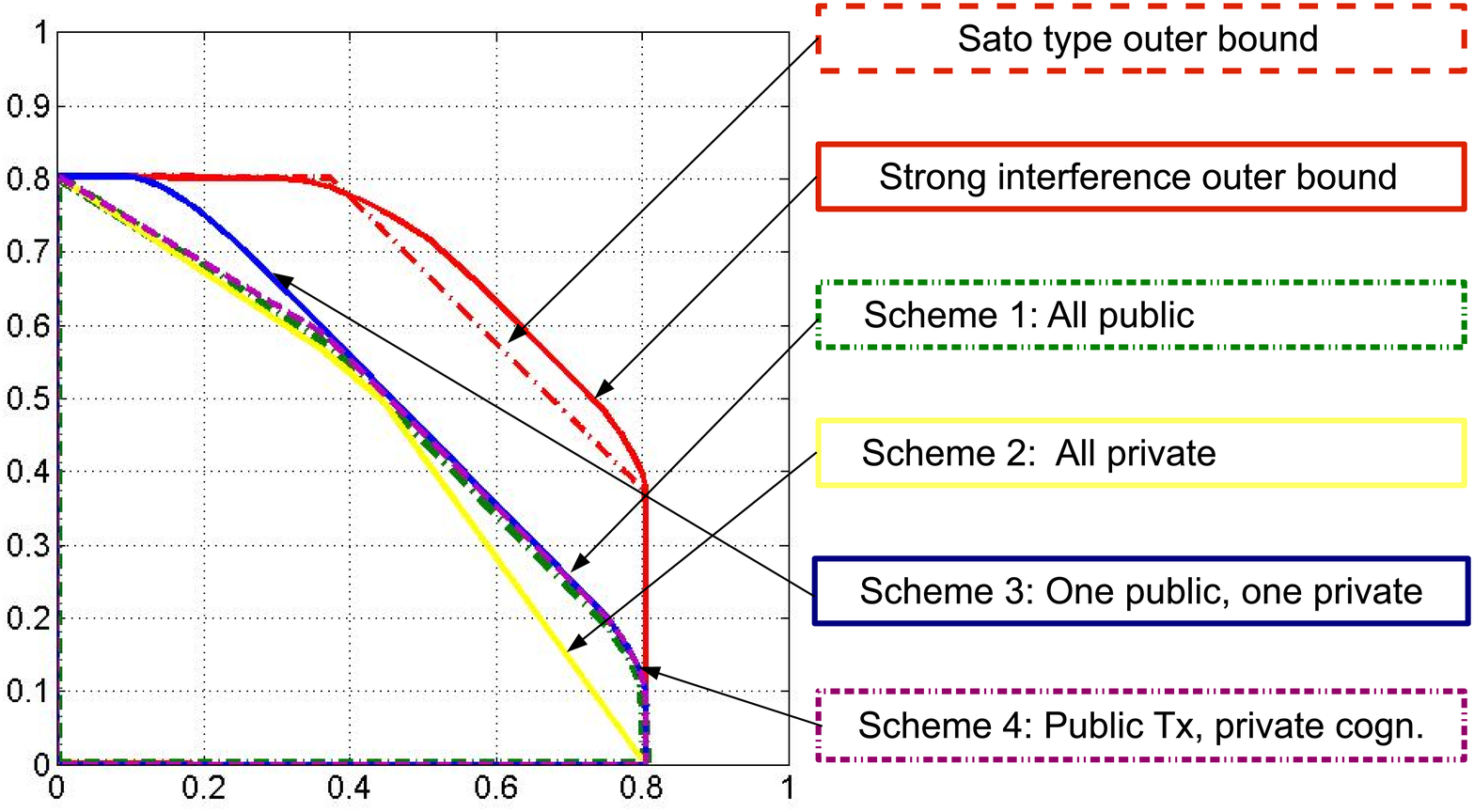}
\caption{A plot for $|h_{11}|=|h_{22}|=|h_{1c}|=|h_{2c}|=1$ and $h_{12}=-2, h_{21}=+1$.}
\label{fig:plot-2+1}
\end{figure*}

\begin{figure*}%[ht]
\centering
\includegraphics[width=16 cm]{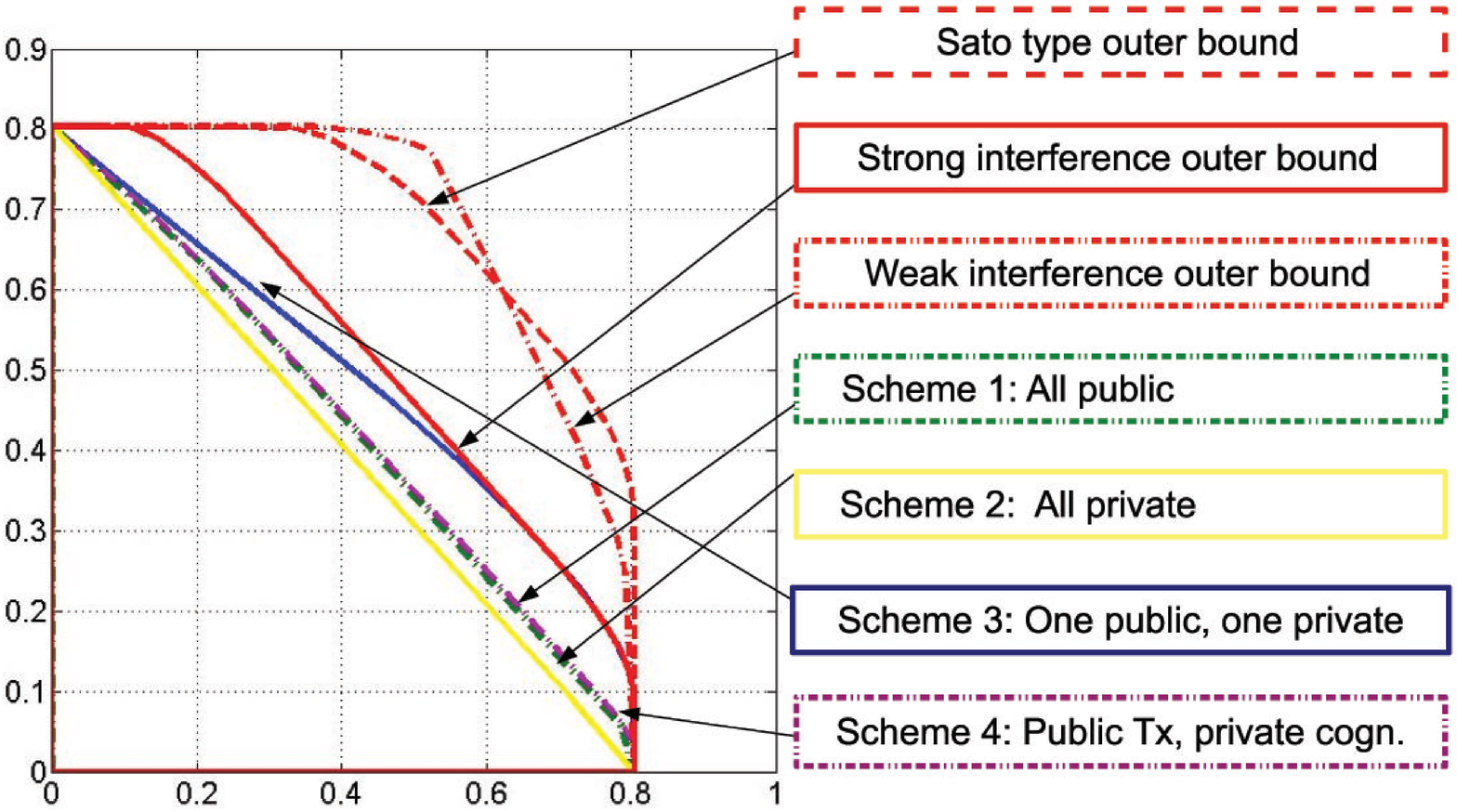}
\caption{A plot for $|h_{11}|=|h_{22}|=|h_{1c}|=|h_{2c}|=1$ and $h_{12}=.5, h_{21}=+1$.}
\label{fig:plot05+1}
\end{figure*}

In Figs.~\ref{fig:plot-2-2},~\ref{fig:plot-2+1}  and~\ref{fig:plot05+1}  we  compare inner and outer bounds for three points in the plane   $[h_{12},h_{21}] \in [-10:10]\times[-10,10]$  for fixed $|h_{11}|=|h_{22}|=|h_{1c}|=|h_{2c}|=1$ :
\begin{itemize}
  \item Fig.~\ref{fig:plot-2-2}: $(h_{12},h_{21})=(-2,-2)$, where the Sato type outer bound of Thm.~\ref{thm:general outer ITW dublin} holds, but not the outer bounds of Thm.~\ref{thm:strong int. outer bound Gaussian} or Thm.~\ref{th:Weak interference outer bound for the degraded Gaussian IFC-CR};

  \item Fig.~\ref{fig:plot-2+1}: $(h_{12},h_{21})=(-2,+1)$, where the Sato type outer bound of Thm.~\ref{thm:general outer ITW dublin}
  and the ``strong interference at Rx~2'' outer bound of Thm.~\ref{thm:strong int. outer bound Gaussian} hold;
  \item Fig.~\ref{fig:plot05+1}: $(h_{12},h_{21})=(0.5,+1)$, where the Sato type outer bound of Thm.~\ref{thm:general outer ITW dublin},
  the ``strong interference at Rx~2'' outer bound of Thm.~\ref{thm:strong int. outer bound Gaussian}
  and the ``weak interference at Rx~1'' outer bound of Thm.~\ref{th:Weak interference outer bound for the degraded Gaussian IFC-CR} hold.
\end{itemize}

In Fig.~\ref{fig:plot-2-2} we notice that a combination of common and private message, the scheme in Sec.~\ref{sec:OnePrivateOnePublic}, outperforms the schemes that utilize only common or only private messages, the schemes in Sec.~\ref {sec:All private messages} and Sec.~\ref{sec:All public messages}, respectively.
Despite of the good performance of the scheme in Sec.~\ref{sec:OnePrivateOnePublic}, a substantial distance between inner and outer bound can be observed.
The outer bound of Thm.~\ref{thm:more capable BC outer} is known to be capacity for the CIFC in ``weak'' interference, ``very strong'' interference and for the ``primary decodes cognitive'' regime~\cite{RTDjournal2}.
%shows that an outer bound derived from the BC with degraded message set is tight.
This result shows that the outer bound in Thm.~\ref{thm:more capable BC outer} is not tight far all the parameter region.

Fig.~\ref{fig:plot-2+1} shows that the ``strong interference'' outer bound of Cor.~\ref{cor:strong int outer bound} is tighter than the Sato-type outer bound in Thm.~\ref{thm:more capable BC outer} for some rate pairs. The scheme with one common and one private message in Sec.~\ref{sec:OnePrivateOnePublic} outperforms the schemes in Sec.~\ref{sec:All public messages}, Sec.~\ref{sec:All private messages} and Sec.~\ref{sec:OnePrivateOnePublic2} although the performance is comparable for some parameter values.

In Fig.~\ref{fig:plot05+1} we observe that the ``weak interference'' outer bound in~\ref{th:Weak interference outer bound for the degraded Gaussian IFC-CR} is tighter than the Sato-type outer bound in~\ref{thm:general outer ITW dublin} for some rate pairs, although the ``strong interference'' outer bound of~\ref{thm:strong int. outer bound Gaussian} remains the tightest in this case. For this specific choice of parameters the channel is both in ``weak interference'' at destination~1 as well as in
 ``strong interference'' at destination~2.  In this specific regime the scheme in~\ref{sec:OnePrivateOnePublic} approaches the strong interference outer bound for some parameter values. Since $Y_2$ is a degraded  version of $Y_1$ conditioned on $X_2$, loosely speaking, there is no loss of generality in having receiver~1 decode the message in $X_2$; for this reasons one expects the scheme in Sec.~\ref{sec:OnePrivateOnePublic} to perform well in this case.

%\begin{figure}
%\centering
%%\includegraphics[width=10 cm ]{SimulationGaussianIFC-CR}
%\includegraphics[width=10 cm ]{GaussianIFC-CRVSI}
%\vspace{-0.4in}
%\caption{The ``strong interference'' regime of Thm.~\ref{thm:strong int.
%outer bound Gaussian} (light blue) and the ``very strong interference''
%regime of Thm.~\ref{th:Capacity in the ``very strong interference'' regime for the Gaussian IFC-CR} for user~1 as well as the
%``strong interference'' regime (light green) and the ``very strong interference''
%regime (dark green)  for user~2
%for  the Gaussian IFC-CR
%with $h_{11}=h_{22}=h_{1c}=h_{2c}=1$ on the plane $h_{12} \times
%h_{21}=[0:10]\times [0:10]$.
%%{ DT: wrong fig!}
%}
%\vspace{-.5 cm }
%\label{fig:sim}
%\end{figure}

\section{Conclusion and Future Work}
\label{sec:Conclusion and Future Work}

We introduce  new, general outer bounds for the  IFC-CR that are
inspired by capacity results available for the broadcast channel and the cognitive interference channel.
We show the achievability of one outer
bound in the ``very strong interference'' regime by having both decoders decode
both messages as in a compound multiple access channel.  This result
is very similar in nature to the ``very strong interference'' capacity
results for the interference channel and the cognitive interference
channel.
We also derive the provably largest achievable rate region for this channel model by using classical random coding arguments such as rate
 splitting, superposition coding and binning.
This region contains all the key transmission features using in achieving capacity in channels and classes of channels for which capacity is known.
As such, this general achievable rate region is algebraically complex, but fairly general, and is shown to reduce to capacity for
 all sub-channels for which capacity is known.
% This region is capacity when the channel reduces to a simpler model where capacity is known.
%Although algebraically complex, this region contains all the key transmission features to approach capacity.
%
The contributions of this paper are a
first step to a better understanding of the capacity region of the
cognitive interference channel with  a cognitive relay which remains
largely undiscovered.

%{
%
%\begin{itemize}
%  \item change all the $U_{2  {\rm pb}}$ in $U_{2  {\rm pb}}$
%  \item explain the graph on the first figure where it appears
%  \item
%\end{itemize}
%=
%}

\bibliographystyle{IEEEtran}
\bibliography{steBib1}

%\newpage

\begin{appendices}

\section{Proof of Theorem~\ref{thm:general outer ITW dublin}}
\label{app:thm:general outer ITW dublin}

From Fano's inequality, if $P_e \to 0$ as $N\rightarrow \infty$
then
\[
H(W_i|Y_i^N) \leq N \ep_N
\quad \text{with}\quad
\ep_N  \to 0
\quad \text{as}\quad
N\rightarrow \infty,
\]
with $i\in\{1,2\}$ and thus
\[
N (R_i - \ep_N )\leq I(W_i; Y_i^N) \leq I(W_i; Y_i^N|W_{\overline{i}}), \ i\in\{1,2\},\ \overline{i}\not=i,
\]
where the last inequality in the above expression follows from the independence of the source messages.

The rate $R_1$ can be bounded as in~\eqref{eq:gen sato r1} (and similarly
for $R_2$ in~\eqref{eq:gen sato r2}) since
\ean{
  &N(R_1-\ep_N)
\\&\leq I(W_1; Y_1^N|W_2)           \\& \lag \text{Fano's inequality}
%\\& \leq  I(W_1; Y_1^N)            \\& \lag \text{Fano's inequality}
%\\&\leq I(W_1; Y_1^N,W_2)          \\& \lag \text{Non-negativity of mutual information}
%\\&=    I(W_1; Y_1^N| W_2)         \\& \lag \text{Independence of messages}
\\&=    H(Y_1^N| W_2)-H(Y_1^N| W_1 ,W_2) \\& \lag \text{Definition of mutual information}
\\&=    H(Y_1^N| W_2,    X_2^N(W_2))
\\&\quad - H(Y_1^N| W_1,W_2,X_1^N(W_1),X_2^N(W_2),X_c^N(W_1,W_2)) \\& \lag \text{Deterministic encoding}
\\&= H(Y_1^N| W_2, X_2^N)
    - \sum_{t}H(Y_{1,t}| X_{1,t},X_{2,t},X_{c,t})   \\& \lag \text{Memoryless channel}
\\&= \sum_{t}H(Y_{1,t}| W_2, X_2^N,Y_1^{t-1})
    - \sum_{t}H(Y_{1,t}| X_{1,t},X_{2,t},X_{c,t})   \\& \lag \text{Chain rule for entropy}
\\&\leq \sum_{t}H(Y_{1,t}| X_{2,t})
    - \sum_{t}H(Y_{1,t}| X_{1,t},X_{2,t},X_{c,t})   \\& \lag \text{Conditioning reduces entropy}
\\&= \sum_{t}I(Y_{1,t}; X_{1,t},X_{c,t}| X_{2,t})   \\& \lag \text{Definition of mutual information}
\\&= N \ I(Y_1 ; X_1, X_c |X_2, Q),                 \\& \lag \text{Introduction of time-sharing RV}
}
where, in the last equality, $Q$ is a time sharing RV that is independent of all other RVs
and uniformly distributed on $[1:N]$.

\medskip
Next, let $\widetilde{Y}_i^N$ have the same conditional marginal distribution as $Y_i^N$, $i\in\{1,2\}$.
The Sato-type bound~\cite{sato1978outer} sum-rate bounds in~\eqref{eq:gen sato r1+r2 a}
and~\eqref{eq:gen sato r1+r2 b} follow since
\ean{
  &N(R_1+R_2-2\ep_N)
%\\&\leq I(Y_1^N; W_1)+I(Y_2^N; W_2)  \\& \lag \text{Fano's inequality}
\\&\leq I(Y_1^N; W_1|W_2)+I(Y_2^N; W_2)                      \\& \lag \text{Fano's inequality}
\\&\leq I(Y_1^N,\widetilde{Y}_2^N; W_1| W_2) + I(Y_2^N; W_2) \\& \lag \text{Non-negativity of mutual information}
\\&= I(Y_1^N; W_1| W_2,\widetilde{Y}_2^N) + I(Y_2^N; W_1,W_2)\\& \lag \text{$\widetilde{Y}_2^N$ and $Y_2^N$ have the same marginal cdf}
\\&\leq I(Y_1^N;  X_1^N,X_c^N|\widetilde{Y}_2^N, X_2^N)
  + I(Y_2^N; X_1^N,X_2^N,X_c^N)                              %\\& \lag \text{Steps similar to the $R_1$-bound above}
\\&\leq N\Big(I(Y_1;  X_1,X_c|\widetilde{Y}_2, X_2, Q)+
I(Y_2; X_1,X_2,X_c|Q)\Big),  %\\& \lag \text{Introduction of time-sharing RV}
}
and where the last two inequalities follows from steps similar to the derivation of the
bound on $R_1$ above.

\section{Proof of Theorem~\ref{thm:more capable BC outer}}
\label{app:thm:more capable BC outer}

%The single-rate bounds in~\eqref{eq:gen r1 again} and~\eqref{eq:gen r2 again},
%as well as the sum-rate bounds in~\eqref{eq:gen r1+r2 a again} and~\eqref{eq:gen r1+r2 b again},
%also appear in Thm.~\ref{thm:general outer ITW dublin}.
%%
The bound in~\eqref{eq:gen r2 2nd}, and similarly for~\eqref{eq:gen r1 2nd}
but with the role of the users swapped, is obtained as follows
\ean{
  &N(R_2 -\epsilon_N)
\\& \leq  I(Y_2^N ; W_2)          \\& \lag \text{Fano's inequality}
\\&  =  \sum_{i=1}^N H(Y_{2,i} | Y_{2,i+1}^{N})
 -H(Y_{2,i} | Y_{2,i+1}^{N}, W_2) \\& \lag \text{Chain rule for entropy}
%%%\\&  =  \sum_{i=1}^N H(Y_{2,i} | Y_{2,i+1}^{N})
%%% -H(Y_{2,i} | Y_{2,i+1}^{N}, W_2, { X_{2,i}}) \\& \lag \text{Deterministic encoding}
\\&  \leq  \sum_{i=1}^N H(Y_{2,i})
 -H(Y_{2,i} | Y_{1}^{i-1}, Y_{2,i+1}^{N}, W_2, { X_{2,i}}) \\& \lag \text{Conditioning reduces entropy}
%%%\\&= \sum_{i=1}^N I(Y_{2,i}; U_{2,i}, { X_{2,i}}), \\& \lag \text{$U_{2,i}:=[W_2,Y_{2,i+1}^N,Y_1^{i-1}]$}
%%%\\&= N \ I(Y_2; U_2, { X_2}). \\& \lag \text{Introduction of time-sharing RV}
%%%\\&\text{ DT: SAME BOUND:}
\\&= \sum_{i=1}^N I(Y_{2,i}; V_{i}, U_{2,i}, { X_{2,i}}), %\\&\text{definition of $U_{2,u}:=[W_u], u\in\{1,2\}$ and $V_{i}:=[Y_{2,i+1}^N,Y_1^{i-1}]$}
%\\&= N \ I(Y_2; V, U_2, { X_2}|Q)   \\& \lag \text{Introduction of time-sharing RV}
%\\&\leq N \ I(Y_2; V, U_2, { X_2}). \\& \lag \text{Replace $(V,Q)$ by a new $V$}
}
where we defined
\ean{
  &U_{u,i}:=[W_u], u\in\{1,2\},
\\&V_{i}:=[Y_{2,i+1}^N,Y_1^{i-1}].
}

\medskip
The bound of~\eqref{eq:gen r2 1st}, and similarly for~\eqref{eq:gen r1 1st}
but with the role of the users swapped, is obtained as follows
\ean{
  &N(R_2 -\epsilon_N)
\\& \leq  I(Y_2^N ; W_2 |W_1)          %&&\text{Fano's inequality}
\\&  =  \sum_{i=1}^N H(Y_{2,i} | Y_{2,i+1}^{N}, W_1, X_{1,i})
\\&\lag -H(Y_{2,i} | Y_{2,i+1}^{N}, W_2, { X_{2,i}}, W_1, X_{1,i}, X_{c,i})
\\&  \leq  \sum_{i=1}^N H(Y_{2,i}| W_1, X_{1,i})
\\&\lag -H(Y_{2,i} | Y_{1}^{i-1}, Y_{2,i+1}^{N}, W_2, { X_{2,i}}, W_1, X_{1,i}, X_{c,i})
\\&=  \sum_{i=1}^N I(Y_{2,i}; V_{i},  U_{2,i}, { X_{2,i}},  X_{c,i}| U_{1,i}, X_{1,i})
\\&=\sum_{i=1}^N I(Y_{2,i};X_{2,i},X_{c,i}|U_{1,i},X_{1,i}).
%\\&\leq N \ I(Y_2; V, U_2, { X_2, X_c}|U_1, { X_1})
}

\medskip
The sum-rate bound in~\eqref{eq:gen r1+r2 1st}, and similarly for~\eqref{eq:gen r1+r2 2nd}
but with the role of the users swapped, is obtained as
\ean{
  &N(R_1+R_2 - 2\epsilon_{N})
\\& \leq  I(Y_1^N ; W_1 | W_2) +I(Y_2^N ; W_2)
\\& \leq   \sum_{i=1}^N
     I(Y_{1,i} ; W_1,Y_{2,i+1}^N | Y_1^{i-1},W_2,{ X_{2,i}})
\\&\lag  +I(Y_{2,i} ; W_2,{ X_{2,i}},Y_{2,i+1}^N)
\\& =   \sum_{i=1}^N
     I(Y_{1,i} ;Y_{2,i+1}^N |Y_1^{i-1},W_2,{ X_{2,i}})
\\&\lag -I(Y_{2,i} ;Y_1^{i-1}| W_2,{ X_{2,i}},Y_{2,i+1}^N)
\\&\lag +I(Y_{1,i} ;W_1 | Y_1^{i-1},Y_{2,i+1}^N,W_2,{ X_{2,i}})
\\&\lag +I(Y_{2,i} ;W_2,{ X_{2,i}},Y_{2,i+1}^N,Y_1^{i-1})
\\&\stackrel{\rm(a)}{=} \sum_{i=1}^N
    I(Y_{1,i} ;W_1 | Y_1^{i-1},Y_{2,i+1}^N,W_2,{ X_{2,i}})
\\&\lag +I(Y_{2,i} ;Y_{2,i+1}^N,Y_1^{i-1},W_2,{ X_{2,i}})
\\&=  \sum_{i=1}^N
         I(Y_{1,i} ;U_{1,i}, X_{1,i}, X_{c,i} |V_{i}, U_{2,i}, { X_{2,i}}),
\\&\lag +I(Y_{2,i} ;V_{i}, U_{2,i}, { X_{2,i}}),
%\\&=  N \ \Big( I(Y_{1} ;U_{1},  X_{1}, X_{c}|V, U_{2}, { X_{2}})
%                  +I(Y_{2} ;V, U_{2}, { X_{2}})  \Big)
\\&= \sum_{i=1}^N  I(Y_{1,i};X_{1,i},X_{c,i}|V_i,U_{2,i},X_{2,i})
+ I(Y_{2,i};V_i,U_{2,i},X_{2,i})
}
where the equality in (a)  follows from the ``Csisz\'{a}r's  sum identity''~\cite{csiszar1982information}.
Note that the Markov chain in~\eqref{eq:markov chain thm:generalBC outer} holds since for all $i\in[1:N]$
we have
\ean{
V_i
 \to (U_{1,i},U_{2,i})
 \to (X_{1,i},X_{2,i},X_{c,i})
 \to (Y_{1,i},Y_{2,i})
}
owing to the cognition structure and the memoryless channel that imply
\ean{
  &P_{W_1,W_2,X_1^N,X_2^N,X_c^N,Y_1^N,Y_2^N}
\\&= P_{W_1} P_{W_2}
\prod_{i=1}^{N}\delta(W_1-U_{1,i})\delta(W_2-U_{2,i})
P_{X_{1,i}|U_{1,i}}
P_{X_{2,i}|U_{2,i}}
\\& \ \
P_{X_{c,i}|U_{1,i},U_{2,i}}
P_{Y_{1,i},Y_{2,i}|X_{1,i},X_{2,i},X_{c,i}},
}
from which the factorization in~\eqref{eq:factorization thm:generalBC outer} also follows.

Note that we do not need a time sharing RV here since $Q$ can be incorporated in the RV $V$ without loss of generality.

\section{Proof of Corollary~\ref{cor:strong int outer bound}}
\label{app:cor:strong int outer bound}

Similar to \cite[Lem. 4]{MaricYatesKramer07} and \cite[Lem. 1]{CostaElGamal87},
if the condition in~\eqref{eq:strong int. cond. at Rx1}
holds for all distributions in~\eqref{eq:factorization condition cor:strong int outer bound},
then
\ea{
I(Y_2; X_2, X_c | X_1, U ) \leq I(Y_1 ; X_2, X_c | X_1, U ),
\label{eq:consequence of eq:strong int. cond. at Rx1}
}
for all  $P_{X_1,X_2,X_c,U}=P_{X_1}P_{X_2} P_{X_c| X_1, X_2} P_{U|X_1,X_2,X_c}$.
From this, it follows that when condition~\eqref{eq:strong int. cond. at Rx1} holds,
the bound in~\eqref{eq:gen r1+r2 2nd} may be upper bounded as:
\begin{align*}
  & I(Y_1;V,U_1,X_1)+I(Y_2;U_2,X_2,X_c|V,U_1,X_1)
\\&=I(Y_1;V,U_1,X_1)+I(Y_2;X_2,X_c|V,U_1,X_1)
\\&\leq I(Y_1;V,U_1,X_1)+I(Y_1;X_2,X_c|V,U_1,X_1)
\\&= I(Y_1;X_2,X_c, V,U_1,X_1)
%\\
%  & I(Y_2; X_2, X_c | X_1 , U_2,Q) + I(Y_1; U_2, X_1|Q)
%\\& \leq I(Y_1; X_2, X_c |X _1, U_2,Q)+ I(Y_1; U_2, X_1|Q)
%\\& \leq  I(Y_1; X_1, X_2, X_c,U_2|Q)
\\& = I(Y_1; X_1, X_2, X_c),
\end{align*}
where the last equality  follows from the Markov chain in~\eqref{eq:markov chain thm:generalBC outer}
%$Y_1-(X_1,X_2,X_c)-U_2$ which is readily established by using the memoryless property of the channel to write
%\[ P_{Y_1,Y_2 X_1,X_2,X_c U_2} =P_{Y_1,Y_2|X_1,X_2,X_c} P_{X_1,X_2,X_c,U_2} = P_{Y_1,Y_2|X_1,X_2,X_c} P_{X_1,X_2,X_c} P_{U_2|X_1,X_2,X_c}.\]

%{ DT: THE FOLLOWING IS NOT TRUE SINCE FOR~\eqref{eq:consequence of eq:strong int. cond. at Rx1}
%WE NEED $U \to X \to Y$; SO, EVEN WITH $U \sim Y$ THE `NOISE' IN U MUST BE INDEPENDENT OF THE NOISE IN 'Y'.
%MY BAD, I CAME UP WITH THE ARGUMENT AND I ALSO ADDED AS A REMARK IN THE ISIT PAPER ... WHICH IS NOT CORRECT UNLESS
%THERE IS A DIFFERENT WAY TO FIX IT.
%
%The same can be proved starting from Thm.~\ref{thm:general outer ITW dublin}.
%By choosing $U=Y_1$ in~\eqref{eq:consequence of eq:strong int. cond. at Rx1}
%we have
%\[
%0\leq I(Y_2; X_2, X_c  | X_1, Y_1 )  \leq I(Y_1 ; X_2, X_c | X_1, Y_1) =0
%%I(Y_2; X_2, X_c  | X_1, Y_1 ) = 0.
%\]
%and thus the bound in~\eqref{eq:cor:strong int outer bound R1+R2 1} follows from the one in~\eqref{eq:gen r1+r2 2nd}.
%%\ea{
%%& I(Y_2; X_2, X_c  | X_1, Y_1 )  \leq I(Y_1 ; X_2, X_c | X_1, Y_1)=0, \implies  \nonumber \\
%%& I(Y_2; X_2, X_c  | X_1, Y_1 ) = 0. \ \ \ \ \ \
%% \label{eq:sate type}
%%}
%}

\section{Proof of Corollary~\ref{cor:weak int outer bound}}
\label{app:cor:weak int outer bound}
Consider dropping from the outer bound in Thm.~\ref{thm:more capable BC outer}
all rate constraints but~\eqref{eq:gen r1 1st},~\eqref{eq:gen r2 2nd} and~\eqref{eq:gen r1+r2 1st}, i.e.,
consider the outer bound
\eas{
R_1    &\leq I(Y_1;X_1,X_c| U_2,X_2 ),\label{eq:gen r1 2nd again}\\
R_2    &\leq I(Y_2;V,U_2,X_2),\label{eq:gen r2 2nd again}\\
R_1+R_2&\leq I(Y_2;V,U_2,X_2)
    +I(Y_1;X_1,X_c|V,U_2,X_2),\label{eq:gen r1+r2 1st again}
}{\label{eq:gen 1st}}
We intend to show that when the condition in~\eqref{eq:weak int. cond. at Rx1}
holds for all distributions in~\eqref{eq:factorization condition cor:weak int outer bound},
the region in~\eqref{eq:gen 1st} can be rewritten as
\eas{
R_1&\leq I(Y_1;X_1,X_c|V,U_2,X_2),\label{eq:gen r1+r2 1st again NOTsame}\\
R_2&\leq I(Y_2;V,U_2,X_2),        \label{eq:gen r2 2nd again same}
}{\label{eq:gen 1st rewritten}}
which is equivalent to the region in~\eqref{eq:DM ifc-cr weak int outer bound R1}--\eqref{eq:DM ifc-cr weak int outer bound R2} by defining $U=[V,U_2]$.
Successively we show how the rate bound in \eqref{eq:DM ifc-cr weak int outer bound R2-2} can be added to the region in \eqref{eq:gen 1st rewritten} to obtain a
tighter outer bound.
%{\red DT: THIS REMINDS ME OF TRYING TO SHOW THAT THE MORE-CAPABLE-BC REGION
%CAN BE SIMPLIFIED WHEN THE BC IS LESS NOISY. IT SHOULD BE AN EXERCISE IN THE CIRZAR-KORNER
%BOOK ....}
%STE: is a very close call. It
%By defining $U=[V,U_2]$, we obtain the result claimed in Cor.~\ref{cor:weak int outer bound}.

For any fixed $P_{V,U_2,X_1,X_2,X_c}$, the region in \eqref{eq:gen 1st} has three Pareto optimal points:
\begin{itemize}
\item $P_1 = \lb 0, \eqref{eq:gen r2 2nd again}\rb$,
\item $P_2 = \lb \eqref{eq:gen r1+r2 1st again}-\eqref{eq:gen r2 2nd again}, \eqref{eq:gen r2 2nd again}\rb$,
\item $P_3 = \lb \eqref{eq:gen r1 2nd again}, \eqref{eq:gen r1+r2 1st again}-\eqref{eq:gen r1 2nd again}\rb$.
\item $P_4 = \lb  \eqref{eq:gen r1 2nd again},0 \rb$.
\end{itemize}
%
%  \item $P_1=\lb 0,\min\{\eqref{eq:gen r2 2nd again},\eqref{eq:gen r1+r2 1st again}\} \rb $,
%    \item $P_2=\lb [\eqref{eq:gen r1+r2 1st again}-\eqref{eq:gen r2 2nd again}]^+,\eqref{eq:gen r2 2nd again} \rb$,
%  \item $P_3=\lb \eqref{eq:gen r1+r2 1st again},0 \rb$.
%\end{itemize}
We now show that the outer bound in \eqref{eq:gen 1st rewritten} contains each of these points. By considering the union over all the possible distributions $P_{V,U_2,X_1,X_2,X_c}$ we can conclude that the outer bound in \eqref{eq:gen 1st rewritten} is looser than \eqref{eq:gen 1st}.
The corner points $P_1$,$P_2$ and $P_4$ are also corner points of the region in \eqref{eq:gen 1st rewritten} for the same $P_{V,U_2,X_1,X_2,X_c}$.
Consider the region of \eqref{eq:gen 1st rewritten} for $V=\emptyset$, then the corner point $P_3$ is included in such region when
\ea{
&I(Y_2; U_2, X_2 ) \geq I(Y_2; V, U_2, X_2) + I(Y_1; X_1 , X_c | V, U_2, X_2) \nonumber \\
&\quad \quad \quad -I(Y_1 ; X_1, X_c| U_2, X_2) \nonumber  \\
&I(Y_2;V| U_2, X_2 ) \geq I(Y_1;V | U_2, X_2)-I(Y_1 ;V| U_2, X_2,X_1, X_c) \nonumber  \\
&I(Y_2;V| U_2, X_2 ) \geq I(Y_1;V | U_2, X_2),
\label{eq:last passage weak}
}
where \eqref{eq:last passage weak} follows from the Markov chain in   \eqref{eq:markov chain thm:generalBC outer}.
As for the  App. \ref{app:cor:strong int outer bound}, the result of \cite[Lem. 4]{MaricYatesKramer07} and \cite[Lem. 1]{CostaElGamal87}
assures that condition in \eqref{eq:weak int. cond. at Rx1}  for $U=V$ implies that
$$
I(Y_2;V| U_2, X_2 ) \geq I(Y_1;V | U_2, X_2),
$$
for any $P_{X_2, U_2,V}$, from which it follows that \eqref{eq:gen 1st} is contained into \eqref{eq:gen 1st rewritten}
when \eqref{eq:weak int. cond. at Rx1} holds.
Finally the rate bound in \eqref{eq:DM ifc-cr weak int outer bound R2-2} is obtained from \eqref{eq:DM ifc-cr weak int outer bound R2} by noticing that
\ea{
R_2 \leq I(Y_2;X_2,X_c|U_1,X_1) \leq I(Y_2;X_2,X_c|X_1)
\label{eq:weak interference Gaussian R2 bound 2}
}
The bound in \eqref{eq:weak interference Gaussian R2 bound 2} is not required to prove capacity for the CIFC in ``weak interference''
\cite{WuDegradedMessageSet,JovicicViswanath06} but it can be tighter than \eqref{eq:DM ifc-cr weak int outer bound R2} for Gaussian IFC-CR in ``weak interference'' of Sec. \ref{sec:Gaussian Channel under ``weak interference''}.
%
%This is  not the case for the Gaussian IFC-CR, as , for example, when $\al$ in \eqref{eq:weak int outer bound expression degraded Gaussian IFC-CR}, \eqref{eq:weak int outer bound expression degraded Gaussian IFC-CR R2-2} is tighter than \eqref{eq:weak int outer bound expression degraded Gaussian IFC-CR R2-1}.

\section{Proof of Theorem~\ref{thm:general inner bound}}
\label{app:inner bound}

For easy of notation we omit the time sharing RV $Q$ in the following.
The coding scheme is as follows.
\begin{itemize}

\item {\bf Class of input distributions}

Consider a distribution from~\eqref{eq:inner bound distirbution}.

\item {\bf Rate-splitting}

Each independent message $W_i$,  $i\in\{1,2\}$,
uniformly distributed on $[1:2^{N R_i}]$, %\times [1:2^{N R_2}]$,
is split into four sub-messages:
\begin{itemize}
  \item $W_{i  {\rm c}}$:  a common  message transmitted by source $i$ for both destinations,
  \item $W_{i  {\rm p}}$:  a private message transmitted by source $i$ for destination $i$,
  \item $W_{i  {\rm cb}}$: a common  message transmitted by the cognitive relay to both destinations,
  \item $W_{i  {\rm pb}}$: a private message transmitted by the cognitive relay to destination $i$.
\end{itemize}

The sub-messages $\{W_{k}\}_{k \in \{1{\rm c},2{\rm c},1{\rm p},2{\rm p}, 1{\rm cb}, 2{\rm cb},1{\rm pb}, 2{\rm pb} \} }$,
are independent with $W_{k}$ uniformly distributed on $[1:2^{N R_{k}}]$
so that
\eas{
W_1 &=(W_{1  {\rm c}},W_{1  {\rm p}},W_{1  {\rm cb}},W_{1  {\rm pb}}),
\\&\quad  R_1=R_{1  {\rm c}}+R_{1  {\rm p}}+R_{1  {\rm cb}}+R_{1  {\rm pb}}, \\
W_2 &=(W_{1  {\rm c}},W_{2  {\rm p}},W_{2  {\rm cb}},W_{2  {\rm pb}}),
\\&\quad  R_2=R_{2  {\rm c}}+R_{2  {\rm p}}+R_{2  {\rm cb}}+R_{2  {\rm pb}}.
}{\label{eq:rate split}}

\item {\bf Code-book generation}

Given any distribution in~\eqref{eq:inner bound distirbution},
the sources and the cognitive relay generate the following codebooks:
    \begin{itemize}
      \item
      Common message: $w_{i  {\rm c}}\in[1:2^{N R_{i  {\rm c}}}]$ is encoded into $U_{i  {\rm c}}^N(w_{i  {\rm c}})$ with iid distribution $P_{U_{i  {\rm c}}}$, $i\in\{1,2\}$.

      \item
      Private message: for a given $w_{i  {\rm c}}$, $w_{i  {\rm p}}\in[1:2^{N R_{i  {\rm p}}}]$ is encoded into $X_i^N(w_{i  {\rm p}}|w_{i  {\rm c}})$ with iid distribution $P_{X_i|U_{i  {\rm c}}}$, $i\in\{1,2\}$.

      \item
      Common broadcasted messages: for a given pair $(w_{1  {\rm c}},w_{2  {\rm c}})$, the pair $w_{1  {\rm cb}}\in[1:2^{N R_{1  {\rm cb}}}], \ w_{2  {\rm cb}}\in[1:2^{N R_{2  {\rm cb}}}]$ is encoded into $U_{0  {\rm cb}}^N(w_{1  {\rm cb}},w_{2  {\rm cb}},b_{0  {\rm cb}}| w_{1  {\rm c}},w_{2  {\rm c}})$, $b_{0  {\rm cb}}\in[1:2^{N R_{0  {\rm cb}}'}]$, with iid distribution $P_{U_{0  {\rm cb}}|U_{1  {\rm c}},U_{2  {\rm c}}}$.

      \item
      Private broadcasted message: for a given $(w_{1  {\rm c}},w_{2  {\rm c}},w_{1  {\rm cb}},w_{2  {\rm cb}},b_{0  {\rm cb}},w_{i  {\rm p}})$, $w_{i  {\rm pb}}\in[1:2^{N R_{i  {\rm pb}}}]$ is encoded into $U_{i  {\rm pb}}^N(w_{i  {\rm pb}},b_{i  {\rm pb}}|w_{1  {\rm c}},w_{2  {\rm c}},w_{1  {\rm cb}},w_{2  {\rm cb}},b_{0  {\rm cb}}, w_{i  {\rm p}})$, $b_{i  {\rm pb}}\in[1:2^{N R_{i  {\rm pb}}'}]$, with distribution  $P_{U_{i  {\rm pb}}|U_{1  {\rm c}},U_{2  {\rm c}},U_{0  {\rm cb}}, X_i}^N$, $i\in\{1,2\}$.

    \end{itemize}

\item {\bf Encoding}

Given $w_1=(w_{1  {\rm p}},w_{1  {\rm c}},w_{1  {\rm cb}},w_{1  {\rm pb}})$
  and $w_2=(w_{2  {\rm p}},w_{2  {\rm c}},w_{2  {\rm cb}},w_{2  {\rm pb}})$:

    \begin{itemize}
      \item
%Given the message  $w_1=(w_{1  {\rm p}},w_{1  {\rm c}},w_{1  {\rm cb}},w_{1  {\rm pb}})$,
source~1  sends $X_1^N(w_{1  {\rm p}}|w_{1  {\rm c}})$.

      \item
%Given the message  $w_2=(w_{2  {\rm p}},w_{2  {\rm c}},w_{2  {\rm cb}},w_{2  {\rm pb}})$,
source~2  sends $X_2^N(w_{2  {\rm p}}|w_{2  {\rm c}})$.

      \item
  First binning step:
  the cognitive relay looks for an index $b_{0  {\rm cb}}$  such that
%DT:WRONG, IT WAS ALREADY GENERATED WITH THAT DISTRIBUTION.
%$U_{0  {\rm cb}}^N(w_{1  {\rm cb}},w_{1  {\rm cb}},b_{0  {\rm cb}}|w_{1  {\rm c}},w_{2c})$
%appear to be generated according to the distribution $P_{U_{0  {\rm cb}}|U_{1  {\rm c}},U_{2  {\rm c}}}$.
\ea{
(
&U_{1  {\rm c}}^N(w_{1  {\rm c}}),X_1^N(w_{1  {\rm p}}|w_{1  {\rm c}}),
U_{2  {\rm c}}^N(w_{2  {\rm c}}),X_2^N(w_{2  {\rm p}}|w_{2  {\rm c}}),
\nonumber\\
&U_{0  {\rm cb}}^N(w_{1  {\rm cb}},w_{1  {\rm cb}},b_{0  {\rm cb}}|w_{1  {\rm c}},w_{2c})
)
\nonumber\\
&\in T_\ep^N(P_{U_{0  {\rm cb}},X_1,X_2,U_{1  {\rm c}},U_{2  {\rm c}}})
\label{eq:cr binning step 1}
}
If more than one such index satisfies the relationship in~\eqref{eq:cr binning step 1}, it selects one uniformly at random;
if no such index exists, it sets $b_{0  {\rm cb}}=1$ and in this case we say that a encoding error at the first binning step has occurred.

      \item
Second binning step:
Let $b_{0  {\rm cb}}^*$ be the index determined at the first binning step.
The cognitive relay looks for  a pair of indexes $(b_{1  {\rm pb}}, b_{2  {\rm pb}})$  such that
\ea{
(
&U_{1  {\rm c}}^N(w_{1  {\rm c}}),X_1^N(w_{1  {\rm p}}|w_{1  {\rm c}}),
U_{2  {\rm c}}^N(w_{2  {\rm c}}),X_2^N(w_{2  {\rm p}}|w_{2  {\rm c}}),
\nonumber\\
&U_{0  {\rm cb}}^N(w_{1  {\rm cb}},w_{1  {\rm cb}},b_{0  {\rm cb}}^*|w_{1  {\rm c}},w_{2c}),
\nonumber\\
&U_{1  {\rm pb}}^N(w_{1  {\rm pb}},b_{1  {\rm pb}}|w_{1  {\rm c}},w_{2  {\rm c}},w_{1  {\rm cb}},w_{2  {\rm cb}},b_{0  {\rm cb}}^*, w_{1  {\rm p}}),
\nonumber\\
&U_{2  {\rm pb}}^N(w_{2  {\rm pb}},b_{2  {\rm pb}}|w_{1  {\rm c}},w_{2  {\rm c}},w_{1  {\rm cb}},w_{2  {\rm cb}},b_{0  {\rm cb}}^*, w_{2  {\rm p}}))
\nonumber\\
& \in T_\ep^N(P_{U_{1  {\rm pb}},U_{2  {\rm pb}},U_{0  {\rm cb}},X_1,X_2,U_{1  {\rm c}},U_{2  {\rm c}}}).
\label{eq:cr binning step 2}
}
If more than one such pair of indices satisfies the relationship in~\eqref{eq:cr binning step 2}, it selects one uniformly at random;
if no such pair exists, it sets $(b_{1  {\rm pb}}, b_{2  {\rm pb}})=(1,1)$ and in this case we say that a encoding error at the second binning step has occurred.
%$U_{1  {\rm pb}}(w_{1  {\rm pb}},b_{1  {\rm pb}},w_{1  {\rm c}},w_{2c},w_{1  {\rm p}},w_{1  {\rm cb}},w_{2  {\rm cb}},b_{0  {\rm cb}})$
%and
%$U_{2  {\rm pb}}(w_{1  {\rm pb}},b_{1  {\rm pb}},w_{1  {\rm c}},w_{2c},w_{2  {\rm p}},w_{1  {\rm cb}},w_{2  {\rm cb}},b_{0  {\rm cb}})$ appear to be generated according to the distribution
%$P_{U_{1  {\rm pb}},U_{2  {\rm pb}}|U_{1  {\rm c}},U_{2  {\rm c}},X_1,X_2,U_{0  {\rm cb}}}$.  If more that two codewords satisfy this relationship, it picks one pair at random.  If no such pair exists, it sets $b_{1  {\rm pb}}=b_{2  {\rm pb}}=1$.

\item
For the found triplet $(b_{0  {\rm cb}}^*,b_{1  {\rm pb}}^*,b_{2  {\rm pb}}^*)$ the cognitive relay
sends a codeword
\ean{
X_c^N
(&w_{1  {\rm pb}},b_{1  {\rm pb}}^*,
 w_{2  {\rm pb}},b_{2  {\rm pb}}^*,
\\
&w_{1  {\rm cb}},w_{2  {\rm cb}},b_{0  {\rm cb}}^*,
 w_{1  {\rm c}},w_{2  {\rm c}},
 w_{1  {\rm p}},w_{2  {\rm p}})
 }
jointly typical with all the selected codewords.

\end{itemize}

\item {\bf Encoding error analysis}

%Given the symmetry in the achievable scheme, we only focus on the decoding errors at destination~1.
%The error events at destination~2 are obtained by swapping the role of the users.
Given the symmetry of the codebook generation, we can assume
without loss of generality that the messages
\ean{
W_1 &=(W_{1  {\rm c}},W_{1  {\rm p}},W_{1  {\rm cb}},W_{1  {\rm pb}})=(1,1,1,1), \\
W_2 &=(W_{2  {\rm c}},W_{2  {\rm p}},W_{2  {\rm cb}},W_{2  {\rm pb}})=(1,1,1,1),
}
were transmitted. We now derive the conditions under which encoding is successful with high probability.
Let also $(B_{0  {\rm cb}}^*,B_{1  {\rm pb}}^*,B_{2  {\rm pb}}^*)$ be the triplet found by
the cognitive relay during the two binning steps of the encoding process.

%\smallskip
%{\bf Encoding Errors:}
    %We have two Marton-like regions, one for the common broadcasted messages $(U_{1  {\rm cb}},U_{2  {\rm cb}})$ and one for the private broadcasted messages
%     $(U_{1  {\rm pb}},U_{2  {\rm pb}})$. We do a sequential encoding, as joint encoding seems to be pretty complicated for this scheme.
%Encoding is performed in two steps: in the first steps we perform binning of the common broadcasted messages $(U_{1  {\rm cb}},U_{2  {\rm cb}})$ and in the second step of the private broadcasted messages.
%An encoding error is committed if we cannot find a $(b_{1  {\rm c}},b_{2c})$ in the first
%    step or if, upon finding  suitable $(b_{1  {\rm c}},b_{2c})$  in the first encoding
%    step, we cannot find the correct $(b_{1  {\rm p}},b_{2  {\rm p}})$  in  the second step.
    Let $E_{{\rm cb}}$, resp. $E_{{\rm pb}}$,  denote the event that the first binning step in~\eqref{eq:cr binning step 1},
    resp. the second binning step in~\eqref{eq:cr binning step 2}, is not successful. The probability of encoding error is bounded by:
    \ean{
    \Pr[{\rm encoding\,error}] %= \Pr[E_{{\rm enc}}]
     \leq \Pr[E_{{\rm cb}}]+\Pr[E_{{\rm pb}}| E_{{\rm cb}}^c]
    }
    where $E_{{\rm cb}}^c$ denotes the complement of the event $E_{{\rm cb}}$.

We start by noting that the encoded sequences are generated iid according to
\ea{
P^{ {\rm (gen)}}
&\triangleq
P_{U_{1  {\rm c}},X_1}
P_{U_{2  {\rm c}},X_2}
P_{U_{0  {\rm cb}}|U_{2  {\rm c}},U_{1  {\rm c}}}
\nonumber\\&\lag
P_{U_{1  {\rm pb}}|U_{2  {\rm c}},U_{1  {\rm c}},U_{0  {\rm cb}},X_1}
P_{U_{2  {\rm pb}}|U_{2  {\rm c}},U_{1  {\rm c}},U_{0  {\rm cb}},X_2}
\label{eq:pdf gen tx rx 1}
}
but after binning they look as if generated iid according to
\ea{
P^{ {\rm (enc)}}
&\triangleq
P_{U_{1  {\rm c}},X_1}
P_{U_{2  {\rm c}},X_2}
P_{U_{0  {\rm cb}}|U_{2  {\rm c}},U_{1  {\rm c}},X_1,X_2}
\nonumber\\&\lag
P_{U_{1  {\rm pb}},U_{2  {\rm pb}}|U_{2  {\rm c}},U_{1  {\rm c}},U_{0  {\rm cb}},X_1,X_2};
\label{eq:pdf enc tx rx 1}
}
we thus expect the encoding error probability to be of the form
\begin{align}
  &\EE\left[\log\frac{P^{ {\rm (gen)}}}{P^{ {\rm (enc)}}}\right]
\nonumber\\&=I(U_{0  {\rm cb}}; X_1,X_2 | U_{1  {\rm c}},U_{2  {\rm c}})
\nonumber\\&+I(U_{1  {\rm pb}}; X_2| U_{1  {\rm c}},U_{2  {\rm c}},U_{0  {\rm cb}})
+I(U_{2  {\rm pb}}; X_1| U_{1  {\rm c}},U_{2  {\rm c}},U_{0  {\rm cb}})
\nonumber\\&+I(U_{1  {\rm pb}}; U_{2  {\rm pb}}| U_{1  {\rm c}},U_{2  {\rm c}},U_{0  {\rm cb}},X_1,X_2).
\label{eq:mutual info 0}
\end{align}
The rigorous error analysis is as follows.
%\begin{table}
%\centering
%\label{tab:encoding errors}
%\caption{Possible encoding errors at the cognitive relay}
%\begin{tabular}{|l|lll|}
%\hline
% & $b_{0  {\rm cb}}$ & $b_{1  {\rm pb}}$ & $b_{2  {\rm pb}}$ \\
%\hline
%$E_1$ & X & $\ldots$ & $\ldots$ \\
%$E_2$ & $\checkmark$ & X & X \\
%$E_3$ & $\checkmark$ & X & $\checkmark$ \\
%$E_4$ & $\checkmark$ & $\checkmark$ & X \\
%\hline
%\end{tabular}
%\end{table}

 \begin{itemize}
 \item   {First binning step.}
$E_{{\rm cb}}$ is the event that
for all $b_{0  {\rm cb}}\in[1:2^{N R_{0  {\rm c}}'}]$
\ean{
&(U_{1  {\rm c}}^N(1),X_1^N(1|1),
 U_{2  {\rm c}}^N(1),X_2^N(1|1),
 U_{0  {\rm cb}}^N(1,1,b_{1  {\rm c}}|1,1)
\\&\not\in T_\epsilon^N(P_{U_{1  {\rm c}},X_1,U_{2  {\rm c}},X_2,U_{0  {\rm cb}}}),
}
% For $E_{0  {\rm cb}}$ we have:
%    \ean{
%    P [E_{bc }] &= (1-P[\lb U_{1  {\rm c}}^N(1),X_{1}^N(1,1),U_{2  {\rm c}}^N(1),X_{2}^N(1,1),U_{0  {\rm cb}}^N(1,1,1,1b_{c})\rb  \notin
%    T_\epsilon^N (p_{U_{1  {\rm c}},X_1,U_{2  {\rm c}},X_2,U_{0  {\rm cb}}})])^{2^{N R'_{0  {\rm cb}} }},
%    }
%    where
%    \ean{
%    P[\lb U_{1  {\rm c}}^N(1), X_{1}^N(1,1),U_{2  {\rm c}}^N(1),X_2(1,1),U_{0  {\rm cb}}(1,1,1,1,b_{0  {\rm cb}}) \rb
%    \notin T_\epsilon^N (p_{U_{1  {\rm c}},X_{1},U_{2  {\rm c}},X_2,U_{0  {\rm cb}}})] \geq\\
%    (1-\ep) 2^{-N(I(X_2,X_1; U_{0  {\rm cb}}| U_{1  {\rm c}},U_{2  {\rm c}})+\de)}.
%    }
By standard arguments, $\Pr[E_{{\rm cb}}] \to 0$ as $N \to \infty$ if
\[
R_{0  {\rm cb}} \geq I(X_1,X_2; U_{0  {\rm cb}}| U_{1  {\rm c}},U_{2  {\rm c}}),
\]
as in~\eqref{eq:binning rates 0}.

\item    {Second binning step.} Let $b_{0  {\rm cb}}^*$ be the index that was found
to satisfy~\eqref{eq:cr binning step 1} at the first decoding step.
We bound the probability of error in the second encoding step as
    \ean{
    &\Pr[E_{{\rm pb}}|E_{{\rm cb}}^c]
    = \Pr\Big[ \bigcap_{b_{1}=1}^{2^{N R_{1  {\rm pb}}'}} \bigcap_{b_{2}=1}^{2^{N R_{2  {\rm pb}}'}}
    (U_{1  {\rm c}}^N(1),X_1^N(1|1),
  \\&U_{2  {\rm c}}^N(1),X_2^N(1|1),
     U_{0  {\rm cb}}^N(1,1,b_{0  {\rm cb}}^*|1,1),
  \\&U_{1  {\rm pb}}^N(1,b_{1}|1,1,1,1,b_{0  {\rm cb}}^*),
  \\&U_{2  {\rm pb}}^N(1,b_{2}|1,1,1,1,b_{0  {\rm cb}}^*)
     )
     \not\in T_{\ep}^N (P^{\rm(enc)}) \Big]
  \\& = \Pr[K=0] \leq \f {\var[K] }{\EE[K]^2},
    }
    where $P^{\rm(enc)}$ is given in~\eqref{eq:pdf enc tx rx 1}, where
    \[
    K=\sum_{b_{1}=1}^{2^{N R_{1  {\rm pb}}'}} \sum_{b_{2}=1}^{2^{N R_{2  {\rm pb}}'}} K_{b_{1},b_{2}},
    \]
    with $K_{b_{1},b_{2}}$ the indicator function of the event
    \ean{
%      &K_{b_{1},b_{2}}=1_{\{
   &(U_{1  {\rm c}}^N(1),X_1^N(1|1),
     U_{2  {\rm c}}^N(1),X_2^N(1|1),
     U_{0  {\rm cb}}^N(1,1,b_{0  {\rm cb}}^*|1,1),
 \\& U_{1  {\rm pb}}^N(1,b_{1}|1,1,1,1,b_{0  {\rm cb}}^*),
     U_{2  {\rm pb}}^N(1,b_{2}|1,1,1,1,b_{0  {\rm cb}}^*)
     )
 \\&  \in T_{\ep}^N (P^{\rm(enc)}) %\} }.
    }

    The mean value of $K$ (neglecting all terms that depend on $\epsilon$ and that eventually go to zero as $N\to\infty$) is:
    \ean{
    \EE[K]
    & = \sum_{b_{1}=1}^{2^{N R'_{1  {\rm pb}} }}  \sum_{b_{2}=1}^{2^{N R'_{2  {\rm pb}}} } \Pr[K_{b_{1},b_{2}}=1] \nonumber \\
    & = 2^{N(R'_{1  {\rm pb}} +R'_{2  {\rm pb}}-A)}
    }
    with
    \ean{
    &2^{-N A} =\Pr[K_{b_{1},b_{2}}=1]=\EE[K_{b_{1},b_{2}}]
%    \\&=\Pr[ \lb U_{1  {\rm c}}^N(1), X_1^N(1,1),U_{2  {\rm c}}^N(1), X_{2}^N(1,1),U_{0  {\rm cb}}(1,1,1,1,1,b_{0  {\rm cb}}),U_{1  {\rm pb}}^N(1,1,1,b_{1}) ,U_{2  {\rm pb}}^N(1,1,1,b_{2}) \rb
%    \in T_\epsilon^N (P^{\rm(enc)}) ]
    \\&=\sum_{ (u^N_{1  {\rm pb}},u^N_{2  {\rm pb}})
    \in T_\epsilon^N (P^{\rm(enc)}|u^N_{1  {\rm c}},x^N_{1},u^N_{2c},x^N_{2})}
   \\&P^N_{U_{1  {\rm pb}}|U_{1  {\rm c}},U_{2  {\rm c}},X_1,U_{0  {\rm cb}}}
      P^N_{U_{2  {\rm pb}}|U_{1  {\rm c}},U_{2  {\rm c}},X_2,U_{0  {\rm cb}}}
    \\&= 2^{-N [I(U_{1  {\rm pb}}; X_2 | U_{1  {\rm c}},U_{2  {\rm c}},X_1,U_{0  {\rm cb}})
                  +I(U_{2  {\rm pb}}; X_1 | U_{1  {\rm c}},U_{2  {\rm c}},X_2,U_{0  {\rm cb}})]}
    \\&\lag 2^{-N I(U_{1  {\rm pb}}; U_{2  {\rm pb}}| U_{1  {\rm c}}, X_1,U_{2  {\rm c}},X_2,U_{0  {\rm cb}})]}.
    }
    The variance of $K$ (neglecting all terms that depend on $\epsilon$ and that eventually go to zero  as $N\to\infty$) is:
    \ean{
    &\var[K]
      = \sum_{b_{1}=1}^{2^{N R'_{1  {\rm pb}}}} \sum_{b_{2}=1}^{2^{N R'_{2  {\rm pb}}}} \sum_{b_{1}'=1}^{2^{N R'_{1  {\rm pb}} }} \sum_{b_{2}'=1}^{2^{N R'_{2  {\rm pb}}}}
  \\&\lb \Pr[K_{b_1,b_2}=1,K_{b'_1,b'_2}=1]  -\Pr[K_{b_1,b_2}=1]\Pr[K_{b'_1,b'_2}=1]\rb
    %\end{align*}
    %We can write:
    %\begin{align*}
    %&\var[K]
    \\&\leq \underbrace{\sum_{\,b_{1}=b_{1}',\, b_{2}=b_{2}'}
     \Pr[K_{b_{1},b_{2}}=1]}_{=\EE[K]}
    \\&+ \underbrace{\sum_{b_{1}=b_{1}',\, b_{2}\not=b_{2}'}
     \Pr[K_{b_{1},b_{2}}=1]\Pr[K_{b_{1},b_{2}'}=1|K_{b_{1},b_{2}}=1]}_{=\EE[K]\, 2^{N (R'_{2  {\rm pb}}-B)}}
    \\&+ \underbrace{\sum_{b_{1}\not=b_{1}',\, b_{2}=b_{2}'}
     \Pr[K_{b_{1},b_{2}}=1]\Pr[K_{b_{1}',b_{2}}=1|K_{b_{1},b_{2}}=1]}_{=\EE[K]\, 2^{N (R'_{1  {\rm pb}}-C)}}
    \\&+ \underbrace{\sum_{b_{1}\not=b_{1}',\, b_{2}\not=b_{2}'}
     \Pr[K_{b_{1},b_{2}}=1]\Pr[K_{b_{1}',b_{2}'}=1|K_{b_{1},b_{2}}=1]}_{=\EE[K]\,2^{N (R'_{1  {\rm pb}}+N R'_{2  {\rm pb}}-D)}}
    }
    with
    \ean{
    2^{-N B}
    &= \Pr[K_{b_{1},b_{2}'}=1|K_{b_{1},b_{2}}=1]
%    \\&= \Pr[\lb U_{1  {\rm c}}^N(1),X_1^N(1,1),U_{2  {\rm c}}^N(1),X_{2}^N(1,1),U_{0  {\rm cb}}(1,1,1,1,1,b_{0  {\rm cb}}),U_{1  {\rm pb}}^N(1,1,1,b_{1}),U_{2  {\rm pb}}^N(1,1,1,b_{2}') \rb \in T_\epsilon^N (P^{\rm(enc)}) |
%    \\& \quad\lb U_{1  {\rm c}}^N(1),X_1^N(1,1),U_{2  {\rm c}}^N(1),X_{2}^N(1,1),U_{0  {\rm cb}}(1,1,1,1,1,b_{0  {\rm cb}}),U_{1  {\rm pb}}^N(1,1,1,b_{1}),U_{2  {\rm pb}}^N(1,1,1,b_{2} ) \rb \in T_\epsilon^N (P^{\rm(enc)}) ]
    \\&= \sum_{ u^N_{2  {\rm pb}} \in T_\epsilon^N (P^{\rm(enc)}|u^N_{1  {\rm c}},x_1^N,u^N_{2c},x^N_{2},u^N_{0  {\rm cb}},u^N_{1  {\rm pb}})}
  \\&P^N_{U_{2  {\rm pb}}|U_{2  {\rm c}},U_{1  {\rm c}},U_{0  {\rm cb}},X_2}
  \\&= 2^{-N I(U_{2  {\rm pb}};X_1, U_{1  {\rm pb}}|U_{2  {\rm c}},U_{1  {\rm c}},X_{2},U_{0  {\rm cb}})},
    }
    and similarly, i.e., swap the role of the users in the expression above,
    \ean{
    2^{-N C}
    %= \Pr[K_{b_{1}',b_{2}}=1|K_{b_{1},b_{2}}=1]
    %\\&= \Pr[  \lb U_{1  {\rm c}}^N(1),X_1^N(1,1),U_{2  {\rm c}}^N(1), X_{2}^N(1,1),U_{1  {\rm pb}}^N(1,1,1,b_{1}'),  U_{2  {\rm pb}}^N(1,1,b_{2}) \rb \in T_\epsilon^N (p_{U_{1  {\rm c}},X_1,U_{2  {\rm c}},X_{2},U_{1  {\rm pb}},U_{2  {\rm pb}}}) |
    %\\& \quad \lb U_{1  {\rm c}}^N(1), X_1^N (1,1), U_{2  {\rm c}}^N(1), X_{2}^N(1,1),U_{1  {\rm pb}}^N(1,1,1,b_{1}), \rnone
    %\\& \lnone \quad \quad    U_{2  {\rm pb}}^N(1,1,1,b_{2}) \rb
    %\in T_\epsilon^N (p_{U_{1  {\rm c}},X_1,U_{2  {\rm c}},X_{2},U_{1  {\rm pb}},U_{2  {\rm pb}}}) ]
    %\\&= \sum_{ u^N_{1  {\rm pb}}
    %\in T_\epsilon^N (p_{U_{1  {\rm c}},X_1,U_{2  {\rm c}},X_{2},U_{1  {\rm pb}},U_{2  {\rm pb}}}|u^N_{1  {\rm c}},x_1^N,u^N_{2c},x^N_{2},u^N_{2  {\rm pb}})}
    %p^N_{U_{1  {\rm pb}}|U_{2  {\rm c}},U_{1  {\rm c}},X_1}
    %\\&
    &= 2^{-N I(U_{1  {\rm pb}}; X_{2}, U_{2  {\rm pb}}| U_{1  {\rm c}}, X_1,U_{2  {\rm c}},U_{0  {\rm cb}})},
    }
    and finally
    \ean{
    &2^{-N D}
      =\Pr[K_{b_{1}',b_{2}'}=1|K_{b_{1},b_{2}}=1]
%    \\&= \Pr[\lb U_{1  {\rm c}}^N(1),X_1^N(1,1),U_{2  {\rm c}}^N(1),X_{2}^N(1,1),U_{1  {\rm pb}}^N(1,1,1,b_{1}'),U_{2  {\rm pb}}^N(1,1,1,b_{2}') \rb \in T_\epsilon^N (P^{\rm(enc)}) |
%    \\& \quad\lb U_{1  {\rm c}}^N(1),X_1^N(1,1),U_{2  {\rm c}}^N(1),X_{2}^N(1,1),U_{1  {\rm pb}}^N(1,1,1,b_{1} ),U_{2  {\rm pb}}^N(1,1,1,b_{2} ) \rb \in T_\epsilon^N (P^{\rm(enc)}) ]
    \\&= \sum_{ (u^N_{1  {\rm pb}},u^N_{2  {\rm pb}})
    \in T_\epsilon^N (P^{\rm(enc)}|u^N_{1  {\rm c}},x_1^N,u^N_{2c},x^N_{2},u_{0  {\rm cb}}^N)}
    \\&P^N_{U_{2  {\rm pb}}|U_{1  {\rm c}},U_{2  {\rm c}},X_{2},U_{0  {\rm cb}}}
       P^N_{U_{1  {\rm pb}}|U_{1  {\rm c}},U_{2  {\rm c}},X_1,U_{0  {\rm cb}}}
    %p^N_{U_{2  {\rm pb}}|U_{1  {\rm c}},X_1,U_{2  {\rm c}},X_{2}} \ p^N_{U_{1  {\rm pb}}|U_{1  {\rm c}}, X_1 , U_{2  {\rm c}}, X_2}
%    \\&=2^{-N (I(U_{1  {\rm pb}}; X_{2} | U_{1  {\rm c}}, U_{2  {\rm c}}, X_1,U_{0  {\rm cb}}) + I(U_{2  {\rm pb}}; X_{1} | U_{1  {\rm c}}, U_{2  {\rm c}}, X_2,U_{0  {\rm cb}}) + I(U_{1  {\rm pb}}; U_{2  {\rm pb}}| U_{1  {\rm c}}, X_1 , U_{2  {\rm c}},X_2,,U_{0  {\rm cb}}))}
    \\&=2^{- N A}.
    }

    Hence, we can bound $\Pr[K=0]$ as:
    \begin{align*}
    &\Pr[K=0]
  \\&\leq \displaystyle\frac{
    1
    +2^{N( R'_{1  {\rm pb}}            -C)}
    +2^{N(            R'_{2  {\rm pb}} -B)}
    +2^{N( R'_{1  {\rm pb}} +R'_{2  {\rm pb}} -A)}
    }{
     2^{N(R'_{1  {\rm pb}} +R'_{2  {\rm pb}}-A)}
    }
    \end{align*}
    and $\Pr[K=0]\to 0$ if
    \begin{align*}
      &R'_{1  {\rm pb}} +R'_{2  {\rm pb}}-A                           >0
    \\&R'_{1  {\rm pb}} +R'_{2  {\rm pb}}-A -(            R'_{2  {\rm pb}} -B)>0
    \\&R'_{1  {\rm pb}} +R'_{2  {\rm pb}}-A -( R'_{1  {\rm pb}}            -C)>0
%    \\&R'_{1  {\rm pb}} +R'_{2  {\rm pb}}-A -( R'_{1  {\rm pb}} +R'_{2  {\rm pb}}  -A)>0,
    \end{align*}
    that is, if
    \ean{
      R'_{1  {\rm pb}} +R'_{2  {\rm pb}}
               > A
%               =I(U_{1  {\rm pb}}; X_2| U_{1  {\rm c}},U_{2  {\rm c}},X_1,U_{0  {\rm cb}})
%               +I(U_{2  {\rm pb}}; X_1 | U_{1  {\rm c}},U_{2  {\rm c}},X_2,U_{0  {\rm cb}})
%       \\&\quad+I(U_{1  {\rm pb}}; U_{2  {\rm pb}}| U_{1  {\rm c}}, X_1,U_{2  {\rm c}},X_2,U_{0  {\rm cb}})
       =\text{eq.\eqref{eq:binning rates 12}}
    \\R'_{1  {\rm pb}} > A-B %=I(U_{1  {\rm pb}}; X_2| U_{1  {\rm c}},U_{2  {\rm c}},X_1,U_{0  {\rm cb}})
    =\text{eq.\eqref{eq:binning rates 1}}
    \\R'_{2  {\rm pb}} > A-C %=I(U_{2  {\rm pb}}; X_1| U_{1  {\rm c}},U_{2  {\rm c}},X_2,U_{0  {\rm cb}})
    =\text{eq.\eqref{eq:binning rates 2}}
    }
    since
    \ean{
    A&= I(U_{1  {\rm pb}}; X_2 | U_{1  {\rm c}},U_{2  {\rm c}},X_1,U_{0  {\rm cb}})
    \\&+I(U_{2  {\rm pb}}; X_1 | U_{1  {\rm c}},U_{2  {\rm c}},X_2,U_{0  {\rm cb}})
    \\&+I(U_{1  {\rm pb}}; U_{2  {\rm pb}}| U_{1  {\rm c}}, X_1,U_{2  {\rm c}},X_2,U_{0  {\rm cb}})
    \\&= I(U_{1  {\rm pb}}; X_2 | U_{1  {\rm c}},U_{2  {\rm c}},X_1,U_{0  {\rm cb}})+B
%    \\&+\underbrace{I(U_{2  {\rm pb}}; U_{1  {\rm pb}},X_1 | U_{1  {\rm c}},U_{2  {\rm c}},X_2,U_{0  {\rm cb}})}_{=B}
    \\&= I(U_{2  {\rm pb}}; X_1 | U_{1  {\rm c}},U_{2  {\rm c}},X_2,U_{0  {\rm cb}})+C.
%    \\&+\underbrace{I(U_{1  {\rm pb}}; U_{2  {\rm pb}},X_2 | U_{1  {\rm c}},U_{2  {\rm c}},X_1,U_{0  {\rm cb}})}_{=C}
    }
\end{itemize}

\item
{\bf Decoding.}
We only describe the decoding at destination~1 as the same applies to destination~2 with the role of the users swapped.
Destination~1 looks
for a unique quadruplet $(w_{1  {\rm p}},w_{1  {\rm c}},w_{1  {\rm cb}},w_{1  {\rm pb}})$
and
for some  quadruplet $(w_{2  {\rm c}},w_{2  {\rm cb}},b_{0  {\rm cb}},b_{1  {\rm pb}})$
such that
\ea{
(
&U_{1  {\rm c}}^N(w_{1  {\rm c}}),X_1^N(w_{1  {\rm p}}|w_{1  {\rm c}}),
 U_{2  {\rm c}}^N(w_{2  {\rm c}}), %X_2^N(w_{2  {\rm p}}|w_{2  {\rm c}}),
\nonumber\\
&U_{0  {\rm cb}}^N(w_{1  {\rm cb}},w_{1  {\rm cb}},b_{0  {\rm cb}}|w_{1  {\rm c}},w_{2c}),
\nonumber\\
&U_{1  {\rm pb}}^N(w_{1  {\rm pb}},b_{1  {\rm pb}}|w_{1  {\rm c}},w_{2  {\rm c}},w_{1  {\rm cb}},w_{2  {\rm cb}},b_{0  {\rm cb}}, w_{1  {\rm p}}),
\nonumber\\
&)\in T_\epsilon^N (P^{\rm(dest.1)})
\label{eq:dec at dest1}
}
where
\ea{
P^{\rm(dest.1)}
&=
\sum_{X_2,U_{2  {\rm pb}},X_c}
P_{U_{1  {\rm c}},X_1}
P_{U_{2  {\rm c}},X_2}
\nonumber\\&=
P_{U_{1  {\rm pb}},U_{2  {\rm pb}},U_{0  {\rm cb}},X_c|U_{1  {\rm c}},X_1, U_{2  {\rm c}},X_2}
\nonumber\\&=
P_{U_{1  {\rm c}},X_1}
P_{U_{2  {\rm c}}}
P_{U_{1  {\rm pb}},U_{0  {\rm cb}}|U_{1  {\rm c}},X_1, U_{2  {\rm c}}}
\label{eq:dec at dest1 prob}
}
If none or more than one quadruplet $(w_{1  {\rm p}},w_{1  {\rm c}},w_{1  {\rm cb}},w_{1  {\rm pb}})$ is found
an error has occurred.

\item
{\bf Decoding Error Analysis.}

    Let $E_{{\rm dest.1}}$ denote the event that the relationship in~\eqref{eq:dec at dest1} is not satisfied
    by any $(w_{1  {\rm p}},w_{1  {\rm c}},w_{1  {\rm cb}},w_{1  {\rm pb}})$ or that is it satisfied by more than
    one such a quadruplet.
    We have
    \ean{
    \Pr[{\rm decoding\,error}]
      &\leq \Pr[{\rm encoding\,error}]
    \\&+\Pr[E_{{\rm dest.1}}| {\rm encoding\,successful}],
    }
    where $\Pr[{\rm encoding\,error}]\to 0$ if the rates
    are chosen form the ``binning rate region'' $\Rcal_{0}$ defined by~\eqref{eq:total inner bound}.
    Hence we only need to analyze the probability of decoding error assuming the encoding was successful.

    \begin{table*}
    \centering
    \caption{Possible decoding errors at destination~1. Legend:
    a ``0'' means that the corresponding message is in error,
    a ``$\checkmark$'' that the corresponding message is correct, and
    the ``$\ldots$'' that is does not matter whether the corresponding message is correct or not as in either case the
    joint density needed to evaluate the error event probability factorizes as if the message were in error (because of superposition
    to at least one message in error). The event $E_8$ is ``special'' in that the term
    $I(U_{0  {\rm cb}}; X_1 | U_{1  {\rm c}},U_{2  {\rm c}})$ in~\eqref{eq:mutual info 1} must be omitted.}
\begin{tabular}{|l|lllll|l|}
\hline
      & $U_{1  {\rm c}}$ & $U_{2  {\rm c}}$ & $X_1$            & $U_{0  {\rm cb}}$                                   & $U_{1  {\rm pb}}$                   & Set $\star$ to be used in~\eqref{eq:mutual info 1} \\
      & $w_{1  {\rm c}}$ & $w_{2  {\rm c}}$ & $w_{1  {\rm p}}$ & $(w_{1  {\rm cb}},w_{2  {\rm cb}},b_{0  {\rm cb}})$ & $(w_{1  {\rm pb}},b_{1  {\rm pb}})$ & \\
\hline
\hline
$E_1$ & 0            & 0            & $\ldots$     & $\ldots$     & $\ldots$ & $\emptyset$      \\ %$P_{Y_1}$ \\
$E_2$ & 0            & $\checkmark$ & $\ldots$     & $\ldots$     & $\ldots$ & $U_{2  {\rm c}}$ \\ %$P_{Y_1|U_{2  {\rm c}}}$ \\
\hline
$E_3$ & $\checkmark$ & 0            & 0            & $\ldots$     & $\ldots$ & $U_{1  {\rm c}}$ \\ %$P_{Y_1|U_{1  {\rm c}}}$ \\
$E_4$ & $\checkmark$ & $\checkmark$ & 0            & 0            & $\ldots$ & $U_{1  {\rm c}},U_{2  {\rm c}}$ \\ %$P_{Y_1|U_{1  {\rm c}},U_{2  {\rm c}}}$ \\
$E_5$ & $\checkmark$ & $\checkmark$ & 0            & $\checkmark$ & $\ldots$ & $U_{1  {\rm c}},U_{2  {\rm c}},U_{0  {\rm cb}}$ \\ %$P_{Y_1|U_{1  {\rm c}},U_{2  {\rm c}},U_{0  {\rm cb}}}$ \\
\hline
$E_6$ & $\checkmark$ & 0            & $\checkmark$ & $\ldots$     & $\ldots$ & $U_{1  {\rm c}},X_1$ \\ %$P_{Y_1|U_{1  {\rm c}},X_1}$ \\
$E_7$ & $\checkmark$ & $\checkmark$ & $\checkmark$ & 0            & $\ldots$ & $U_{1  {\rm c}},X_1,U_{2  {\rm c}}$ \\ %$P_{Y_1|U_{1  {\rm c}},X_1,U_{2  {\rm c}}}$ \\
\hline
$E_8$ & $\checkmark$ & $\checkmark$ & $\checkmark$ & $\checkmark$ & 0        & $U_{1  {\rm c}},X_1,U_{2  {\rm c}},U_{0  {\rm cb}}$ (special) \\ %not applicable ($U_{0  {\rm cb}}$ is correct)\\ %$P_{Y_1|U_{1  {\rm c}},U_{2  {\rm c}},X_1,{ U_{0  {\rm cb}}}}$ \\
\hline
%      & $\checkmark$ & $\checkmark$ & $\checkmark$ & $\checkmark$ & $\checkmark$  & (this is NOT an error) \\
%\hline
\end{tabular}
\label{tab:decoding errors at dest1}
\end{table*}

Table~\ref{tab:decoding errors at dest1} summarizes the possible error events at destination~1, where
    a ``0'' means that the corresponding message index is in error,
    a ``$\checkmark$'' that the corresponding message index, and bin index if any, is correct, and
    the ``$\ldots$'' that is does not matter whether the corresponding message index is correct or not as in either case the
    joint density needed to evaluate the error event probability factorizes as if the message were in error (because
    of superposition to at least one codeword with a message index in error).

For the cases where $U_{0  {\rm cb}}$
does not have the correct dependency on $(U_{1  {\rm c}},U_{2  {\rm c}},X_1)$,
i.e., for all cases listed in Table~\ref{tab:decoding errors at dest1} but for event $E_8$
which is marked as ``special'', an intuitive analysis of the probability of error is as follows.
Depending on which messages are wrongly decoded at destination~1,
and assuming the encoding steps were successful,
the decoded codewords and the received $Y_1^N$ are iid jointly distributed according to
    \begin{align}
    P_{1|\star}
    \triangleq P_{U_{1  {\rm c}},X_1} P_{U_{2  {\rm c}}} P_{U_{0  {\rm cb}}|U_{2  {\rm c}},U_{1  {\rm c}}}P_{U_{1  {\rm pb}}|U_{2  {\rm c}},U_{1  {\rm c}},X_1,U_{0  {\rm cb}}}
    P_{Y_1|\star},
    \label{eq:pdf tx rx 1}
    \end{align}
    where ``$\star$'' in~\eqref{eq:pdf tx rx 1} indicates the set of correctly decoded messages.
    However, the actual transmitted codewords and the received $Y_1^N$
    considered at destination~1 look as if they were generated iid according to
    \begin{align}
    P_{1}
    \triangleq P_{U_{1  {\rm c}},X_1}P_{U_{2  {\rm c}}} P_{U_{0  {\rm cb}},U_{1  {\rm pb}}|U_{2  {\rm c}},U_{1  {\rm c}},X_1}
    P_{Y_1|U_{2  {\rm c}},U_{1  {\rm c}},X_{1},U_{0  {\rm cb}},U_{1  {\rm pb}}}.
    \label{eq:pdf dec 1}
    \end{align}
    Hence we expect the probability of error at destination~1 to depend on
    terms of the type
    \begin{align}
    &I_{1|\star}
     =\EE\left[\log\f{P_{1}}{P_{1|\star}}\right] \nonumber\\
    &=\EE\left[\log\f{P_{U_{0  {\rm cb}}| U_{1  {\rm c}},U_{2  {\rm c}},X_1} P_{Y_1|U_{2  {\rm c}},U_{1  {\rm c}},X_1,U_{0  {\rm cb}},U_{1  {\rm pb}}}}{P_{U_{0  {\rm cb}}| U_{1  {\rm c}},U_{2  {\rm c}}} P_{Y_1|\star}}\right] \nonumber\\
    &=I(U_{0  {\rm cb}}; X_1 | U_{1  {\rm c}},U_{2  {\rm c}})+I(Y_1; U_{1  {\rm c}},U_{2  {\rm c}},X_1,U_{0  {\rm cb}},U_{1  {\rm pb}}| \star).
    \label{eq:mutual info 1}
    \end{align}
When $U_{0  {\rm cb}}$ has the correct dependency on $(U_{1  {\rm c}},U_{2  {\rm c}},X_1)$,
i.e., only for the ``special'' event $E_8$ in Table~\ref{tab:decoding errors at dest1},
the density $P_{1|\star}$ in~\eqref{eq:pdf tx rx 1} must be modified as follows.
We must use $P_{U_{0  {\rm cb}}|U_{2  {\rm c}},U_{1  {\rm c}},X_1}$ (i.e., correct dependency on $(U_{1  {\rm c}},U_{2  {\rm c}},X_1)$)
rather than $P_{U_{0  {\rm cb}}|U_{2  {\rm c}},U_{1  {\rm c}}}$. This results in the absence of the term
$I(U_{0  {\rm cb}}; X_1 | U_{1  {\rm c}},U_{2  {\rm c}})$ in~\eqref{eq:mutual info 1}.
% DT NOTE: for event E5 we cannot remove the term $I(U_{0  {\rm cb}}; X_1 | U_{1  {\rm c}},U_{2  {\rm c}})$
%because the taxed  U0 depends on the taxed X1 which, for events E5, is different from the one picked at the rx.

The rigorous analysis of the error probability is as follows.
\begin{itemize}
\item $\Pr[E_1]$ and $\Pr[E_2]$: $U_{1  {\rm c}}$ is in error.

If the decoding of $U_{1  {\rm c}}$ fails, the codewords $(X_1,U_{1  {\rm cb}},U_{2  {\rm cb}},U_{1  {\rm pb}})$
cannot be successfully decoded since they are superposed to a wrong $U_{1  {\rm c}}$.
$U_{2  {\rm c}}$, which is generated independently of $U_{1  {\rm c}}$, can be in error or not and we shall
distinguish the two cases in the following.

Event $E_1$ in Table~\ref{tab:decoding errors at dest1}
corresponds to the case where both $U_{1  {\rm c}}$ and $U_{2  {\rm c}}$ are in error
(and thus all the messages superposed to them are in error too);
its probability can be bounded as
    \begin{align*}
    &\Pr[E_1]
     = \Pr\lsb \bigcup_{
     \wt_{1  {\rm c}}\neq 1,
     \wt_{2  {\rm c}}\neq 1,
     \wt_{1  {\rm p}},
     \wt_{1  {\rm cb}},
     \wt_{2  {\rm cb}},
     \wt_{1  {\rm pb}},
     \bt_{0  {\rm cb}},
     \bt_{1  {\rm pb}}}\rnone
    \\
    &\quad (Y_1^N,
    U_{1  {\rm c}}^N (\wt_{1  {\rm c}}),
    X_1^N            (\wt_{1  {\rm p}}|\wt_{1  {\rm c}}),
    U_{2  {\rm c}}^N (\wt_{2  {\rm c}}),
  \\&\quad U_{0  {\rm cb}}^N(\wt_{1  {\rm cb}},
                      \wt_{2  {\rm cb}},\bt_{0  {\rm cb}}|\wt_{1  {\rm c}},\wt_{2  {\rm c}}),
    \\&\quad
    U_{1  {\rm pb}}^N(\wt_{1  {\rm pb}},\bt_{1  {\rm pb}}| \wt_{1  {\rm c}}, \wt_{1  {\rm p}},
                      \wt_{1  {\rm cb}},\wt_{2  {\rm cb}},\bt_{0  {\rm cb}})
    \\&\quad \lnone \in T_\epsilon^N (P^{\rm(dest.1)}) \rsb
    \\
    &\leq  2^{N(R_{1  {\rm c}}+R_{1  {\rm p}}+R_{2  {\rm c}}+L_{0  {\rm cb}}+L_{1  {\rm pb}})}
    \\& \sum_{(y_1^N,u^N_{1  {\rm c}},u^N_{2c},x^N_{1},u^N_{0  {\rm cb}},u^N_{1  {\rm pb}})\in T_\epsilon^N (P^{\rm(dest.1)})}
    P_{1|\star}|_{\star=\emptyset}
    \\
    &\leq  2^{N(R_{1  {\rm c}}+R_{1  {\rm p}}+R_{2  {\rm c}}+L_{0  {\rm cb}}+L_{1  {\rm pb}}-
    I_{1|\star}|_{\star=\emptyset})},
    \end{align*}
    for $P_{1|\star}$ given in~\eqref{eq:pdf dec 1} and $I_{1|\star}$ given
    in~\eqref{eq:mutual info 1} evaluated for $\star=\emptyset$.
    Hence $\Pr[E_1]\to 0$ as $N \to \infty$ if~\eqref{eq:inner bound 1} holds.
%    $$
%    R_{1  {\rm c}}+R_{2  {\rm p}}+R_{2  {\rm c}}+L_{0  {\rm cb}}+L_{1  {\rm pb}} \leq  I(U_{0  {\rm cb}}; X_1 | U_{1  {\rm c}},U_{2  {\rm c}})+I(Y_1; U_{1  {\rm c}},U_{2  {\rm c}},X_1,U_{0  {\rm cb}},U_{1  {\rm pb}}).
%    $$

Event $E_2$ in Table~\ref{tab:decoding errors at dest1}
corresponds to the case where $U_{1  {\rm c}}$ is in error
(and thus all the messages superposed to it are in error too)
and $U_{2  {\rm c}}$ is correctly decoded.
Similarly to what done for event $E_1$, the probability of event $E_2$
goes to zero if~\eqref{eq:inner bound 2} holds.

\item $\Pr[E_3]$, $\Pr[E_4]$ and $\Pr[E_5]$: $X_1$ is in error.

Similarly to what done for event $E_1$,
the probability of event $E_3$ goes to zero if~\eqref{eq:inner bound 3} holds,
the probability of event $E_4$ goes to zero if~\eqref{eq:inner bound 4} holds, and
the probability of event $E_5$ goes to zero if~\eqref{eq:inner bound 5} holds.

\item $\Pr[E_6]$ and $\Pr[E_7]$: $U_{0  {\rm cb}}$ is in error.

Similarly to what done for event $E_1$,
the probability of event $E_6$ goes to zero if~\eqref{eq:inner bound 6} holds, and
the probability of event $E_7$ goes to zero if~\eqref{eq:inner bound 7} holds.

\item $\Pr[E_8]$: $U_{1  {\rm pb}}$ is in error.

Similarly to what done for event $E_1$,
the probability of event $E_8$ goes to zero if~\eqref{eq:inner bound 8} holds.

\end{itemize}

\end{itemize}

\section{Proof of Thm.~\ref{thm:inclusion Jiang}}
\label{app:inclusion of jiang region in our region}

Without loss of generality we may introduce in Thm.~\ref{thm:general inner bound}
a new RV $U_{i  {\rm p}}$ and let $X_i$ be a deterministic function of $(U_{i  {\rm c}},U_{i  {\rm p}})$,
i.e.  $X_i=X_i(U_{i  {\rm c}},U_{i  {\rm p}})$, $i\in\{1,2\}$.

With
\[
R_{0  {\rm cb}}'=R_{1  {\rm cb}}=R_{2  {\rm cb}}=R_{1  {\rm pb}}=R_{2  {\rm pb}}=0, \quad U_{0  {\rm cb}}=\emptyset,
\]
the achievable rate region in Thm.~\ref{thm:general inner bound} given by~\eqref{eq:total inner bound} becomes
\eas{
R_{1  {\rm pb}}' & \geq  I(U_{1  {\rm pb}}; U_{2  {\rm p}} | U_{1  {\rm c}},U_{2  {\rm c}},U_{1  {\rm p}}) \label{eq:jiang comparison achievable region 2 b1}\\
R_{2  {\rm pb}}' & \geq  I(U_{2  {\rm pb}}; U_{1  {\rm p}} | U_{1  {\rm c}},U_{2  {\rm c}},U_{2  {\rm p}}) \label{eq:jiang comparison achievable region 2 b2} \\
R_{1  {\rm pb}}'
+R_{2  {\rm pb}}'& \geq  I(U_{1  {\rm pb}}; U_{2  {\rm p}} | U_{1  {\rm c}},U_{2  {\rm c}},U_{1  {\rm p}})  \nonumber \\
              & \quad  + I(U_{2  {\rm pb}}; U_{1  {\rm p}} | U_{1  {\rm c}},U_{2  {\rm c}},U_{2  {\rm p}})  \nonumber \\
              & \quad  + I(U_{1  {\rm pb}}; U_{2  {\rm pb}}| U_{1  {\rm c}},U_{1  {\rm p}},U_{2  {\rm c}},U_{2  {\rm p}})
              \label{eq:jiang comparison achievable region 2 b12}\\
%\nonumber \\
R_{1  {\rm c}}+R_{2  {\rm c}}+L_{1  {\rm p}}  & \leq  I(Y_1 ; U_{1  {\rm c}},U_{2  {\rm c}}, U_{1  {\rm p}},U_{1  {\rm pb}}|Q)
\label{eq:jiang comparison achievable region 2 dec1 1} \\
               R_{2  {\rm c}}+L_{1  {\rm p}}  & \leq  I(Y_1 ; U_{2  {\rm c}}, U_{1  {\rm p}},U_{1  {\rm pb}}|U_{1  {\rm c}},Q)
\label{eq:jiang comparison achievable region 2 dec1 2} \\
R_{1  {\rm c}}               +L_{1  {\rm p}}  & \leq  I(Y_1 ; U_{1  {\rm c}}, U_{1  {\rm p}},U_{1  {\rm pb}}|U_{2  {\rm c}},Q)
\label{eq:jiang comparison achievable region 2 dec1 3} \\
                              L_{1  {\rm p}}  & \leq  I(Y_1 ; U_{1  {\rm p}},U_{1  {\rm pb}}| U_{1  {\rm c}},U_{2  {\rm c}},Q)
\label{eq:jiang comparison achievable region 2 dec1 4} \\
  & L_{1  {\rm p}}  = R_{1  {\rm p}}  + R_{1  {\rm pb}}' \nonumber  \\
R_{1  {\rm c}}+R_{2  {\rm c}}+L_{2  {\rm p}}  & \leq  I(Y_2 ; U_{1  {\rm c}},U_{2  {\rm c}}, U_{2  {\rm p}}, U_{2  {\rm pb}}|Q)
\label{eq:jiang comparison achievable region 2 dec2 1} \\
               R_{2  {\rm c}}+L_{2  {\rm p}}  & \leq  I(Y_2 ; U_{2  {\rm c}}, U_{2  {\rm p}}, U_{2  {\rm pb}}|U_{1  {\rm c}},Q)
\label{eq:jiang comparison achievable region 2 dec2 2} \\
R_{1  {\rm c}}               +L_{2  {\rm p}}  & \leq  I(Y_2 ; U_{1  {\rm c}}, U_{2  {\rm p}}, U_{2  {\rm pb}}|U_{2  {\rm c}},Q)
\label{eq:jiang comparison achievable region 2 dec2 3} \\
                              L_{2  {\rm p}}  & \leq  I(Y_2 ; U_{2  {\rm p}}, U_{2  {\rm pb}}| U_{1  {\rm c}},U_{2  {\rm c}},Q)
\label{eq:jiang comparison achievable region 2 dec2 4} \\
  & L_{2  {\rm p}}  = R_{2  {\rm p}}  + R_{2  {\rm pb}}' \nonumber
}{\label{eq:jiang comparison achievable region 2}}
for all distributions that factors as
\ea{
&P_{Q}
P_{U_{1  {\rm c}},U_{1  {\rm p}},X_1|Q}
P_{U_{2  {\rm c}},U_{2  {\rm p}},X_2|Q}
\nonumber\\&
P_{U_{1  {\rm pb}},U_{2  {\rm pb}},X_c| U_{1  {\rm c}}, U_{1  {\rm p}},U_{2  {\rm c}},U_{2  {\rm p}}, X_1, X_2, Q}.
\label{eq:factorization jiang comparison achievable region 2}
}

%In the region of \cite[(21)-(31)]{jiang-achievable-BCCR}  we can take $X_1=V_1$ and $X_2=V_2$  without loss of generality.
%With this consideration we may establish a correspondence between the RVs described Table~\ref{tab:jiang RVs 1}:
%the resulting regions are identical but for the binning rates in  \cite[(21)-(31)]{jiang-achievable-BCCR}  which  are more restrictive than~\eqref{eq:jiang comparison achievable region b1},\eqref{eq:jiang comparison achievable region b2} and~\eqref{eq:jiang comparison achievable region b12}.

\begin{table}
\centering
\caption{The correspondence of RVs in the comparison between the region
in~\cite{jiang-achievable-BCCR} and the region in~\eqref{eq:jiang comparison achievable region 2}. }
\label{tab:jiang RVs 1}
\begin{tabular}{|l|l|}
\hline
Region in~\cite{jiang-achievable-BCCR} & Region in~\eqref{eq:jiang comparison achievable region 2} \\%{eq:total inner bound}  \\
\hline
$U_1$ & $U_{1  {\rm c}}$ \\
$U_2$ & $U_{2  {\rm c}}$ \\
$V_1$ & $U_{1  {\rm p}}$ \\ %$X_{1}$ &  \\
$V_2$ & $U_{2  {\rm p}}$ \\ %$X_{2}$ &  \\
$W_1$ & $U_{1  {\rm pb}}$\\
$W_2$ & $U_{2  {\rm pb}}$\\
\hline
\end{tabular}
\end{table}

In order to compare the special case of our achievable rate region given by~\eqref{eq:jiang comparison achievable region 2}
with the region in~\cite{jiang-achievable-BCCR}, consider the correspondence of RVs in Table~\ref{tab:jiang RVs 1}.
With this correspondence we see that the regions in~\cite[(20)-(31)]{jiang-achievable-BCCR}
and~\eqref{eq:jiang comparison achievable region 2}
have the same rate bounds and holds for the same set of input distributions.
Since the region in~\eqref{eq:jiang comparison achievable region 2} is a special case of our
general achievable rate region, we conclude that the region in~\eqref{eq:jiang comparison achievable region 2}
contains the region in~\cite{jiang-achievable-BCCR}.

\section{Proof of Corollary~\ref{cor:all private messages}}
\label{app:cor:all private messages}

Let
$R_1=R_{1  {\rm p}}$ and $R_2 = R_{2  {\rm p}}$, i.e.,
\[
R_{1  {\rm c}}=R_{2  {\rm c}}
= R_{0  {\rm cb}}'
= R_{1  {\rm cb}} = R_{2  {\rm cb}}
= R_{1  {\rm pb}} = R_{2  {\rm pb}}=0.
\]
The region in~\eqref{eq:inner bound} becomes
    \eas{
    R_{1  {\rm pb}}' &\geq I(X_2; U_{1  {\rm pb}}| X_1) \\
    R_{2  {\rm pb}}' &\geq I(X_1; U_{2  {\rm pb}}| X_2) \\
    R_{1  {\rm pb}}' +R_{2  {\rm pb}}' & \geq  I(X_2; U_{1  {\rm pb}}| X_1)+I(X_1; U_{2  {\rm pb}}| X_2)
    \nonumber \\\lag
    &+I(U_{1  {\rm pb}};U_{2  {\rm pb}}|X_1,X_2)\\
    R_{1  {\rm p}}+R_{1  {\rm pb}}' & \leq I(Y_1; X_1,U_{1  {\rm pb}}) \\
    R_{2  {\rm p}}+R_{2  {\rm pb}}' & \leq I(Y_2; X_2,U_{2  {\rm pb}})
    }
    {\label{eq:inner bound scheme 2 a}}
With
    \ean{
    R_{1  {\rm pb}}' &= I(X_2; U_{1  {\rm pb}}| X_1) + a_1, \ a_1\geq 0, \\
    R_{2  {\rm pb}}' &= I(X_1; U_{2  {\rm pb}}| X_2) + a_2, \ a_2\geq 0, \\
    a_1+a_2 &= I(U_{1  {\rm pb}};U_{2  {\rm pb}}|X_1,X_2),
    }
the achievable rate region in~\eqref{eq:inner bound scheme 2 a} becomes
    \ean{
   &\bigcup%{a_1+a_2 = I(U_{1  {\rm pb}};U_{2  {\rm pb}}|X_1,X_2)}
    \left\{\begin{array}{l}
    R_{1  {\rm p}} \leq I(Y_1; X_1,U_{1  {\rm pb}})-I(X_2; U_{1  {\rm pb}}| X_1) - a_1, \\
    R_{2  {\rm p}} \leq I(Y_2; X_2,U_{2  {\rm pb}})-I(X_1; U_{2  {\rm pb}}| X_2) - a_2, \\
    \end{array}\right.
%\\&=\left\{\begin{array}{ll}
%    R_{1  {\rm p}} &\leq I(Y_1; X_1,U_{1  {\rm pb}})-I(X_2; U_{1  {\rm pb}}| X_1), \\
%    R_{2  {\rm p}} &\leq I(Y_2; X_2,U_{2  {\rm pb}})-I(X_1; U_{2  {\rm pb}}| X_2), \\
%    R_{1  {\rm p}} + R_{2  {\rm p}} &\leq
%       I(Y_1; X_1,U_{1  {\rm pb}})-I(X_2; U_{1  {\rm pb}}| X_1)
%     + I(Y_2; X_2,U_{2  {\rm pb}})-I(X_1; U_{2  {\rm pb}}| X_2)\\
%     &- I(U_{1  {\rm pb}};U_{2  {\rm pb}}|X_1,X_2)
%    \end{array}\right.
    }
    where the union is over all $(a_1,a_2)\in\RR^2_+$ such that ${a_1+a_2 = I(U_{1  {\rm pb}};U_{2  {\rm pb}}|X_1,X_2)}$,
which coincides with~\eqref{eq:all private messages}.

Interestingly,
we point out that the Fourier-Motzkin elimination of the region
with only $(X_1,X_2,U_{1  {\rm pb}},U_{2  {\rm pb}})$ and %in Thm.~\ref{thm:general inner bound}
with $R_{1  {\rm pb}}\geq 0$ and $R_{2  {\rm pb}}\geq 0$
is the same as with $R_{1  {\rm pb}}= 0, R_{2  {\rm pb}}= 0$.
%without loss of generality when considering all private messages.
%for the sub-scheme in Cor.~\ref{cor:all private messages}.

\section{Proof of Corollary~\ref{cor:all public messages}}
\label{app:cor:all public messages}

Let
$R_1=R_{1  {\rm c}}$ and $R_2 = R_{2  {\rm c}}$, that is
\[
R_{1  {\rm p}}=R_{2  {\rm p}}=L_{1  {\rm pb}}=L_{2  {\rm pb}}=L_{0  {\rm cb}}=0.
\]
The region in~\eqref{eq:inner bound} with $U_{1  {\rm pb}}=U_{2  {\rm pb}}=\emptyset$ and
$I(X_1, X_2; U_{0  {\rm cb}}| U_{1  {\rm c}},U_{2  {\rm c}})=0$
becomes
\eas{
R_{1  {\rm c}}+R_{2  {\rm c}}   & \leq   I(Y_1; U_{1  {\rm c}},U_{2  {\rm c}},U_{0  {\rm cb}},X_1) \\
R_{1  {\rm c}}                  & \leq   I(Y_1; U_{1  {\rm c}},U_{0  {\rm cb}},X_1 | U_{2  {\rm c}})  \\
R_{2  {\rm c}}+R_{1  {\rm c}}   & \leq   I(Y_2; U_{1  {\rm c}},U_{2  {\rm c}},U_{0  {\rm cb}},X_2) \\
R_{2  {\rm c}}                  & \leq   I(Y_2; U_{2  {\rm c}},U_{0  {\rm cb}},X_1 | U_{1  {\rm c}})
}{\label{eq:inner bound scheme 2}}
which coincides with the region in~\eqref{eq:all public messages} by choosing
$X_1=U_{1  {\rm c}}, X_2=U_{2  {\rm c}}, X_c=U_{0  {\rm cb}}$.

\section{Proof of Corollary~\ref{cor:OnePrivateOnePublic}}
\label{app:cor:OnePrivateOnePublic}
Let
$R_1=R_{1  {\rm p}}+R_{1  {\rm pb}}$ and $R_2 = R_{2  {\rm c}}$, that is
\[
R_{1  {\rm c}}=R_{2  {\rm p}}=L_{2  {\rm pb}}=L_{0  {\rm cb}}=0.
\]
The region in~\eqref{eq:inner bound} with $U_{1  {\rm c}}=\emptyset, X_2=U_{2  {\rm c}},U_{0  {\rm cb}}=U_{2  {\rm c}}$ and $U_{2  {\rm pb}}=U_{0  {\rm c}}$
becomes
\eas{
R_{2  {\rm c}}+R_{1  {\rm p}}+R_{1  {\rm pb}} & \leq I(Y_1; X_1,U_{2  {\rm c}},U_{1  {\rm pb}}) \\
R_{2  {\rm c}}+\ \ \  \ \ \ \quad R_{1  {\rm pb}} & \leq I(Y_1; U_{2  {\rm c}},U_{1  {\rm pb}}|U_{1  {\rm p}}) \\
      R_{1  {\rm p}}+R_{1  {\rm pb}} & \leq I(Y_1; U_{1  {\rm p}},U_{1  {\rm pb}}|U_{2  {\rm c}}) \\
R_{1  {\rm pb}} & \leq I(Y_1; U_{1  {\rm pb}}| X_1,U_{2  {\rm c}}) \\
%\nonumber \\
R_{2 {\rm c}} &\leq I(Y_2 ; U_{2  {\rm c}}).
}{\label{eq:OnePrivateOnePublic rate-split}}
%The Fourier-Motzkin elimination of the region in~\eqref{eq:OnePrivateOnePublic rate-split} yields the region in~\eqref{eq:OnePrivateOnePublic}
which coincides with the region in~\eqref{eq:OnePrivateOnePublic} by choosing
$X_2=U_{2  {\rm c}}, X_c=U_{1  {\rm pb}}$.

%
%{\red DT: maybe add the major FME steps;
%the region in~\eqref{eq:OnePrivateOnePublic} is expressed as a function
%of the inputs while the one in~\eqref{eq:OnePrivateOnePublic rate-split}
%as a function of the aux.RVs. Add here the assignment of teh aux.RVs.}

%which can be FME as
%\eas{
%R_1 & \leq  I(Y_1 ; U_{1  {\rm c}}, U_{1  {\rm pb}}| U_{2  {\rm c}}) \\
%R_2 & \leq   I(Y_2; X_2) \\
%R_2 & \leq   I(Y_1; U_{1  {\rm pb}}, U_{2  {\rm c}} | U_{1  {\rm c}}) \\
%R_1+R_2  & \leq  I(Y_1 ; U_{2  {\rm c}}, U_{1  {\rm p}}, U_{1  {\rm pb}})
%}

%{\red DT: SHOW HERE THAT THE R2-BOUND THAT DEPENDS ON Y1 CAN BE DROPPED.}
%

\section{Proof of Corollary~\ref{cor:OnePrivateOnePublic2}}
\label{app:cor:OnePrivateOnePublic2}
Let
$R_1=R_{1  {\rm c}}+R_{1  {\rm pb}}$ and $R_2 = R_{2  {\rm c}}$, that is
\[
R_{1  {\rm p}}=R_{2  {\rm p}}=L_{2    {\rm pb}}=L_{2  {\rm pb}}=L_{0  {\rm cb}}=0.
\]
The region in~\eqref{eq:inner bound} with $X_1=U_{1  {\rm c}}, X_2=U_{2  {\rm c}},U_{0  {\rm cb}}=U_{2  {\rm c}}$ and $U_{2  {\rm pb}}=U_{0  {\rm c}}$
becomes
\eas{
R_{1  {\rm c}}+R_{2  {\rm c}}+R_{1  {\rm pb}} &\leq  I(Y_1; U_{1  {\rm c}},U_{2  {\rm c}},U_{1  {\rm pb}})\\
R_{2  {\rm c}}+R_{1  {\rm pb}} &\leq  I(Y_1; U_{2  {\rm c}},U_{1  {\rm pb}}|U_{1  {\rm c}})\\
R_{1  {\rm c}}\quad \ \ \ \ \ \ +R_{1  {\rm pb}} &\leq  I(Y_1; U_{1  {\rm c}},U_{1  {\rm pb}}|U_{2  {\rm c}})\\
 R_{1  {\rm pb}} &\leq  I(Y_1; U_{1  {\rm pb}}| U_{1  {\rm c}},U_{2  {\rm c}})\\
% \nonumber \\
R_{1  {\rm c}}+R_{2  {\rm c}}&\leq  I(Y_2; U_{1  {\rm c}},U_{2  {\rm c}})\\
R_{2  {\rm c}} &\leq  I(Y_2; U_{2  {\rm c}}| U_{1  {\rm c}})\\
R_{1  {\rm c}}\ \ \ \ \ \quad   &\leq  I(Y_2; U_{1  {\rm c}}| U_{2  {\rm c}}).
}{\label{eq:OnePrivateOnePublic2 rate-split2}}
%The Fourier-Motzkin elimination of the region in~\eqref{eq:OnePrivateOnePublic2 rate-split2} yields the region of~\eqref{eq:OnePrivateOnePublic2}
which coincides with the region in~\eqref{eq:OnePrivateOnePublic2} by choosing $X_1=U_{1  {\rm c}}, X_2=U_{2  {\rm c}}, X_c=U_{1  {\rm pb}}$ .

%{\red DT: maybe add the major FME steps;
%the region in~\eqref{eq:OnePrivateOnePublic} is expressed as a function
%of the inputs while the one in~\eqref{eq:OnePrivateOnePublic rate-split}
%as a function of the aux.RVs. Add here the assignment of teh aux.RVs.}

%\eas{
%R_1 & \leq &  I(Y_1; U_{1  {\rm c}},U_{1  {\rm pb}}|U_{2  {\rm c}}) \\
%R_1 & \leq &  I(Y_1; U_{1  {\rm pb}}| U_{1  {\rm c}},U_{2  {\rm c}})+I(Y_2; U_{1  {\rm c}}| U_{2  {\rm c}}) \\
%\nonumber \\
%R_2 & \leq &  I(Y_1; U_{2  {\rm c}},U_{1  {\rm pb}}|U_{1  {\rm c}})\\
%R_2 & \leq &  I(Y_2; U_{2  {\rm c}}| U_{1  {\rm c}}) \\
%\nonumber \\
%R_1+R_2 & \leq &  I(Y_1; U_{1  {\rm c}},U_{2  {\rm c}},U_{1  {\rm pb}}) \\
%R_1+R_2 & \leq &  I(Y_1; U_{2  {\rm c}},U_{1  {\rm pb}}|U_{1  {\rm c}})+I(Y_2; U_{1  {\rm c}}| U_{2  {\rm c}}) \\
%R_1+R_2 & \leq &  I(Y_1; U_{1  {\rm pb}}|U_{1  {\rm c}},U_{2  {\rm c}})+ I(Y_2; U_{2  {\rm c}},U_{2  {\rm c}})\\
%\nonumber \\
%R_1 + 2 R_2 & \leq & I(Y_1; U_{2  {\rm c}},U_{1  {\rm pb}}|U_{1  {\rm c}})+I(Y_2; U_{1  {\rm c}},U_{2  {\rm c}})
%}
%

\section{The IFC-CR in standard form}
\label{app:standard form}

A general IFC-CR is expressed as
\eas{
\Yt_1=\htt_{11} \Xt_1+\htt_{1c} \Xt_c+\htt_{12} \Xt_2+\Zt_1 ,\\
\Yt_2=\htt_{22} \Xt_1+\htt_{2c} \Xt_c+\htt_{21} \Xt_1+\Zt_2 ,
}{\label{eq:general IFC-CR}}
for $\htt_i, \ i \in \{11,22,1c,2c,12,21\}$, $\EE[ |\Xt_j|^2]\leq \Pt_j, \ j \in \{1,2,c\}$ and  $\EE[ |\Zt_k|^2]=\sgs_k$,
$k \in \{1,2\}$.
Assuming without loss of generality that all the entries of $(\Pt_1,\Pt_2,\Pt_c,\sgs_1,\sgs_2)$ are strictly positive,%
\footnote{
If $\Pt_1=0,\Pt_2=0,\Pt_c=0$ the channel capacity is trivially $R_1=R_2=0$.
If $\Pt_1=0,\Pt_2=0,\Pt_c>0$ the channel is equivalent to a Gaussian BC with input $X_c$ whose capacity is known~\cite{bergmans1973random}.
If $\Pt_1=0,\Pt_2>0,\Pt_c=0$, and similarly
if $\Pt_1>0,\Pt_2=0,\Pt_c=0$,
 the channel is a Gaussian point-to-point channel whose capacity is known~\cite{shannon48}.
If $\Pt_1=0,\Pt_2>0,\Pt_c>0$, and similarly
if $\Pt_1>0,\Pt_2=0,\Pt_c>0$,
 the channel is equivalent to a Gaussian CIFC whose capacity is known to within 1~bit~\cite{RTDjournal2}.
If $\Pt_1>0,\Pt_2>0,\Pt_c=0$,
 the channel is a Gaussian IFC whose capacity is known to within 1~bit~\cite{etkin_tse_wang}.
If either of the noise variances is zero, the corresponding channel has infinite capacity,
which does not have any physical meaning.
}
consider now the transformation
\ean{
& Y_1 =\f {\Yt_1} {\sigma_1} \eu^{-j \angle \htt_{1  {\rm c}} }
&&Y_2 =\f {\Yt_2} {\sigma_2} \eu^{-j \angle \htt_{2c} }
\\
& X_1 =\f {\Xt_1} { \sqrt{\Pt_1} } \eu^{-j (\angle \htt_{11}+\angle \htt_{1c}) }
&&X_2 =\f {\Xt_2} { \sqrt{\Pt_2} } \eu^{-j (\angle \htt_{22}+\angle \htt_{2c}) }
\\
&X_c =\f {\Xt_c}  {\sqrt{\Pt_c}}
&&
\\
& |h_{11}| =\f{\sqrt{\Pt_1} |\htt_{11}|} {\sigma_1}
&&|h_{22}| =\f{\sqrt{\Pt_2} |\htt_{22}|} {\sigma_2}
\\
& |h_{1c}| =\f{\sqrt{\Pt_c} |\htt_{1c}|} {\sigma_1}
&&|h_{2c}| =\f{\sqrt{\Pt_c} |\htt_{2c}|} {\sigma_2}
\\
& |h_{12}| =\f{\sqrt{\Pt_2} \htt_{12} } {\sigma_1} \eu^{-j \angle \htt_{11}}
&&|h_{21}| =\f{\sqrt{\Pt_1} \htt_{21} } {\sigma_2} \eu^{-j \angle \htt_{22}}.
%\label{eq:transformation standard form}
}
Since the above transformation is invertible, the channel in~\eqref{eq:general IFC-CR} is equivalent to the channel in~\eqref{eq:standard form IFC-CR}.

\section{Proof of Theorem~\ref{thm:strong int. outer bound Gaussian}}
\label{app:proof eq:strong int. conditions Gaussian}

Given the ``Gaussian maximizes entropy'' property~\cite{ThomasCoverBook}
we have that the union over all the
distributions in~\eqref{eq:factorization thm:general outer ITW dublin}
of the region in~\eqref{eq:cor:strong int outer bound} is equal
to the union over all distributions with $Q=\emptyset$ and
$[X_1, X_2, X_c]$ zero-mean proper-complex Gaussian %random vectors
with covariance matrix
\ea{
\cov (X_1,X_2,X_c)=
\begin{pmatrix}
1  & 0 &  \beta_1  \\
0  & 1 &  \beta_2  \\
\beta_1^* & \beta_2^* & 1
\end{pmatrix}
:= \Sm,
\label{eq:covariance matrix Gaussian IFC-CR}
}
for  $(\beta_1,\beta_2)\in \CC^2$ such that $|\beta_1|^2+|\beta_2|^2 \leq 1$.
%When considering the parametrization in~\eqref{eq:covariance matrix Gaussian IFC-CR} for the outer bound region in~\eqref{eq:cor:strong int outer bound}, we note that the choice $ |\beta_1|^2+|\beta_2|^2=1 $ yields the largest region.
%
%%%With~\eqref{eq:covariance matrix Gaussian IFC-CR} we can rewrite the condition in~\eqref{eq:strong int. cond. at Rx1} as
%%%\ea{
%%%\max_{|\beta_1|^2+|\beta_2|^2 \leq 1} \Big\{
%%%&\big||h_{22}|+\beta_2^*|h_{2c}|\big|^2  -\big| h_{12} +\beta_2^*|h_{2c}|\big|^2
%%%\nonumber \\
%%%&+ \lb |h_{2c}|^2-|h_{2c}|^2 \rb \big(1-|\beta_1|^2- |\beta_2|^2\big) \Big\} \leq 0
%%%\label{eq:strong int. cond. at Rx1 Gaussian}
%%%}
%%%The solution of the maximization problem in~\eqref{eq:strong int. cond. at Rx1 Gaussian}
%%%is given by~\eqref{eq:max solution strong int} as shown next.
%
With~\eqref{eq:covariance matrix Gaussian IFC-CR} we write:
\[
X_c = \be_{1}^* X_1 + \be_{2}^* X_2 +\sqrt{1-|\be_{1}|^2+ |\be_{2}|^2} X_{c, in.},
%|\be_{1}|^2+ |\be_{2}|^2\leq 1.
\]
for $X_1,X_2,X_{c, in.}$ iid $\Nc(0,1)$, from which
\ean{
Y_j
%= h_{1j} X_1 + h_{2j} X_2 + h_{jc}(\be_{1}^* X_1 + \be_{2}^* X_2 +\sqrt{1-|\be_{1}|^2+ |\be_{2}|^2} X_{c, in.})
&= \big[h_{j1}+ \be_{1}^*|h_{jc}|\big] X_1
 + \big[h_{j2}+ \be_{2}^*|h_{jc}|\big] X_2
\\&+                     |h_{jc}|\sqrt{1-|\be_{1}|^2+ |\be_{2}|^2} X_{c, in.} +Z_{j}, \ j\in\{1,2\}.
}
and thus, conditioned on $X_1$, we have that $Y_j$ is distributed as
\[
%Y_j|X_1
%\sim
  \big[h_{j2}+ \be_{2}^*|h_{jc}|\big] X_2
+                       |h_{jc}|\sqrt{1-|\be_{1}|^2+ |\be_{2}|^2} X_{c, in.} +Z_{j}, \ j\in\{1,2\}.
\]
%from which
%\[
%I(X_2,X_c; Y_j|X_1) = \Cc ( | h_{j2}+ \be_{2}^*|h_{jc}| |^2 + |h_{jc}|^2(1-|\be_{1}|^2+ |\be_{2}|^2) ).
%\]
Since the condition in~\eqref{eq:strong int. cond. at Rx1} must hold
for all $(\beta_1,\beta_2)\in \CC^2$ such that $|\beta_1|^2+|\beta_2|^2 \leq 1$,
we obtain
\ean{
  &\text{for all Gaussian inputs}  I(Y_2 ; X_2 , X_c |  X_1 ) \leq I(Y_1 ; X_2 , X_c | X_1 )
\\&\Longleftrightarrow
   \forall (\beta_1,\beta_2)\in \CC^2 : |\be_{1}|^2+ |\be_{2}|^2  \leq 1
\\&\quad     \Cc ( ||h_{22}|+ \be_{2}^*|h_{2c}| |^2 + |h_{2c}|^2(1-|\be_{1}|^2+ |\be_{2}|^2) )
\\&\quad \leq\Cc ( | h_{12} + \be_{2}^*|h_{1c}| |^2 + |h_{1c}|^2(1-|\be_{1}|^2+ |\be_{2}|^2) )
%\\&\forall~\eqref{eq:be conditions 1}:
%   \quad I(|h_{22}| X_2+ |h_{2c}| X_c + h_{21} X_1 + Z_2; X_2 , X_c | X_1 )
%   \leq  I(|h_{11}| X_1+ |h_{1c}| X_c + h_{12} X_2 + Z_1; X_2 , X_c | X_1 )
%\\&\Longleftrightarrow
%\\&\forall~\eqref{eq:be conditions 1}:
%   \quad I( |h_{2c}| X_c' + |h_{22}| X_2 + Z_2; X_2 , X_c' )
%   \leq  I( |h_{1c}| X_c' +  h_{12}  X_2 + Z_1; X_2 , X_c' )
\\&\Longleftrightarrow
   \forall (\beta_1,\beta_2)\in \CC^2 : |\be_{1}|^2+ |\be_{2}|^2  \leq 1
\\&\quad |h_{2c}|^2 (1-|\be_{1}|^2) + |h_{22}|^2 +2|h_{2c}||h_{22}|\Re\{\be_{2}\}
\\&\quad \leq  |h_{1c}|^2 (1-|\be_{1}|^2) + |h_{12}|^2 +2|h_{1c}|\Re\{h_{12}\be_{2}\}
\\&\Longleftrightarrow
 \max_{|\be_{1}|^2+ |\be_{2}|^2  \leq 1}\{(|h_{2c}|^2-|h_{1c}|^2) (1-|\be_{1}|^2)
\\&\quad + 2\Re\{(|h_{2c}||h_{22}|-|h_{1c}|h_{12})\be_{2}\} \}
   \leq  |h_{12}|^2 -|h_{22}|^2
\\&\Longleftrightarrow
 \max_{|\be_{1}|^2  \leq 1}
 \{(|h_{2c}|^2-|h_{1c}|^2) (1-|\be_{1}|^2)
\\&\quad + 2\Big||h_{2c}||h_{22}|-|h_{1c}|h_{12}\Big|\sqrt{1-|\be_{1}|^2}\}
   \leq  |h_{12}|^2 -|h_{22}|^2,
}
where in the last step the optimal $\be_{2}$ %\in \CC: \ |\be_{2}|^2  \leq 1-|\be_{1}|^2$
is
\[
\be_{2} = \eu^{-\jj \measuredangle{(|h_{2c}||h_{22}|-|h_{1c}|h_{12})}} \sqrt{1-|\be_{1}|^2}.
\]
Let now
\ean{
  &\sqrt{1-|\be_{1}|^2}=x,
\\&|h_{2c}|^2-|h_{1c}|^2=a,
\\&\Big||h_{2c}||h_{22}|-|h_{1c}|h_{12}\Big|=|b|.
}
The quadratic function
$f(x)=a x^2 +2|b| x$ is non-decreasing in $x\in[0,1]$ if $a x + |b|\geq 0$.
If $a\geq 0$: $+|a| x + |b|\geq 0$ for all $x\in[0,1]$ hence $x=1$ is optimal.
Else (i.e., if $a< 0$):    $-|a| x + |b|\geq 0$ for $x\leq |b|/|a|$. Thus,
if $a< 0, |b|/|a|\leq 1$:  $x= |b|/|a|\in[0,1]$ is optimal, and
if $a< 0, |b|/|a|> 1$:     $x= 1$ is optimal. This shows the optimal
$\be_2$ is the one given in~\eqref{eq:max solution strong int}.

\section{Proof of Theorem~\ref{th:Capacity in the ``very strong interference'' regime for the G-IFC-CR}}
\label{app:proof eq:VSI}
With the parameterization in~\eqref{eq:covariance matrix Gaussian IFC-CR} the condition
in~\eqref{eq:very strong int. condition 1 extra} can be rewritten as
\ean{
  &\text{for all Gaussian inputs}:  I(Y_1 ; X_1, X_2, X_c ) \leq I(Y_2 ; X_1, X_2, X_c )
\\&\Longleftrightarrow
   \forall (\beta_1,\beta_2)\in \CC^2 : |\be_{1}|^2+ |\be_{2}|^2  \leq 1
\\&\quad  \Cc ( ||h_{11}|+ \be_{1}^*|h_{1c}||^2+| h_{12} + \be_{2}^*|h_{1c}| |^2
\\&\quad + |h_{1c}|^2(1-|\be_{1}|^2+ |\be_{2}|^2) )
\\&\quad \leq  \Cc ( | h_{21} + \be_{1}^*|h_{2c}||^2+||h_{22}|+ \be_{2}^*|h_{2c}| |^2
\\&\quad + |h_{2c}|^2(1-|\be_{1}|^2+ |\be_{2}|^2) )
\\&\Longleftrightarrow
  (|h_{11}|^2+|h_{1c}|^2+|h_{12}|^2)-(|h_{21}|^2+|h_{2c}|^2+|h_{22}|^2)
\\&\quad
+ \max_{|\beta_1|^2+|\beta_2|^2 \leq 1}
        2\Re (    \beta_1 \lb |h_{1c}||h_{11}|-|h_{2c}| h_{21} \rb
\\&\quad         +\beta_2 \lb |h_{1c}| h_{12} -|h_{2c}||h_{22}|\rb )
         \leq 0
\\&\Longleftrightarrow
  (|h_{11}|^2+|h_{1c}|^2+|h_{12}|^2)-(|h_{21}|^2+|h_{2c}|^2+|h_{22}|^2)
\\&\quad + 2\max_{|\be_{1}|^2+|\be_{2}|^2 \leq 1}
         \{\Big|\be_{1}\Big| \ \Big||h_{1c}||h_{11}|-|h_{2c}| h_{21} \Big|
\\&\quad  +\Big|\be_{2}\Big| \ \Big||h_{1c}| h_{12} -|h_{2c}||h_{22}|\Big| \}
  \leq 0
\\&\Longleftrightarrow
  (|h_{11}|^2+|h_{1c}|^2+|h_{12}|^2)-(|h_{21}|^2+|h_{2c}|^2+|h_{22}|^2)
\\&\quad + 2
    \sqrt{ \Big||h_{1c}||h_{11}|-|h_{2c}| h_{21} \Big|^2
          +\Big||h_{1c}| h_{12} -|h_{2c}||h_{22}|\Big|^2}
  \leq 0.
}
%$\forall \ |\beta_1|^2+|\beta_2|^2 = 1$, which equivalently can be rewritten as
%\ean{
%\max_{|\beta_1|^2+|\beta_2|^2 = 1} \big||h_{22}|+\beta_2^*|h_{2c}|\big|^2 + |h_{2c}|^2\big(1-|\beta_1|^2- |\beta_2|^2\big) -
%\big| h_{12} +\beta_2^*|h_{2c}|\big|^2 + |h_{2c}|^2\big(1-|\beta_1|^2- |\beta_2|^2\big) \leq 0.
%}

We next show  that, given $A\geq 0$ and $B\geq0$:
\[
\sqrt{A^2+B^2}
=\max\{x A + y B\}
\quad
s.t.
\quad
x^2+y^2\leq 1
\]
%is
%\[
%\sqrt{A^2+B^2}
%\]
%achieved by
%\[
%x^2+y^2=1.
%\]
%%
Indeed, for $t\geq 0$ let the Lagrangian be:
\[
L = x A + y B - 2/t(x^2+y^2-1)
\]
then at the optimal point
\ean{
dL/dx &= A - x/t =0 \Longleftrightarrow x=A t \\
dL/dy &= B - y/t =0 \Longleftrightarrow y=B t
}
hence the optimal Lagrangian  multiplier is
\[
x^2+y^2=(A^2+B^2) t^2 = 1
\Longleftrightarrow
t = \frac{1}{\sqrt{A^2+B^2}}.
\]
%and the optimal values are
%\ean{
%x=\frac{A}{\sqrt{A^2+B^2}} \\
%y=\frac{B}{\sqrt{A^2+B^2}}
%}

%----------------------
\section{``Weak interference'' outer bound for the IFC-CR }
\label{app:weak interference outer bound for the degraded Gaussian IFC-CR}

We now evaluate the ``weak interference at Rx~1'' outer bound in Cor.~\ref{cor:weak int outer bound} for the
channel model in~\eqref{eq:degraded Gaussian IFC-CR outputs}. We proceed as in~\cite{WuDegradedMessageSet}.
We must evaluate the region
\ean{
 R_1 & \leq  I(Y_1 ; X_1 , X_c|  X_2, U)
\\     & = h(Y_1- h_{12} X_2|  X_2, U) - \log(\pi \eu),
\\ R_2 & \leq  I(Y_2 ; X_2, U)             = h(Y_2) - h(Y_2|  X_2, U)
\\     & \leq \log({\rm Var}\big[Y_2])-[h(Y_2-|h_{22}|X_2|  X_2, U) - \log(\pi \eu)],
}
for all distribution that factors as in~\eqref{eq:factorization condition cor:weak int outer bound}.
As for the El~Gamal's converse for the degraded BC we have
\ean{
&h(Z_1)=h(Y_1 | X_1 , X_2 , X_c) \leq h(Y_1 | X_2 U) \leq h(Y_1 - h_{12}X_2 | X_2) \\
 &\Longleftrightarrow
 \log( 1 ) \leq h(Y_1 | X_2 U)- \log(\pi \eu)
\\&\quad \leq  \log \lb 1+ {\rm Var}\big[ X_{\rm eq}  \big| X_2\big] \rb,
}
where $X_{\rm eq} := |h_{11}| X_1 + |h_{1c}| X_c$ as defined in~\eqref{eq:degraded Gaussian IFC-CR markov chain}.

Hence there must exist an $\al \in [0,1]$ such that
\ean{
 &h(Y_1 | X_2 U) -  \log(\pi \eu)
\\&=\log \lb 1+\al {\rm Var}\big[ X_{\rm eq} \big| X_2\big] \rb.
}
Moreover, since conditioned on $X_2$
the channel in~\eqref{eq:degraded Gaussian IFC-CR outputs} is degraded,
the (scalar) Entropy Power Inequality (EPI)~\cite{stam1959EntropyPower}
for complex-valued RVs grants
\ean{
 2^{h(Y_2| X_2,  U)}
 &=2^{h(  |\rho|X_{\rm eq}  + \sqrt{1-|\rho|^2} Z_0 | X_2,  U) }
\\&\geq |\rho|^2 2^{h(Y_1| X_2,  U)}
  + (1-|\rho|^2) 2^{h(Z_0)}
}
which implies
\ean{
 &h(Y_2| X_2,  U)- \log(\pi \eu)
 \geq \log \lb 1+\al {\rm Var}\big[ X_{\rm eq} \big| X_2\big] \rb.
}
With this we obtain
\ean{
  R_1 & \leq  \Ccal \lb \al {\rm Var}\big[ X_{\rm eq} \big| X_2\big] \rb,
\\ R_2 & \leq  \Ccal \lb {\rm Var}\big[ |\rho|X_{\rm eq} + |h_{22}| X_2\big]\rb
- \Ccal \lb \al|\rho|^2 {\rm Var}\big[ X_{\rm eq} \big| X_2\big]  \rb.
}
Moreover, from~\eqref{eq:gen sato r2} we also have

\ean{
R_2 & \leq I(Y_2 ; X_2, X_c | X_1 Q) \\
  & \leq \log \lb 1+{\rm Var}\big[Y_2-|\rho||h_{11}| X_1 \ |\ X_1\big]\rb.
%   \\
%   & = \Ccal \lb {\rm Var}\big[|h_{21}||\al_c| X_c + |h_{22}| X_2\ \big|\ X_1\big] \rb.
}
%
%{\red DT: WE MUST EXPLAIN WHY WE NEED TO INTRODUCE A RATE BOUND THAT WAS NOT IN THE
%WEAK INTERF COROLLARY. ALSO PLEASE UPDATE THE FOLLOWING DERIVATIONS BY USING RHO INSTEAD OF ALPHAC.}
%
By considering the input covariance $\Sm$ defined in~\eqref{eq:covariance matrix Gaussian IFC-CR},
for a fixed $(\be_1,\be_2) : |\be_1|^2+|\be_2|^2\leq1$ we obtain
\ean{
& {\rm Var}\big[X_1 + |\rho| X_c \ \big|\ X_2\big]
\\& \lag =     1+|\rho|^2(1-|\be_2|^2)+2|\rho|\Re\{\be_1\}
\\& \lag \leq  1+|\rho|^2 \ |\be_1|^2 +2|\rho|\Re\{\be_1\}
\\& \lag =\big|1+|\rho| \be_1\big|^2
\\
& {\rm Var}\big[|h_{21}|(X_1 + |\rho| X_c) + |h_{22}| X_2\big]
\\& \lag  =    |h_{21}|^2(1+|\rho|^2)+ |h_{22}|^2 + 2|h_{21}||h_{22}||\rho|\Re\{\be_2\}
\\& \lag  \leq |h_{21}|^2(1+|\rho|^2)+ |h_{22}|^2 + 2|h_{21}| \ |h_{22}|\ |\rho|\ |\be_2|
\\& \lag  \leq |h_{21}|^2(1+|\rho|^2)+ |h_{22}|^2 + 2|h_{21}| \ |h_{22}|\ |\rho|\ \sqrt{1-|\be_1|^2}
\\
& {\rm Var}\big[|h_{21}||\rho| X_c + |h_{22}| X_2\ \big|\ X_1\big]
\\& \lag =     |h_{21}|^2 |\rho|^2(1-|\be_1|^2)+|h_{22}|^2+2|h_{21}||\rho||h_{22}|\Re\{\be_2\}
\\& \lag \leq  |h_{21}|^2  |\rho|^2(1-|\be_1|^2)+|h_{22}|^2+2|h_{21}| \ |\rho| \ |h_{22}| \ |\be_2|
\\& \lag \leq  |h_{21}|^2  |\rho|^2(1-|\be_1|^2)+|h_{22}|^2+2|h_{21}| \ |\rho| \ |h_{22}| \ \sqrt{1-|\be_1|^2}
\\& \lag =    (|h_{21}||\rho|\sqrt{1-|\be_1|^2}+|h_{22}|)^2
}
Note that the above shows that we can only consider $|\be_1|^2+|\be_2|^2=1$ without loss of generality.
With this, we obtain the region in~\eqref{eq:weak int outer bound expression degraded Gaussian IFC-CR}.

\end{appendices}

\end{document}